\documentclass[twocolumn,9pt]{extarticle}
\usepackage[letterpaper,top=0.95in,bottom=0.95in,left=0.75in,right=0.75in,columnsep=0.25in,headsep=14pt,footskip=24pt]{geometry}
\usepackage[T1]{fontenc}
\usepackage{mathptmx}
\usepackage[scaled=0.92]{helvet}
\usepackage{courier}
\usepackage{cite}
\usepackage{graphicx}
\usepackage[labelfont=bf,textfont=it,font=small,labelsep=colon]{caption}
\usepackage{flushend}
\usepackage{fancyhdr}
\usepackage{titlesec}
\usepackage{enumitem}
\setlist{itemsep=2pt plus 1pt,parsep=0pt,topsep=4pt plus 1pt,partopsep=0pt}
\setlist[description]{style=unboxed,leftmargin=1em,labelindent=0pt,labelsep=0.4em,itemsep=2pt plus 1pt,parsep=0pt,topsep=4pt plus 1pt}
\titleformat{\section}{\large\bfseries\raggedright\hyphenpenalty=10000\exhyphenpenalty=10000}{\thesection.}{0.6em}{}
\titlespacing*{\section}{0pt}{14pt plus 3pt minus 2pt}{6pt plus 1pt}
\titleformat{\subsection}{\normalsize\bfseries\raggedright\hyphenpenalty=10000\exhyphenpenalty=10000}{\thesubsection.}{0.6em}{}
\titlespacing*{\subsection}{0pt}{10pt plus 2pt minus 1pt}{4pt plus 1pt}
\titleformat{\subsubsection}{\normalsize\bfseries\itshape\raggedright\hyphenpenalty=10000\exhyphenpenalty=10000}{\thesubsubsection.}{0.6em}{}
\titlespacing*{\subsubsection}{0pt}{8pt plus 2pt minus 1pt}{3pt}
\titleformat{\paragraph}[runin]{\normalsize\bfseries}{\theparagraph}{0.5em}{}
\titlespacing*{\paragraph}{0pt}{7pt plus 2pt minus 1pt}{0.6em plus 0.5em minus 0.1em}
\usepackage[colorlinks,linkcolor=blue,citecolor=blue,urlcolor=blue,breaklinks,
  pdftitle={Same Physics, Different Gradients: A State-of-the-Art Report on Differentiable Physics Simulation},
  pdfauthor={Lei Shu, Zizhou Huang, Yunuo Chen, Yifei Li, Tao Du, Minchen Li, Tuur Stuyck, Miles Macklin, Kui Wu, Bo Zhu, Yin Yang, Ming C. Lin, Alberta Longhini, Rika Antonova, Leonidas Guibas, Gordon Wetzstein, Jeannette Bohg, Chenfanfu Jiang}]{hyperref}
\usepackage{amsmath,amssymb}
\usepackage{booktabs}
\usepackage{multirow}
\usepackage{array}
\usepackage{tabularx}
\usepackage{xcolor}
\usepackage{tikz}
\usetikzlibrary{arrows.meta,positioning,fit,backgrounds,calc,shapes.geometric}
\usepackage[normalem]{ulem}
\newcommand{\R}{\mathbb{R}}
\newcommand{\x}{\mathbf{x}}
\newcommand{\vv}{\mathbf{v}}

\newcommand{\bs}{\mathbf{s}}
\newcommand{\bth}{\boldsymbol{\theta}}

\newcommand{\gres}{\mathbf{g}}
\newcommand{\Loss}{\mathcal{L}}
\newcommand{\jac}[2]{\frac{\partial #1}{\partial #2}}
\newcommand{\adj}[1]{\bar{#1}}
\newcommand{\transp}{^{\!\top}}

\newcommand{\etal}{et~al.}

\definecolor{levOne}{HTML}{2A78D6}%
\definecolor{levTwo}{HTML}{EB6834}%
\definecolor{levThree}{HTML}{199E70}%
\definecolor{inkA}{HTML}{0B0B0B}
\definecolor{inkB}{HTML}{52514E}
\definecolor{inkC}{HTML}{8A8983}
\definecolor{rule}{HTML}{E6E5E1}
\definecolor{seqA}{HTML}{86B6EF}
\definecolor{seqB}{HTML}{2A78D6}
\definecolor{seqC}{HTML}{104281}

\newcommand{\cellragged}{\rightskip=0pt plus 2em\relax\spaceskip=.3333em\relax\xspaceskip=.5em\relax}
\newcolumntype{L}[1]{>{\cellragged\arraybackslash}p{#1}}
\newcolumntype{C}[1]{>{\centering\arraybackslash}p{#1}}

\makeatletter
\g@addto@macro\UrlBreaks{\do0\do1\do2\do3\do4\do5\do6\do7\do8\do9}
\makeatother
\fancypagestyle{firstpage}{\fancyhf{}\fancyfoot[L]{\small Preprint, \today.}}

\makeatletter
\AddToHookNext{shipout/after}{\global\let\@textbottom\relax\global\let\@texttop\relax}
\makeatother
\makeatletter
\renewcommand\small{\@setfontsize\small\@viiipt{9.5}%
  \abovedisplayskip 5\p@ \@plus2\p@ \@minus2\p@
  \abovedisplayshortskip \z@ \@plus1\p@
  \belowdisplayshortskip 3\p@ \@plus\p@ \@minus2\p@
  \def\@listi{\leftmargin\leftmargini \topsep 3\p@ \@plus\p@ \@minus\p@
    \parsep 2\p@ \@plus\p@ \@minus\p@ \itemsep \parsep}%
  \belowdisplayskip \abovedisplayskip}
\let\footnotesize\small
\renewcommand\@openbib@code{\setlength{\itemsep}{0pt plus 0.6pt}\setlength{\parsep}{0pt}\clubpenalty=3000 \widowpenalty=3000 }
\makeatother
\begin{document}

\twocolumn[{%
\vspace*{-16pt}
\begin{center}
{\LARGE\bfseries Same Physics, Different Gradients:\\[3pt]
 A State-of-the-Art Report on Differentiable Physics Simulation\par}
\vspace{14pt}
{\normalsize Lei Shu$^{1}$, Zizhou Huang$^{2}$, Yunuo Chen$^{3}$, Yifei Li$^{4}$, Tao Du$^{5}$,
Minchen Li$^{6}$, Tuur Stuyck$^{7}$, Miles Macklin$^{7}$, Kui Wu$^{8}$, Bo Zhu$^{9}$,
Yin Yang$^{10}$, Ming C. Lin$^{11}$, Alberta Longhini$^{1}$, Rika Antonova$^{12}$,
Leonidas Guibas$^{1}$, Gordon Wetzstein$^{1}$, Jeannette Bohg$^{1}$ and Chenfanfu Jiang$^{3}$\par}
\vspace{6pt}
{\footnotesize\centering \mbox{$^{1}$Stanford University} \quad \mbox{$^{2}$Courant Institute, New York University} \quad \mbox{$^{3}$University of California, Los Angeles} \quad \mbox{$^{4}$Massachusetts Institute of Technology} \\
\mbox{$^{5}$Tsinghua University} \quad \mbox{$^{6}$Carnegie Mellon University} \quad \mbox{$^{7}$NVIDIA} \quad \mbox{$^{8}$LightSpeed Studios, Tencent America} \\
\mbox{$^{9}$Georgia Institute of Technology} \quad \mbox{$^{10}$University of Utah} \quad \mbox{$^{11}$University of Maryland, College Park} \quad \mbox{$^{12}$University of Cambridge}\par}
\end{center}
\vspace{10pt}
\begin{minipage}{\textwidth}
\centering
{\begin{minipage}{\textwidth}
\begin{minipage}[t]{0.2336\linewidth}\includegraphics[width=\linewidth]{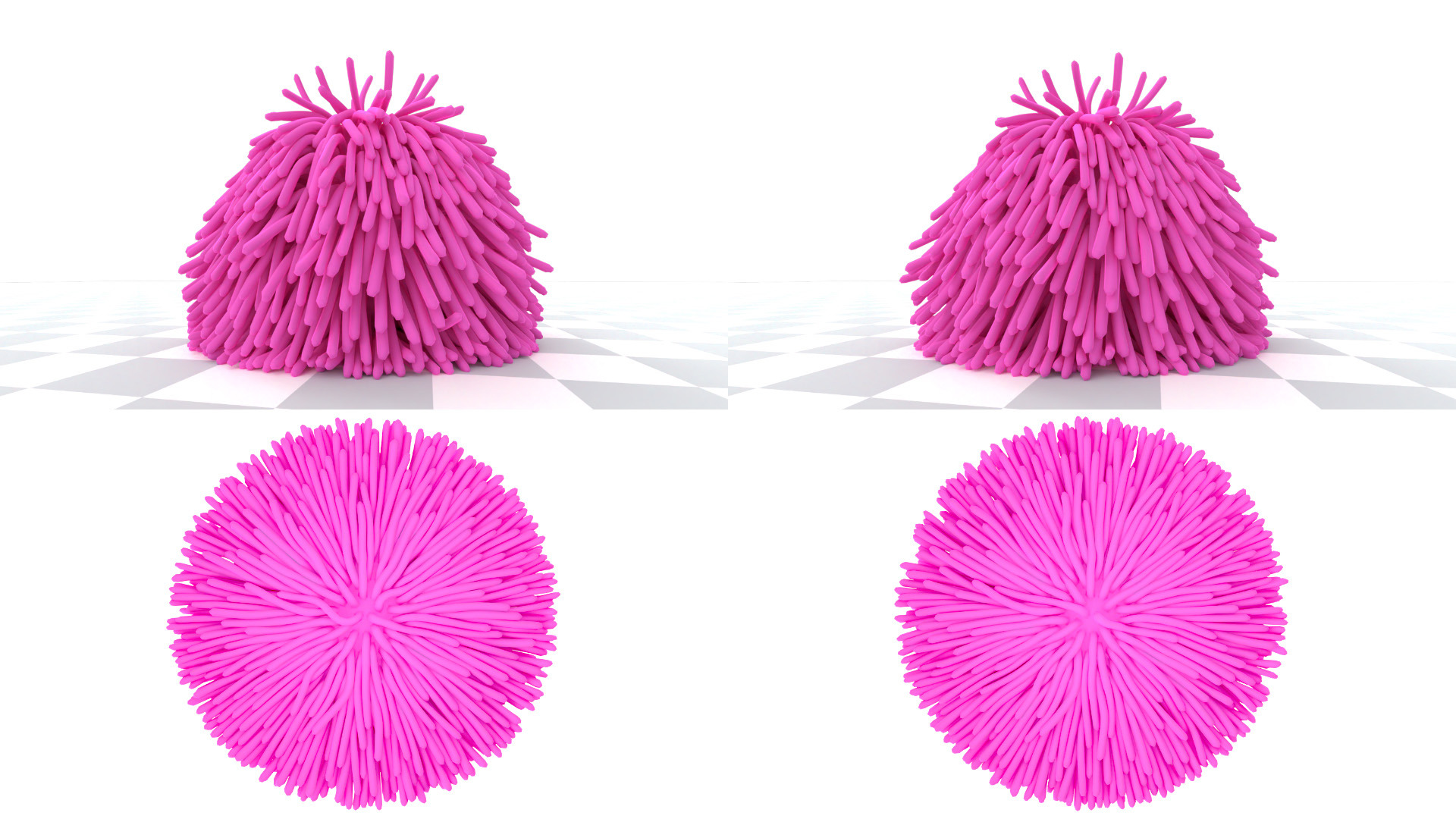}\end{minipage}%
\begin{minipage}[t]{0.1958\linewidth}\includegraphics[width=\linewidth]{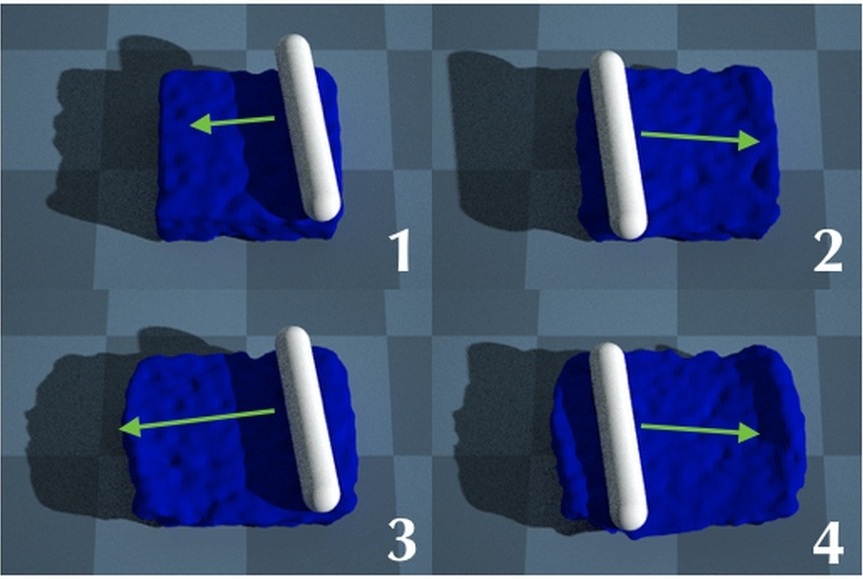}\end{minipage}%
\begin{minipage}[t]{0.4392\linewidth}\includegraphics[width=\linewidth]{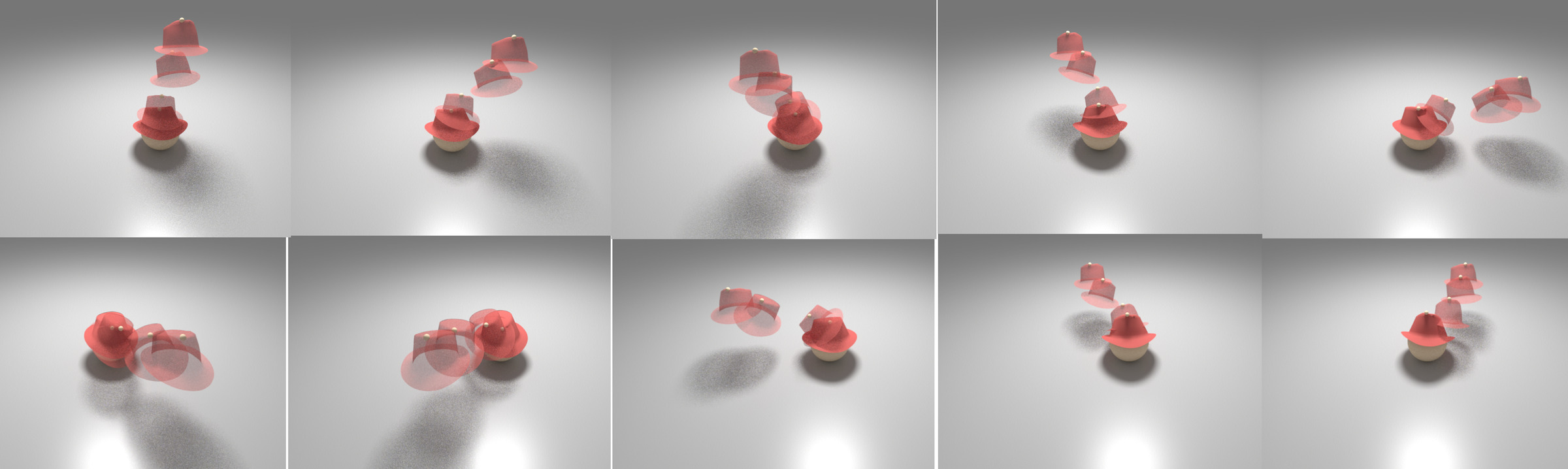}\end{minipage}%
\begin{minipage}[t]{0.1312\linewidth}\includegraphics[width=\linewidth]{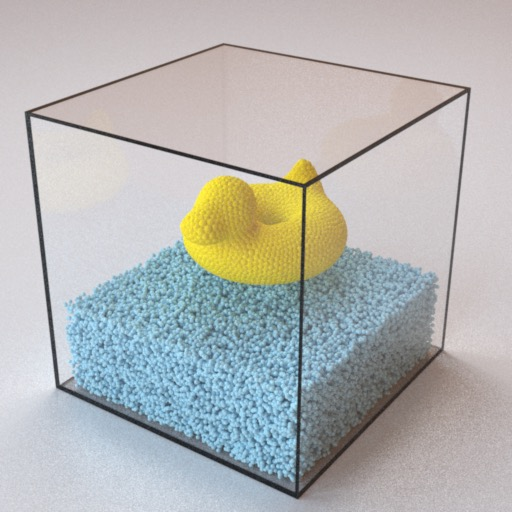}\end{minipage}\\[1pt]
\begin{minipage}[t]{0.2310\linewidth}\includegraphics[width=\linewidth]{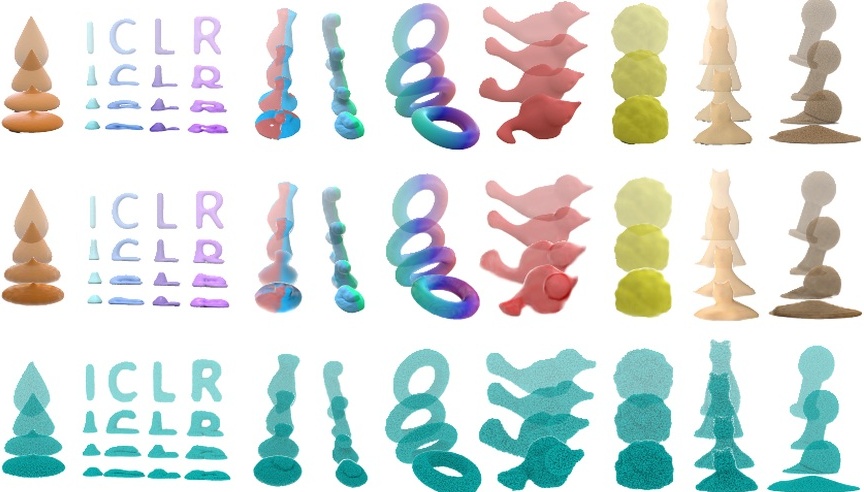}\end{minipage}%
\begin{minipage}[t]{0.1663\linewidth}\includegraphics[width=\linewidth]{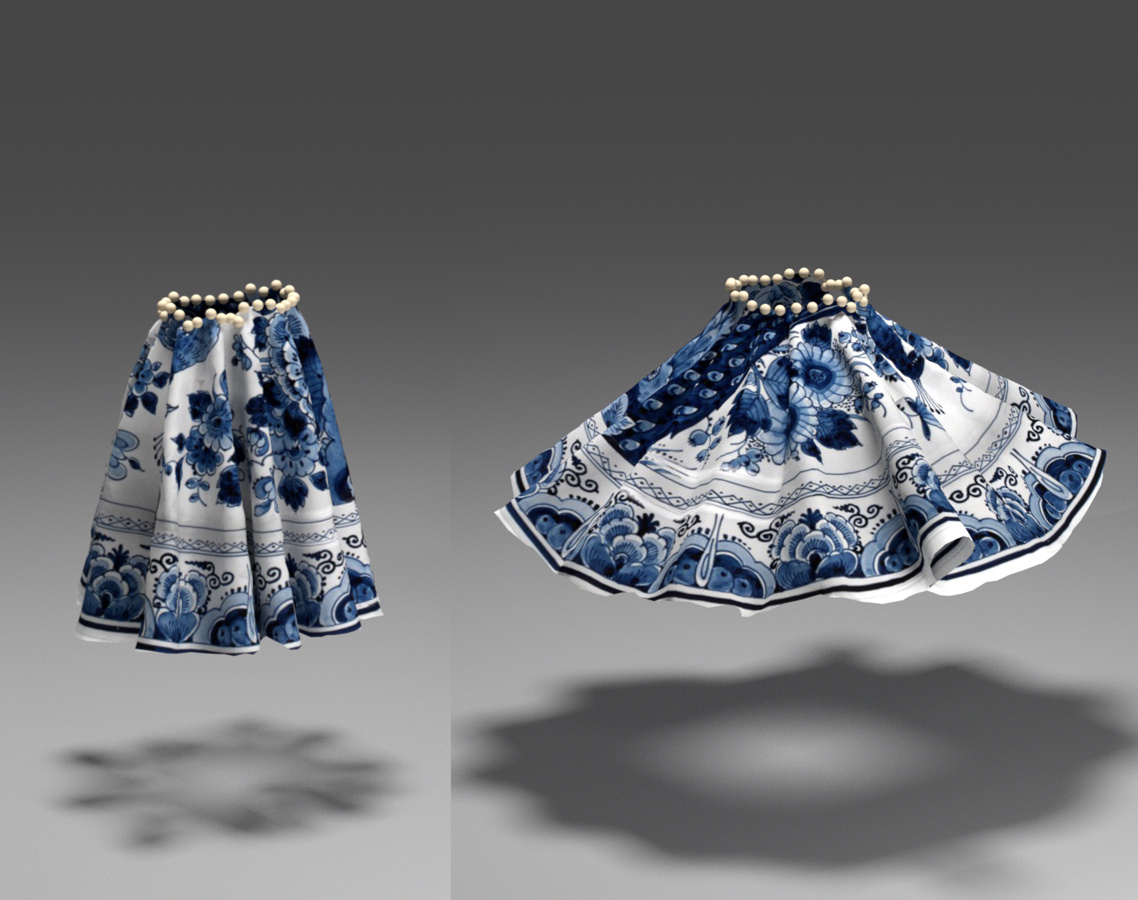}\end{minipage}%
\begin{minipage}[t]{0.2338\linewidth}\includegraphics[width=\linewidth]{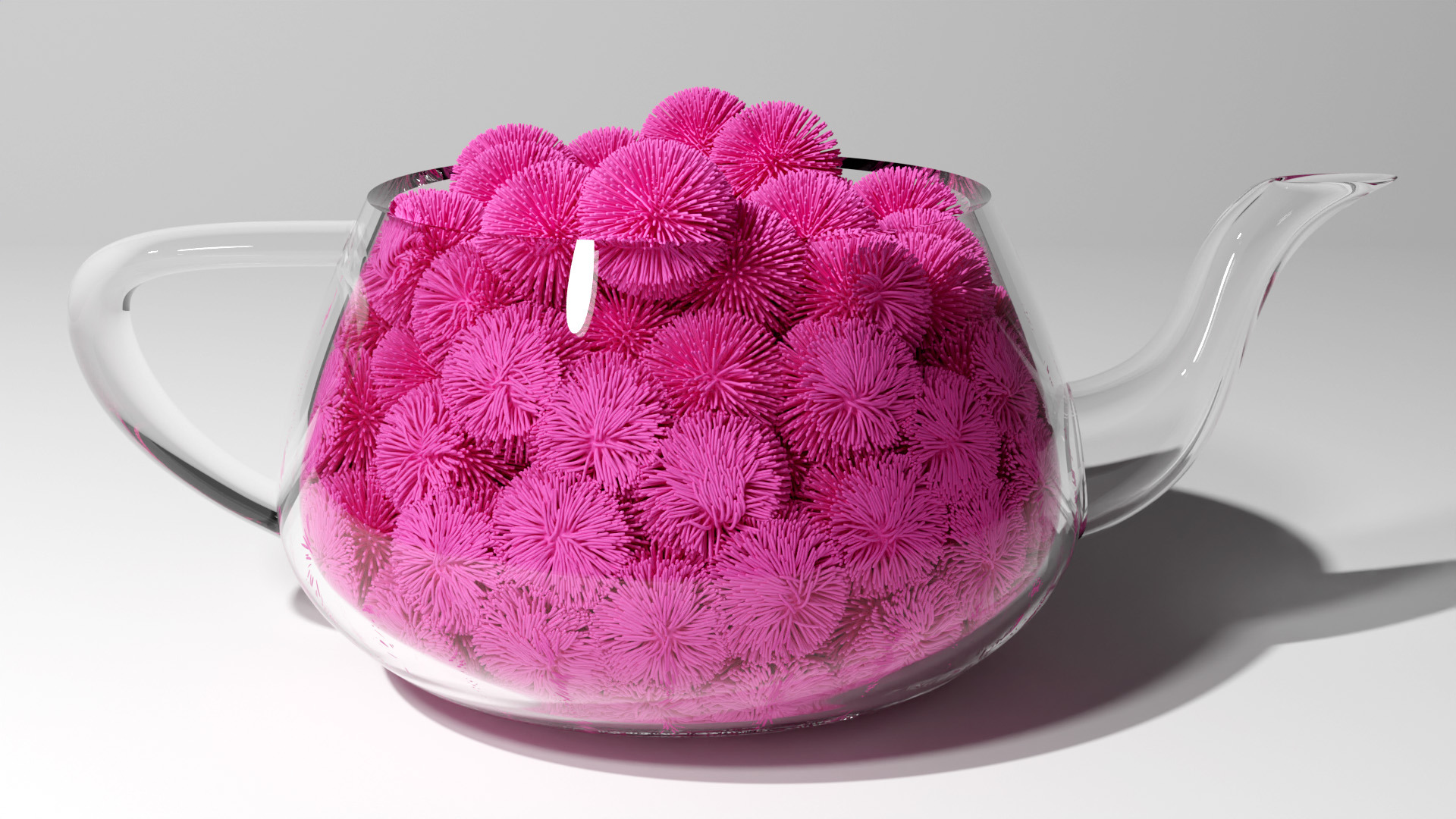}\end{minipage}%
\begin{minipage}[t]{0.3687\linewidth}\includegraphics[width=\linewidth]{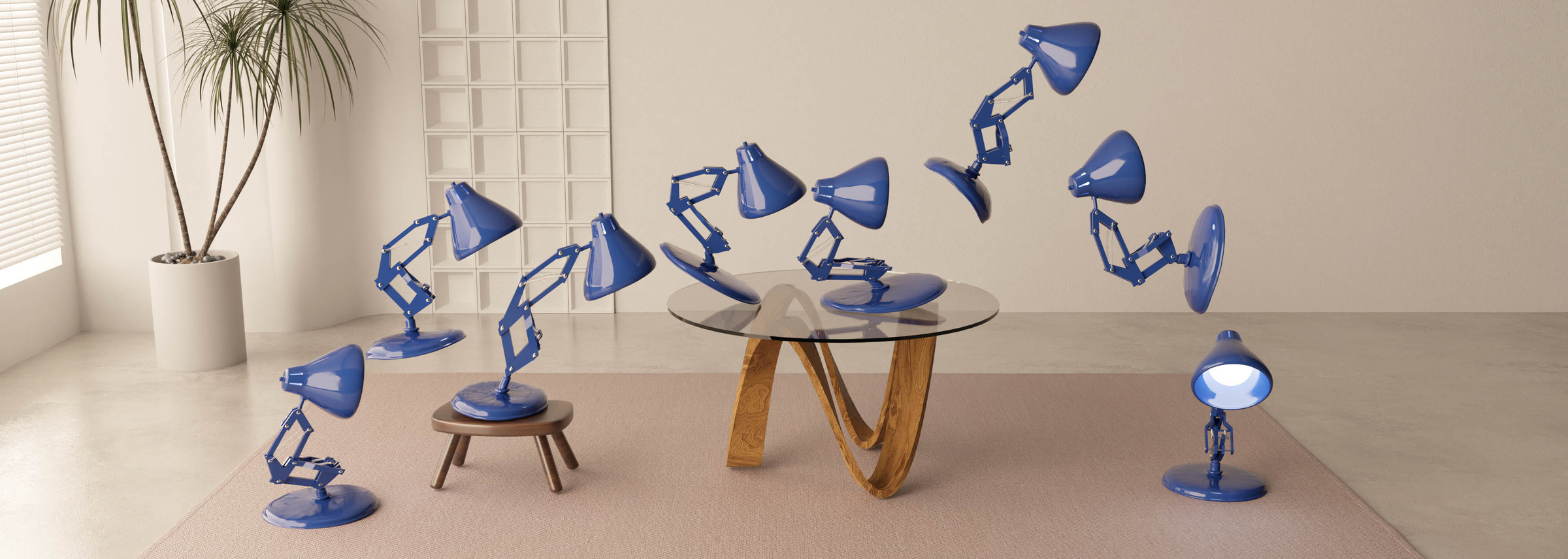}\end{minipage}\\[1pt]
\begin{minipage}[t]{0.1770\linewidth}\includegraphics[width=\linewidth]{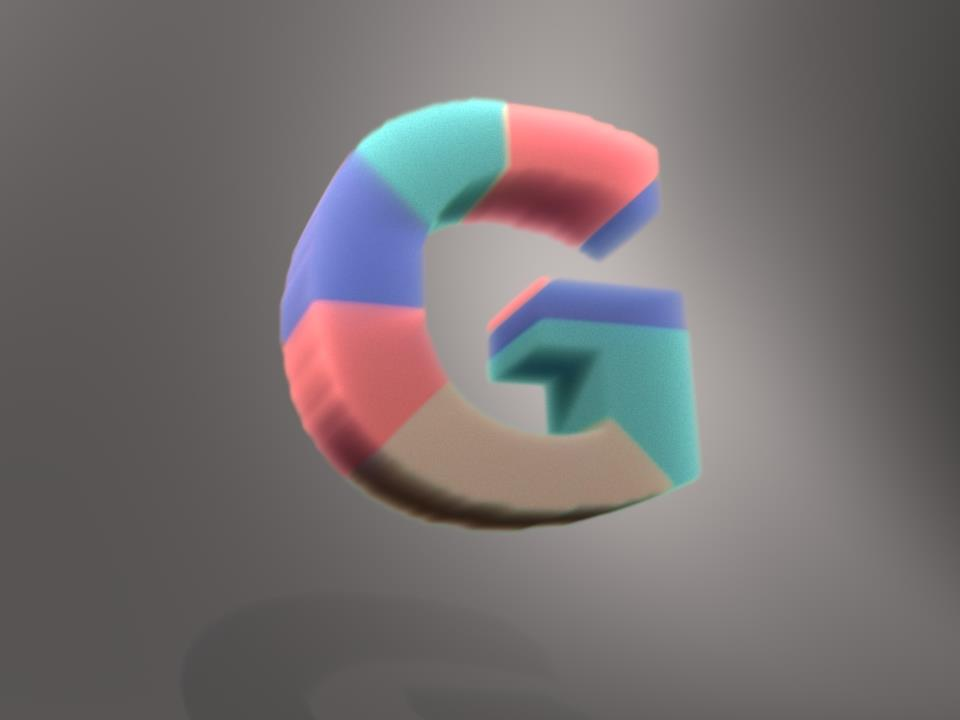}\end{minipage}%
\begin{minipage}[t]{0.3467\linewidth}\includegraphics[width=\linewidth]{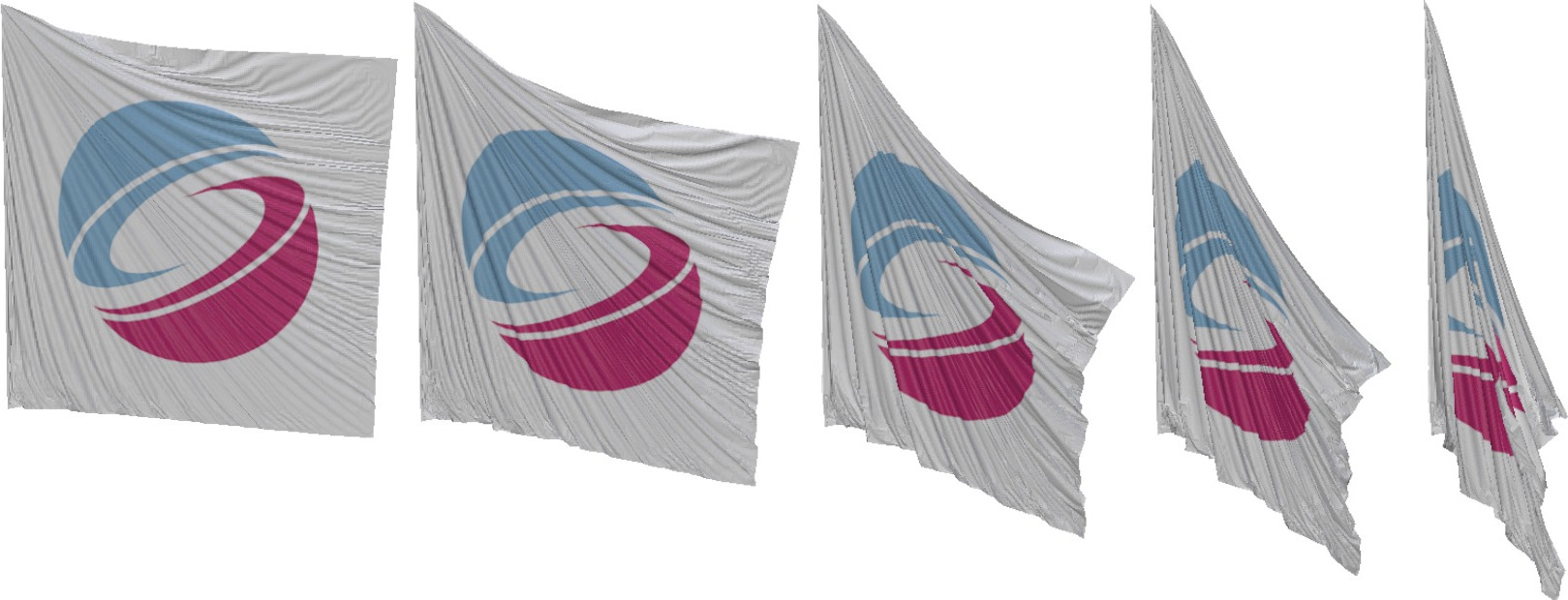}\end{minipage}%
\begin{minipage}[t]{0.2044\linewidth}\includegraphics[width=\linewidth]{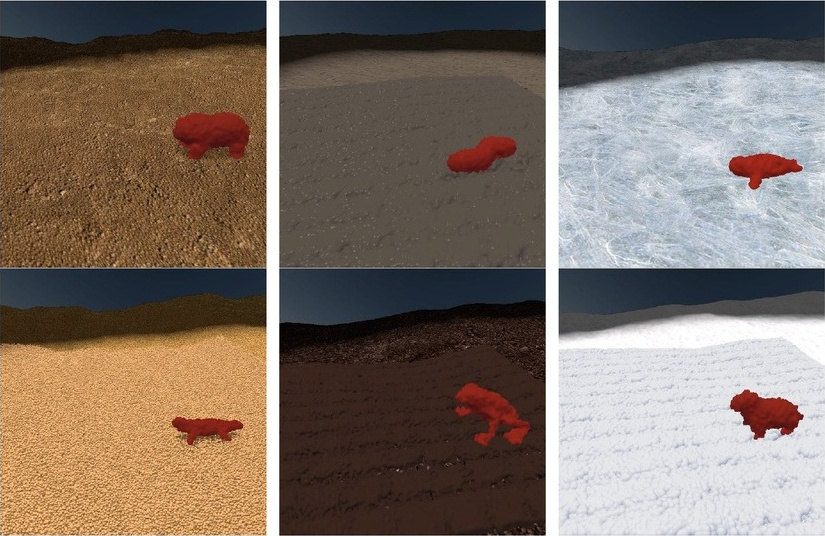}\end{minipage}%
\begin{minipage}[t]{0.1329\linewidth}\includegraphics[width=\linewidth]{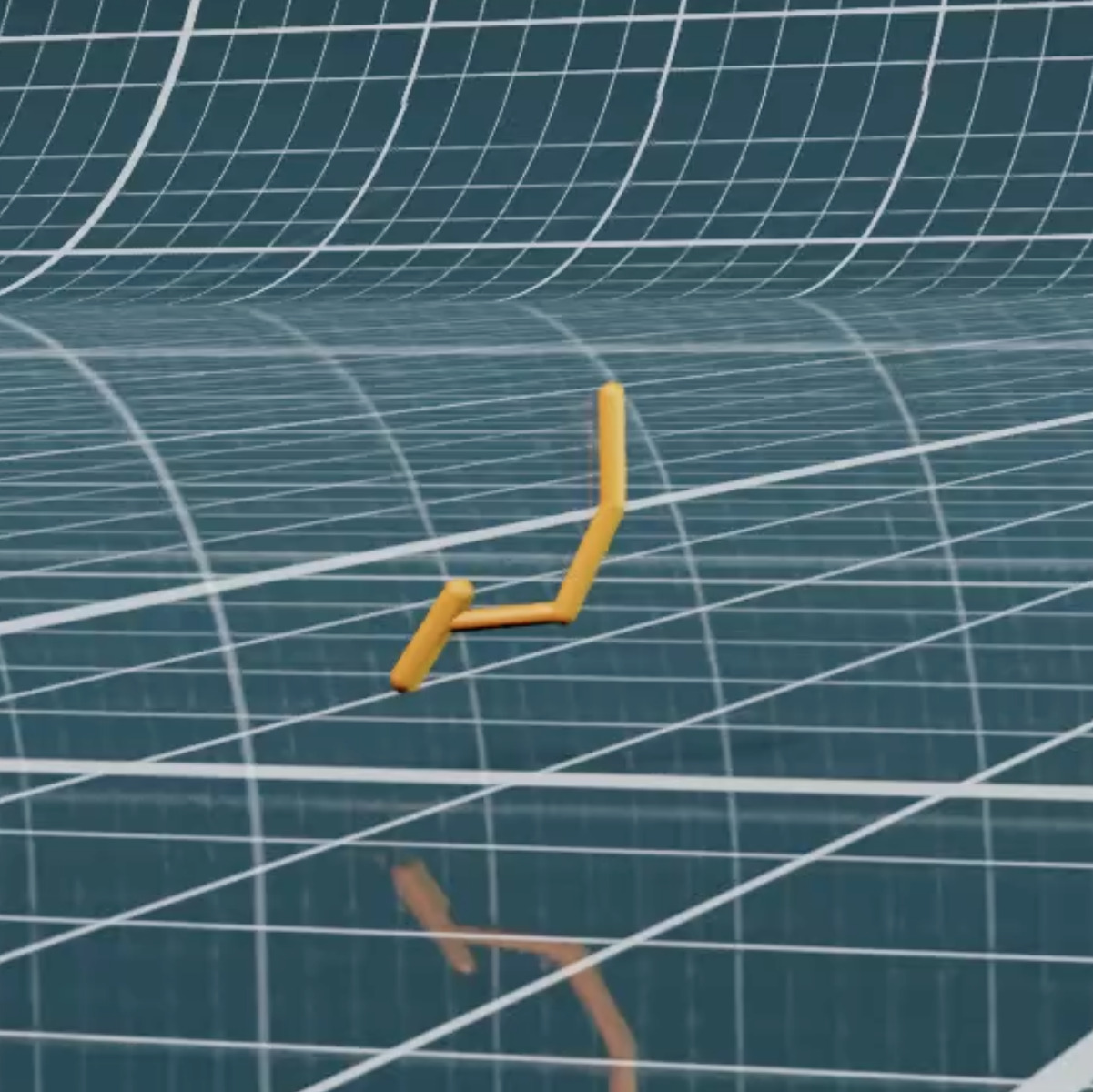}\end{minipage}%
\begin{minipage}[t]{0.1388\linewidth}\includegraphics[width=\linewidth]{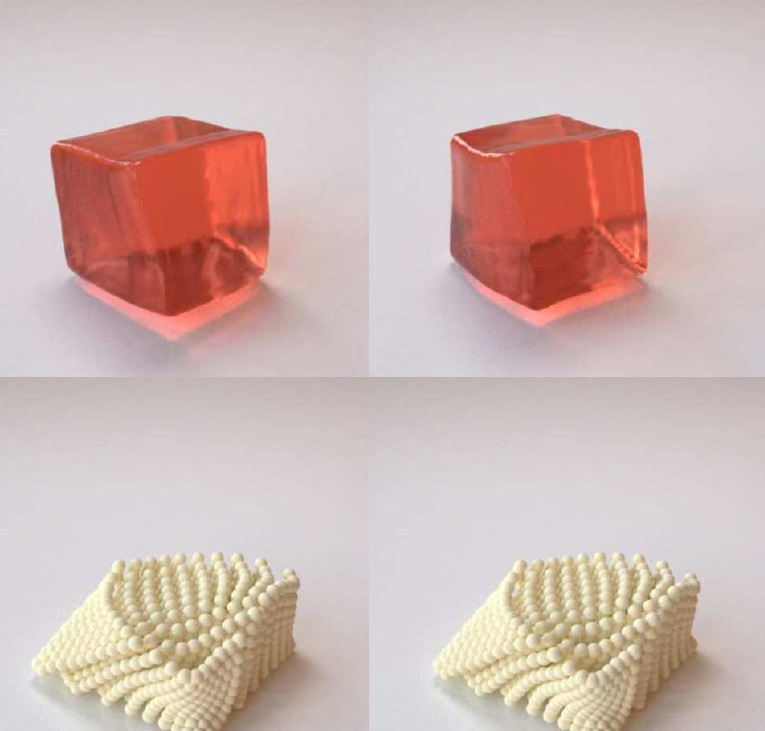}\end{minipage}
\end{minipage}}
\captionof{figure}{Recovering material parameters from video, designing a body and its controller
together, and learning a constitutive law from observed motion are
applications in which gradients propagated through a simulator can support
optimization. This report organizes
the field by two questions that engines answer differently and rarely state
explicitly, rather than by such applications: what the backward pass differentiates
(Section~\ref{sec:taxonomy}), and what it does at contact
(Section~\ref{sec:contact}). Images reproduced, cropped or letterboxed but
otherwise unaltered, from
\cite{Li2021PlasticineLab,Li2022DiffCloth,Chen2024VBD,Ma2023NCLaw,Li2023PACNeRF,Wang2023SoftZoo,Georgiev2024AHAC,ElasticLoco2025,Li2025FlowMapAdjoint,LocalAD2026}.
}
\label{fig:teaser}
\end{minipage}
\vspace{12pt}
\begin{center}
\begin{minipage}{0.92\textwidth}\small\setlength{\parskip}{3pt}
\noindent\textbf{Abstract}\par\smallskip
Differentiable simulation propagates derivatives through a physics simulator,
turning animation, system identification, control, and computational design
into gradient-based optimization. Engines differ in the physics they model and, less visibly, in which
mathematical object they differentiate. We separate two questions the literature conflates. The
\emph{referent} is which object the returned derivative belongs to: the finite
computation that executed, the converged equation it approximates, or the
continuous dynamics. The \emph{realization} describes the organization of the backward pass: automatically
generated differentiation, an equation-level adjoint solve, or a custom
adjoint. The referent defines the target against which exactness is assessed;
the implementation determines how accurately that target is approximated. Cost
depends on the realization together with how the states needed by the backward
pass are retained or recovered.

Crossing the axes
places every method and engine we classify in a cell, separates claims
the literature merges, and shows that the apparent trade-off between exactness
and memory follows from how methods have been built rather than from a
constraint. A second, orthogonal taxonomy covers contact, where the map is
non-smooth.

We then ask whether a correct gradient helps. Four mechanisms by which useful
gradients are lost are cataloged and measured on one testbed, where the
advantage of the exact derivative over the equation-level adjoint depends on
where the fitting target comes from as much as on the solver budget, and we
review the evidence that zeroth-order methods can outperform simulator
gradients in the regimes where those mechanisms act. The model classes,
systems techniques and applications are surveyed through this lens, and we
close with a reporting checklist, a released diagnostic that computes it, and
a research agenda.
\par\medskip
\noindent\textbf{Keywords:} differentiable simulation, adjoint methods, automatic differentiation, physics-based animation, contact, inverse problems, survey.
\end{minipage}
\end{center}
\vspace{16pt}
}]
\thispagestyle{firstpage}

\section{Introduction}
\label{sec:intro}

Physically based animation is built on a forward map. A scene is described by geometry and their
parameters $\bth$ (masses, stiffnesses, rest shapes, friction coefficients, control signals, boundary conditions), and a time integrator advances a state
$\bs_t$ to $\bs_{t+1}$ until a trajectory has been produced. Four decades of
graphics research have made this map fast, stable, and expressive enough to
carry feature films and real-time games.

Most of the questions practitioners ask, however, run in the opposite
direction. An animator wants the wind field that makes a piece of cloth do what the
storyboard says, an engineer wants the gripper geometry that achieves a grasp,
and a vision researcher wants the material parameters that explain a video of
a falling object. Each of these is an inverse problem posed on top of a forward simulator, and
simulator derivatives often make its solution substantially more efficient,
particularly when the number of optimization variables is large.

\emph{Differentiable simulation} is the practice of making those derivatives
available. Spacetime
constraints~\cite{WitkinKass1988,Cohen1992} posed trajectory authoring as a
constrained optimization over a discretized trajectory, and keyframe control of
smoke and of particle systems~\cite{Treuille2003,McNamara2004,Wojtan2006}
differentiated the discrete fluid solvers directly and a cloth solver through a
swapped integrator, both long before
the term existed; the same period produced gradient-driven and sampling-driven
methods for steering rigid-body simulations toward authored
outcomes~\cite{Popovic2000,Popovic2003,Twigg2008}. More recently, general-purpose automatic
differentiation (AD) frameworks~\cite{Paszke2019,Bradbury2018}, GPU-resident
simulation, and domain-specific compilers~\cite{Hu2019Taichi,Hu2020DiffTaichi}
have expanded the scale and range of simulators for which derivatives can be computed efficiently, and a large,
fast-moving literature has grown around the possibility.

Two gaps run through that literature, and together they organize this report.
The first is between the derivative an engine computes and the derivative its
user expects: differentiating the finite solver and differentiating the
equation it approximates target different mathematical objects. Their
agreement requires conditions on both the forward solution and its
sensitivities. The second is between obtaining a derivative and obtaining
useful optimization progress. Even an exact gradient can be ineffective for
finite-step optimization, while pathwise derivatives can be biased estimators
for stochastic objectives with discontinuities. Neither difficulty is
universal; identifying the regimes in which a returned derivative is both
appropriate and useful is a central purpose of this report.

The two gaps are not equally consequential. The
first decides what a method is entitled to claim, and it is always a question of
fact: an engine either returns the derivative of what it ran or it does not. How
much that costs an optimizer is a separate matter, and our own measurement in
Section~\ref{sec:quality} finds the cost concentrated at the tightest solver budgets, and how far above them it persists to depend on where the fitting target comes from. The referent decides what you may claim and whether a gradient
check can pass, while the horizon, the contact treatment and the conditioning
usually decide whether the optimization works.

Beyond its size, the literature is difficult to navigate because two
papers may both claim to present ``a differentiable simulator'' for deformable
bodies with contact, agree on the physical model, agree on the discretization,
and still compute meaningfully different derivatives, because they differentiate different mathematical objects. Neither is thereby incorrect. One
may backpropagate through every Newton iteration that was executed; another may
apply the implicit function theorem to the equation those iterations were trying
to solve. For a smooth, locally regular solution branch, the two sensitivities approach
one another when both the iterates and their parameter sensitivities converge
to that branch. A small forward residual alone does not establish agreement,
and disagreement at a finite iteration budget is a consequence of the
differentiated object rather than necessarily an implementation error.
Comparisons that ignore this distinction end up attributing to the physics, or
to the contact model, differences that originate in the differentiation
strategy.

\subsection{Scope}
\label{sec:scope}

This report surveys \emph{differentiable physics simulation}: methods that
propagate derivatives through the numerical solution of the equations of motion
of rigid bodies, articulated systems, deformable solids, cloth, and fluids. We
treat differentiable rendering as adjacent rather than in scope. It is a mature
subfield with its own surveys and courses~\cite{Kato2020Survey,Zhao2020Course},
its discontinuities are of a different character (visibility rather than contact), and the techniques developed for
them~\cite{Li2018EdgeSampling,Loubet2019,Vicini2022}
address visibility boundaries; their applicability to contact requires separate analysis.
Mitsuba~2~\cite{NimierDavid2019Mitsuba2} provides infrastructure for
differentiable rendering; visibility derivatives are addressed in separate work. We do discuss
the seam where the two meet (Section~\ref{sec:app-video}), because
video-to-physics pipelines require both and because the composition raises
questions neither field answers alone.

We also distinguish differentiable simulation from learned simulation.
Graph-network
surrogates~\cite{Battaglia2016IN,SanchezGonzalez2020GNS,Pfaff2021MGN,Li2019DPINets,grigorev2023hood},
structure-preserving networks~\cite{Cranmer2020LNN}, and physics-informed
networks~\cite{Raissi2019PINN} often expose derivatives through their neural
components, and a learned surrogate can serve as the forward model of a PDE-constrained inverse problem, with the inverse solved by gradient descent through the network~\cite{Zhao2022GNNInverse}. Their rollouts may nevertheless require differentiation through
numerical integration, and discrete structural operations may introduce
additional difficulties. We reference these models where they interact with
numerical solvers, as surrogates, correction terms, or baselines, but do not
survey learned simulators as a separate class.

\subsection{Contributions}

This report contributes:

\begin{itemize}
\item A \textbf{two-axis taxonomy} (Section~\ref{sec:taxonomy}) separating
  the \emph{referent} of a derivative from the \emph{realization} of its
  backward pass. The referent defines the target for an exactness claim;
  the realization and a separate state-recovery policy determine the
  computational and storage costs. Crossing these choices explains why
  exactness and low memory are not fundamentally incompatible, and identifies
  the assumptions and costs of the methods that combine them.

\item An orthogonal \textbf{taxonomy of discontinuity treatments}
  (Section~\ref{sec:contact}) for contact, friction, and collision, classifying
  methods by the mathematical fiction they adopt where the true map is
  non-smooth, and by the bias and variance that fiction introduces.
\item A \textbf{critical account of gradient quality}
  (Section~\ref{sec:quality}) that separates derivative correctness from
  practical optimization progress, identifies four distinct mechanisms that
  impair gradient-based optimization, and maps published mitigations onto them.

\item A \textbf{cross-cutting comparison} of physical model families
  (Section~\ref{sec:models}), systems techniques (Section~\ref{sec:systems}), and
  engines (Table~\ref{tab:engines}) on a common set of axes.
\item A \textbf{reporting checklist} (Section~\ref{sec:eval}) that would make
  differentiable simulators comparable, and a \textbf{research agenda}
  (Section~\ref{sec:open}) organized around the gaps the taxonomy exposes.
\end{itemize}

\section{Related Surveys, Literature Selection, and Organization}
\label{sec:related}

\subsection{Related surveys}
\label{sec:related-surveys}

Differentiable simulation has been reviewed before, most directly by Newbury
\etal~\cite{Newbury2024Review}, who discuss gradient construction, dynamics
and contact models, integrators, engines, and robotics applications, and, as one strand of a broader tutorial treatment of physics-based deep learning, by Thuerey \etal~\cite{Thuerey2021PBDL}.
We refine this mechanism-oriented view by separating derivative referents,
backward realizations, and state recovery. These distinctions help explain
computational cost, failure modes, and differences that are obscured by
engine-level comparisons. They also remain useful as software packages change.
We give particular attention to whether the returned derivatives improve the
optimization for which they are computed, drawing on evidence of challenging
regimes~\cite{Metz2021,Suh2022} rather than assuming that derivative
availability guarantees useful progress.

\subsection{Literature selection}
\label{sec:selection}

Differentiable simulation is developed in parallel in graphics, robotics and
machine learning, so we searched DBLP, Crossref, arXiv, the ACM Digital Library
and Google Scholar for \emph{differentiable}, \emph{gradient-based} or
\emph{adjoint} combined with simulation, physics, solver, contact, cloth, fluid
and related task terms, and then followed the references and citations of each
included work. Expansion stopped when a full round added no new differentiation
strategy, contact treatment or application class. A work was included if it
contributes a differentiable simulator or component, provides evidence about
one, or supplies background the argument depends on. Work that is
differentiable only because it is a neural network, and applications that use
an engine without reporting on its derivatives, were excluded.

\subsection{Organization of this report}
\label{sec:organization}

Section~\ref{sec:background} fixes notation and reviews the material the rest of
the report assumes: the simulation map, forward and reverse mode, and the
distinction between continuous and discrete adjoints.
Section~\ref{sec:taxonomy} develops the first taxonomy, separating the referent
of a derivative from the realization of the backward pass, and is the conceptual
core of the report; Table~\ref{tab:grid} crosses the two axes and
Section~\ref{sec:tax-compare} gives guidance on choosing a cell. Section~\ref{sec:contact} develops the second,
orthogonal taxonomy for contact and other discontinuities, summarized in
Table~\ref{tab:contact}. Readers interested only in the classification may stop
after these two sections.

Section~\ref{sec:models} surveys the physical model families through the lens of
the two taxonomies, and Section~\ref{sec:quality} takes up gradient quality,
which we regard as the most consequential and least settled part of the subject.
Section~\ref{sec:systems} covers systems and implementation and compares the
major engines in Table~\ref{tab:engines}. Section~\ref{sec:applications} surveys
applications, Section~\ref{sec:eval} evaluation practice and our proposed
reporting checklist, and Section~\ref{sec:open} the open problems the taxonomy
exposes.

\section{Background and Notation}
\label{sec:background}

This section fixes the notation used throughout, collected in
Table~\ref{tab:notation}, and the two distinctions the rest of the report rests
on: forward against reverse mode, and the continuous against the discrete
adjoint.

\begin{table}[t]
\centering
\caption{Notation used throughout this report.}
\label{tab:notation}
\small
\begin{tabular}{@{}ll@{}}
\toprule
\textbf{Symbol} & \textbf{Meaning} \\
\midrule
${\bs_t = (\x_t,\vv_t)}$ & simulation state at step $t$: positions, velocities \\
${\bth \in \R^{p}}$      & parameters being optimized (material, shape, \ldots) \\
$\Phi$                 & one time step of the simulator, Eq.~\eqref{eq:step} \\
$\gres$                & residual defining an implicit step, Eq.~\eqref{eq:residual} \\
$E$                    & incremental potential minimized by a step \\
$\Loss$                & scalar objective on the trajectory \\
$T$                    & number of time steps \\
$K$                    & solver iterations executed per step \\
$n$, $p$               & state and parameter dimension \\
$A_t,\,B_t$            & step Jacobians $\partial\Phi/\partial\bs_t$, $\partial\Phi/\partial\bth$ \\
$\adj{\bs}_t$          & adjoint state $(\mathrm{d}\Loss/\mathrm{d}\bs_t)\transp$ \\
$U_j$, $U_j^{*}$       & block update and its Jacobian-transpose action \\
$\phi$, $\lambda$      & contact gap function and normal impulse \\
\bottomrule
\end{tabular}
\end{table}

\subsection{The simulation map}

We write a simulation as a composition of $T$ time steps. A step advances the
state ${\bs_t \in \R^{n}}$, typically positions and velocities ${\bs_t = (\x_t, \vv_t)}$, under parameters ${\bth \in \R^{p}}$. We fold into $\bth$ whatever is being
optimized, including a control sequence or the weights of a controller that
emits one; in the latter case a step depends on $\bth$ through the control it
receives, and the form below is unchanged:
\begin{equation}
\bs_{t+1} = \Phi(\bs_t; \bth), \qquad t = 0,\dots,T-1 .
\label{eq:step}
\end{equation}
Explicit time dependence is suppressed in the notation: $\Phi$ may denote
$\Phi_t$, for example when selecting the control at step $t$ from a sequence.
All step Jacobians below refer to that step's map.
An objective $\Loss$ is evaluated on the trajectory, either terminally or as a
running cost, and the quantity we want is $\mathrm{d}\Loss/\mathrm{d}\bth$.

A central difficulty is that the time step is often specified by equations
that must be solved numerically. Explicit schemes evaluate update formulas
directly, although stiffness can impose restrictive time steps. Many implicit
schemes used for stiff cloth and constrained systems~\cite{Baraff1998}
instead define the next state through a residual equation,
\begin{equation}
\gres(\bs_{t+1}; \bs_t, \bth) = \mathbf{0}.
\label{eq:residual}
\end{equation}
Semi-implicit schemes may use explicit sequential updates or linear solves
and need not require a nonlinear solve. For variational formulations with an
appropriate incremental potential, a step can instead be characterized as a
selected local minimizer of $E(\bs;\bs_t,\bth)$. This characterization is not
equivalent to an arbitrary residual equation.

A simulator runs neither of these directly. It runs a solver for them: Newton with a linear solve, projective
dynamics~\cite{Bouaziz2014}, Gauss--Seidel or Jacobi sweeps, extended
position-based dynamics (XPBD) projections~\cite{Macklin2016XPBD}, block
coordinate descent~\cite{Chen2024VBD}, or, for a grid fluid, a pressure
projection whose Poisson system is solved by conjugate gradients to a
tolerance. It runs for a finite and usually small number of iterations.

The three referents of Section~\ref{sec:taxonomy} diverge in this gap between~\eqref{eq:step} and~\eqref{eq:residual}, between the step we write down and the finite computation we execute.

\subsection{Forward and reverse mode}
\label{sec:fwdrev}

Given the per-step Jacobians, the chain rule assembles the trajectory
derivative. For now, assume that $\bs_0$ is independent of $\bth$ and that
$\Loss$ depends on $\bth$ only through the trajectory. Writing ${A_t = \partial\Phi/\partial\bs_t}$ and
${B_t = \partial\Phi/\partial\bth}$ evaluated at step $t$, the sensitivity of the
final state is
\begin{equation}
\frac{\mathrm{d}\bs_T}{\mathrm{d}\bth} = \sum_{t=0}^{T-1}
  A_{T-1} A_{T-2} \cdots A_{t+1} B_t ,
\label{eq:chain}
\end{equation}
where the factors are ordered by descending index, as the non-commuting product
requires. An empty product is the identity, so the $t=T-1$ summand is
$B_{T-1}$. The two modes of accumulation differ in which end of that product they
start from. Beginning at the parameter end, with $B_t$, and multiplying by each
$A_k$ as $k$ increases is \emph{forward mode}: it propagates an ${n\times p}$
sensitivity matrix alongside the state as time advances, costing $O(p)$
simulation-equivalents but requiring no storage of history. Beginning instead at
the objective end and multiplying backward in time is \emph{reverse mode}, or
the discrete adjoint: one introduces an
adjoint state ${\adj{\bs}_t = (\mathrm{d}\Loss/\mathrm{d}\bs_t)\transp}$, the
\emph{total} derivative of the objective with respect to the state at step $t$
including everything downstream of it. Initialize
$\adj{\bs}_T=(\partial\Loss/\partial\bs_T)\transp$ and $\adj{\bth}=0$,
and sweep backward for $t=T-1,\ldots,0$,
\begin{equation}
\adj{\bs}_t = A_t\transp \adj{\bs}_{t+1} + \left(\jac{\Loss}{\bs_t}\right)\transp,
\qquad
\adj{\bth} \mathrel{+}= B_t\transp \adj{\bs}_{t+1},
\label{eq:adjoint}
\end{equation}
where the second term is the \emph{direct} dependence of a running cost on
$\bs_t$ and vanishes for a terminal objective, costing $O(1)$ simulation-equivalents independent of $p$, but requiring access
to the forward trajectory in reverse order. If the initial state is
parameterized, \eqref{eq:chain} additionally contains
$A_{T-1}\cdots A_0\,\mathrm{d}\bs_0/\mathrm{d}\bth$, and the parameter
gradient includes
$(\mathrm{d}\bs_0/\mathrm{d}\bth)\transp\adj{\bs}_0$. If the objective is
extended to include direct parameter dependence, also add
$(\partial\Loss/\partial\bth)\transp$ to the column gradient.

For the inverse problems that motivate differentiable simulation, $p$ is large
(a per-element material field, a control sequence, a mesh parameterization) and
$\Loss$ is scalar, so reverse mode is the relevant regime and its memory cost is
the binding constraint. This asymmetry between computational work and storage motivates the systems
discussion of Section~\ref{sec:systems}; checkpointing~\cite{Griewank2000} is
a common way to make reverse accumulation feasible when a full tape cannot be
retained (Table~\ref{tab:engines}).

A thorough treatment of AD itself is outside our scope;
see Griewank and Walther~\cite{Griewank2008} for the standard reference, and
Baydin \etal~\cite{Baydin2018} and Margossian~\cite{Margossian2019} for
surveys; for the
broader program of differentiable programming in scientific computing, see
Innes \etal~\cite{Innes2019} and the position piece of Ramsundar
\etal~\cite{Gradientsurvey2023}.

\subsection{Two adjoints: continuous and discrete}
\label{sec:twoadjoints}

The PDE-constrained optimization literature distinguishes two ways to form an adjoint,
and the distinction recurs under other names throughout this report. One
may \emph{differentiate then discretize}: derive the adjoint PDE analytically
from the continuous equations of motion, then discretize that adjoint. Or one may
\emph{discretize then differentiate}: take the discrete forward scheme as the
definition and differentiate it exactly.

A continuous adjoint is derived before choosing its discretization. For a
scalar objective, its evaluation often costs a small number of forward solves
and is largely independent of the number of design parameters, while still
depending on mesh resolution, time stepping, and solver accuracy. A separately
discretized continuous adjoint need not equal the derivative of the chosen
discrete forward map. This mismatch can impair optimization and
finite-difference agreement; it does not necessarily correspond to the
gradient of another scalar objective. Consistent forward--adjoint
discretizations can remove the mismatch in appropriate
settings~\cite{Matsubara2021Symplectic}.

A discrete adjoint instead differentiates a specified discrete object. That
object may be the discrete equations or the finite solver used to approximate
them, which is the distinction developed in Section~\ref{sec:referent}. McNamara
\etal~\cite{McNamara2004} explicitly select a discrete formulation, treating
projection and diffusion as linear operators. Agreement with finite
differences therefore depends on the differentiated object, solver accuracy,
and the usual smoothness and finite-difference conditions.

Most methods surveyed here differentiate discrete objects. In physically based
differentiable rendering, methods such as Li \etal~\cite{Li2018EdgeSampling}
first derive the rendering-integral derivative, including visibility-boundary
terms, and then construct Monte Carlo estimators. This provides a related
example of differentiating before numerical estimation.

\section{A Two-Axis Taxonomy of Backward Passes}
\label{sec:taxonomy}

This section states the organizing claim of this report. Consider a stiff implicit
step: the integrator defines $\bs_{t+1}$ as a root of~\eqref{eq:residual}, and a
solver produces an approximation to that root in $K$ iterations. Two questions
are needed to interpret a returned gradient, and the literature often
answers only the second while appearing to answer the first.

\begin{description}
\item[Which object is the derivative of?] The finite computation the solver
  actually executed, the exact root that computation was approximating, or the
  continuous dynamics the discretization itself approximates. We call this the
  \emph{referent}, and it decides what ``exact'' can mean.
\item[How is the backward pass realized?] By recording the forward computation
  and generating its reverse operations, by solving equation-level adjoint
  systems, or by supplying custom operator adjoints. We call this the
  \emph{realization}; together with state recovery it determines the cost.
\end{description}

These are distinct questions: a referent can admit several implementations,
and a realization can serve more than one referent. Our realization labels
are operational categories rather than an exhaustive Cartesian product of
construction choices, as Section~\ref{sec:realization} makes explicit.
Two methods can target the same derivative by different means, or target
different derivatives while both referring to ``the gradient.''
Table~\ref{tab:grid} crosses the labels; we call this the
\emph{referent/realization distinction}.

\subsection{Axis A: the referent}
\label{sec:referent}

\paragraph*{A1, the executed computation.}
Treat the step as the literal sequence of $K$ solver iterations and
differentiate that computation. On a differentiable execution branch, with
correct derivative rules, this targets the sensitivity of the output produced
by the forward pass whether or not the solve converged. It is the referent
directly tested by finite differences of the production configuration, subject
to truncation error, roundoff, and stability of the execution branch.

One qualification is used throughout the tables. A step often contains a
\emph{linear} solve run to a tolerance, a pressure projection or the inner
solve of a Newton iteration, whose derivative the backward pass takes as that
of the exact linear operator rather than of the iterations that approximated
it. We mark such methods A1*: exact for an idealized program in which the
linear solves are exact operators, which is not the program that ran. How far
the derivative of that idealization is from the derivative of the
finite-tolerance program is not given by the solver tolerance alone, since
the conditioning of the system and the sensitivity of the residual to the
parameters enter as well, and it is to be checked at the production
configuration like any other gap; a finite-difference check of an A1* method
is subject to that gap as well as finite-difference and floating-point error. The unqualified label is kept for programs whose every
solve is differentiated as it ran. The referent distinction of this section
emphasizes truncated \emph{nonlinear} iteration, where the sensitivities can
differ substantially; the asterisk records that reading a linear solve as exact is an
assumption.

The guarantee is narrower than often stated: it is exact for the executed program, not for the physics. A program
containing a branch, a discrete collision-detection query, or a
non-differentiable primitive returns the derivative of the branch that was taken,
or a convention chosen by the framework at a kink. Nor is such an object
guaranteed to be a subgradient.
With the common convention ${\mathrm{ReLU}'(0)=0}$, differentiating the identity
written as ${\mathrm{ReLU}(x)-\mathrm{ReLU}(-x)}$ returns $0$ at the origin,
where the function is differentiable with derivative $1$; the composition rule
is not required to land in the subdifferential of the composite. What AD returns
at a kink depends on local derivative conventions and need not be a derivative
of the composite function; calling it ``the true sensitivity'' without
qualification overstates what was computed.

\paragraph*{A2, the converged equation.}
Treat the step as a selected root of~\eqref{eq:residual}. If the residual is
continuously differentiable near that root and its state Jacobian is
nonsingular, the implicit function theorem gives the local root sensitivity:
\begin{equation}
\jac{\bs_{t+1}}{\bth}
 = - \left( \jac{\gres}{\bs_{t+1}} \right)^{\!-1} \jac{\gres}{\bth},
\label{eq:ift}
\end{equation}
with $\bs_t$ held fixed, so the left-hand side is the step Jacobian $B_t$ of
Section~\ref{sec:fwdrev}. Nothing about the solver trajectory enters. The following distinction is used
throughout the report.
At the exact root, \eqref{eq:ift} gives the sensitivity of that root branch. An
implementation typically evaluates the formula at the state where the solver
stopped and solves the adjoint system only to a finite tolerance. Three
quantities must therefore be distinguished: the converged-root derivative
targeted by A2, the vector returned by the implemented implicit backward pass,
and the derivative of the finite computation targeted by A1. These quantities
generally differ at finite tolerances, although equality is possible in
special cases. Their discrepancy depends on local conditioning and residual
sensitivities, as well as forward and adjoint solve accuracy. Jacobian
modifications introduce another approximation; for example,
DiffXPBD~\cite{Stuyck2023DiffXPBD} projects derivative blocks to be positive
definite. Section~\ref{sec:quality} compares the returned equation-level
vector with the derivative of the executed computation and measures their
optimization consequences.

This is the referent of the
optimization-layer literature, where OptNet~\cite{Amos2017} differentiates a
quadratic program through its Karush--\allowbreak Kuhn--\allowbreak Tucker (KKT) conditions, \texttt{cvxpy\-layers}
\cite{Agrawal2019} generalizes this to convex cones, deep equilibrium
models~\cite{Bai2019} differentiate a fixed point without unrolling it, and
modular libraries~\cite{Blondel2022} package the pattern.

In interactive simulation, the forward solver is often stopped at a small
iteration budget. The resulting state is then the output of the finite solver,
whereas A2 targets a root of its defining equation. Evaluating the implicit
derivative at this state generally approximates that root sensitivity and need
not match the derivative of the finite solver. Section~\ref{sec:quality}
examines this distinction at the iteration budgets used by the forward
simulation.

The dependence on convergence is quantitative. Under appropriate local
regularity conditions, inexact implicit differentiation and
residual-correction analyses can bound the error using forward and adjoint
residuals together with conditioning constants. For a smooth, uniformly
attractive fixed-point map with suitable derivative regularity,
Christianson~\cite{Christianson1994} shows that an adjoint iteration evaluated
at the final forward state converges at a rate controlled by the contraction
factor. This result does not establish derivative accuracy from the forward
stopping tolerance alone. Reporting a bound or an empirical sensitivity check
at the production solver configuration would make the practical approximation
explicit.

\paragraph*{A3, the continuous dynamics.}
Derive the adjoint of the continuous equations of
motion and then discretize that adjoint. A separately chosen adjoint
discretization need not produce the derivative of the chosen discrete forward
map, so agreement with finite differences of that map is not automatic.
Appropriate paired discretizations can recover
consistency~\cite{Matsubara2021Symplectic}. The continuous formulation remains
useful when the discretization changes with the design or when the continuous
problem is the intended optimization target.

The early graphics adjoints should not all be assigned to A3. McNamara
\etal~\cite{McNamara2004} explicitly choose a discrete adjoint, treating their
linear projection and diffusion solves as exact operators; this is A1* under
our convention. Treuille \etal~\cite{Treuille2003} differentiate the steps of
their discrete fluid simulation. Wojtan \etal~\cite{Wojtan2006} use an
integrator-swapping approximation, deriving a fully implicit adjoint for a
linearized-implicit cloth forward solver. Li \etal~\cite{Li2025FlowMapAdjoint}
take a continuous route for flow maps. The classification follows what each
method differentiates rather than its use of the word adjoint.

Outside graphics the row is well populated, and the machine-learning
literature has already compared continuous and discrete adjoints in practice. Neural
ODEs~\cite{Chen2018NODE} are an A3 method in the strict sense: rather than backpropagating through the ODE solver, which the paper considers
and rejects for its memory cost, they treat the solver as a black box and use an
approach that ``computes gradients by solving a second, augmented ODE
backwards in time'', an adjoint derived from the continuous dynamics and
discretized afterward. Later work outside graphics shows the consequences of the referent distinction. Gholaminejad \etal~\cite{Gholaminejad2019} showed that this
``optimize-then-discretize approach may lead to divergent training due to
inconsistent gradients for small time step sizes'', and proposed recovering
consistency by differentiating the discretization instead, at the cost of
storing or recomputing the forward trajectory; Zhuang \etal~\cite{Zhuang2020} reach the same conclusion from the
checkpointing side, finding that ``the adjoint method has numerical errors in
reverse-mode integration'' and recovering accuracy by checkpointing. These studies expose two distinct numerical issues: inconsistency between the
continuous and discrete adjoints, and instability when reconstructing the
forward trajectory by backward integration. Checkpointing addresses state
reconstruction, while differentiating the discrete scheme or using a
consistent forward--adjoint pair addresses the derivative mismatch. A fixed
production discretization makes consistency with that discretization a natural
objective; continuous adjoints remain useful when the continuous model is the
desired target or the mesh changes with the design.

\subsection{Axis B: the realization}
\label{sec:realization}

The realization describes how the backward pass is organized. Two useful
attributes are how derivatives are constructed (automatically, manually, or
by a hybrid of the two) and what computation produces them, such as reversing
operations or solving adjoint equations. Locality and state recovery are
further attributes; neither follows from who writes the code.

We retain three operational categories for the systems comparison:
B1 for automatically generated differentiation, B2 for equation-level
sensitivity or adjoint solves, and B3 for custom operator-level adjoints.
These labels are not a Cartesian product of construction and locality.
B2 includes both implicit derivatives of discrete equations and discretized
continuous sensitivity or adjoint equations. When this equation-level solve
organizes the backward pass, B2 takes precedence regardless of how its
Jacobians are constructed. Exact linear primitives inside an otherwise
differentiated program are recorded by A1*. Locality-aware
AD~\cite{LocalAD2026} illustrates that automatic construction can exploit
mesh structure without by itself establishing a complete block-local
simulator backward pass.

\paragraph*{Labelling rule.}
Tables~\ref{tab:methods} and~\ref{tab:methods2} were labeled by the following
procedure, stated so that it can be applied to an engine we have not seen.
\begin{enumerate}
\item \emph{Unit.} The label describes the organization of the backward pass of
  one time step as a whole, not every operator inside it. A hand-derived
  adjoint that calls automatic differentiation for a local Jacobian-vector
  product, or a tape that wraps a hand-written kernel, keeps the label of its
  organizing structure.
\item \emph{Referent.} Identify the nonlinear solve the step runs and ask whether
  the backward pass differentiates its iterations (A1), the equation it was
  solving, by the implicit function theorem at whatever iterate was reached
  (A2), or a continuous model (A3). A \emph{linear} solve run to tolerance,
  the pressure projection of a fluid step or the inner solve of a Newton step,
  that the backward pass differentiates as the exact operator makes the label
  A1*, exact for the idealized program in which that solve is exact; this is
  how $\Phi$Flow appears in
  Table~\ref{tab:methods2}, its pressure solve differentiated as converged
  and the rest of its step as executed. A backward pass that deliberately
  differentiates a different model or objective is marked \emph{surr.}; its
  surrogate referent is described explicitly rather than assigned to A1, A2,
  or A3 for the production forward map.
\item \emph{Realization.} Use B2 when the backward pass is organized around
  equation-level sensitivity or adjoint solves, including discretized
  continuous adjoint equations. Otherwise, use B1 for a backward pass
  generated by an AD framework or compiler, and B3 for custom operator-level  adjoints. For B3, state whether the operators reverse discrete operations
  or implement a continuous adjoint; a continuous adjoint counts as B3 rather than B2 when its discretization is built on the forward solver's own maps and structures rather than solved as a separate equation. In the A1/A1* operation-mirroring case,
  the backward pass can reuse the forward solver's structure, including
  transposed linear solves; this is not a guarantee for every B3 method.

\item \emph{Components.} If a step combines different referents or
  realizations, identify the component motivating the primary label and
  state the remaining components in the note. For example, de Avila
  Belbute-Peres \etal~\cite{Peres2018}, Liang \etal~\cite{Liang2019}, and
  Qiao \etal~\cite{Qiao2020Scalable} combine framework differentiation with
  an implicitly differentiated contact or collision-response layer. Their
  primary label describes that layer, while the note records the surrounding
  differentiation. Computational cost can motivate which component is
  emphasized, but does not make the remaining components share its
  mathematical referent. List both when a single label would obscure this
  distinction.

\item \emph{State.} Record separately what the backward pass keeps (the State
  column), since the realization does not determine it.
\end{enumerate}

A third question, where the state the backward pass linearizes at comes
from, is kept out of this axis. The state must be held,
recomputed from a checkpoint, or recovered by running the map backward, and
that choice sets the storage asymptotics. It is orthogonal to both of the
questions above (two methods identical in construction and in backward
computation can differ in it by a factor of the solver depth), so it is
reported separately, as its own column in Tables~\ref{tab:methods} and~\ref{tab:methods2}. Memory does
not follow from a realization label.

\paragraph*{B1, recorded.}
Generate the backward pass with an AD framework or compiler, retaining or
recomputing the intermediate values it needs. Fully stored reverse mode is
one implementation, and the category spans a spectrum of engineering.
At one end, simulators are written directly in an array framework such as PyTorch
or JAX and inherit its generated reverse pass and residual-storage rules~\cite{Paszke2019,Bradbury2018};
JAX~MD~\cite{Schoenholz2020} and Brax~\cite{Freeman2021Brax} are prominent
examples, the latter built around simple rigid-body physics so that thousands of
unrolled environments run on an accelerator. At the other end are
domain-specific compilers that generate adjoint code by source transformation.
DiffTaichi~\cite{Hu2020DiffTaichi} is the canonical instance in graphics: it
extends the Taichi language~\cite{Hu2019Taichi} with a ``megakernel'' reverse
mode that preserves the arithmetic intensity and parallelism of the forward
kernels, which operator-level taping can fail to preserve. Its reported gains over an equivalent
formulation in a general-purpose array framework are attributed by its authors
to this structural preservation rather than to any difference in the mathematics. NVIDIA~Warp~\cite{Macklin2022Warp} follows a similar compiled-kernel philosophy,
and Tiny Differentiable
Simulator and its successors~\cite{Heiden2021TDS} take the recorded route while
embedding learned augmentations inside the engine rather than as residual corrections outside it. The most recent engines consolidate: Genesis~\cite{Genesis2024} and Newton~\cite{Newton2025} both build a
multi-physics, GPU-resident stack, differentiable for part of their physics, on a compiled-kernel
foundation, and Newton in particular is built on Warp, one of the
systems in Table~\ref{tab:engines}. That consolidation means the realization axis is being settled by
infrastructure decisions: whichever backward pass the
shared substrate supports is the one most downstream work will use.

Without recomputation, retaining $O(n)$ data for each of $K$ iterations over $T$
steps requires $O(TKn)$ memory. The actual tape constants and problem size
determine whether this exceeds device capacity.
Checkpointing~\cite{Griewank2000} stores selected states and recomputes
intervening segments. The optimality of a binomial schedule assumes a
specified storage and execution-cost model, including a known step count; the
achievable trade-off also depends on recomputation cost, checkpoint size, and
whether the horizon is known in advance. Thus $O(TKn)$ describes fully stored
unrolling, not an unavoidable property of B1.

The same trade-off drives the adjoint formulation of neural
ODEs~\cite{Chen2018NODE}, for which checkpointing variants, one of them adaptive,
exist~\cite{Gholaminejad2019,Zhuang2020}. Compiler-level quantization of
simulation state~\cite{Hu2020Quantaichi} suggests a possible route to
reduced-precision trajectory storage. That work evaluates quantized forward
simulation; using such storage for a backward pass additionally requires
analysis of reconstruction and gradient errors. Checkpointing also
interacts badly with solvers whose iterates are not reproducible, such as
anything using atomics with nondeterministic ordering or randomized sweep orders,
because the recomputed segment is then a slightly different trajectory than the
one differentiated, and keeping the executed-computation guarantee literally
would require reproducing it bit for bit.

Tool-built differentiation can also exploit structure. 
Several works~\cite{fernandez2025symx, herholz2024mesh, Tang2026YASPS} show that derivative
generation for mesh-based computation can be made more efficient, YASPS by
making the mesh relations explicit in a differentiable intermediate
representation. Symbolic and mesh-aware derivative generation can accelerate
forward solvers; extending these structures to trajectory-level sensitivities
requires additional backward-pass machinery.
\etal~\cite{LocalAD2026} demonstrate automatic locality-aware differentiation of mesh-based
computations on the GPU: derivatives are taken per
element by forward-mode AD, confined to registers and shared memory, and
assembled into sparse gradients and Jacobians, so the tool exploits the mesh's
locality rather than treating the computation as a flat trace. That narrows the gap between a
recorded and a rederived realization from the recorded side, and suggests the two
are better seen as ends of a spectrum in how much forward structure the backward
pass is allowed to know about.

\paragraph*{B2, resolved.}
Obtain sensitivities or adjoints by solving linear equations derived from
the defining model, rather than reversing the finite nonlinear solver.
For A2, \eqref{eq:ift} is evaluated at the final state, so the nonlinear
solver's $K$ iterates need not be stored. For A3, continuous sensitivity or
adjoint equations are discretized and solved; their trajectory-state
requirements depend on the time integration and recovery policy.
Real2Sim~\cite{Hahn2019Real2Sim} and neural ODEs~\cite{Chen2018NODE} are examples of this latter organization.

Solving an A2 adjoint system does not require assembling or factorizing it. A matrix-free Krylov method needs only
Jacobian-vector and vector-Jacobian products, which for most simulators are as
cheap as a forward force evaluation, and this is how the approach scales to
problems where an explicit Jacobian would not fit. For the A2 case, the required adjoint solve is defined by the equation or
subsystem differentiated implicitly, and its coupling range follows from
that subsystem, not from the label. When the equation is the whole implicit step, as
for DiffPD, ADD or DiffXPBD, the system~\eqref{eq:ift} couples every degree of
freedom at once, and even matrix-free each Krylov iteration is a product over
the whole mesh. When it is a contact layer, the solve couples the active set:
Qiao \etal~\cite{Qiao2020Scalable} apply it per impact zone, whose extent depends on the connected contact structure. The cost that follows is therefore a property of the
whole-step case, and there it is a loss for forward methods whose
performance argument is that they never form a global system: a
Gauss--Seidel, XPBD or block coordinate descent step visits one vertex or
constraint at a time and never couples the whole mesh in a single operation,
while an equation-level backward pass for the same step does. For a forward
method that already assembles and factorizes a global matrix, as Newton or
projective dynamics does, the backward can reuse factorization work, as in DiffPD~\cite{Du2021DiffPD}, while its additional cost depends on the adjoint iterations and contact treatment; for one
designed to avoid the global system, the backward pass introduces a coupling
and a cost profile that the forward pass avoids. B2 also
requires local regularity for classical implicit differentiation. Buckling
and degenerate constraints can produce singular or ill-conditioned Jacobians,
while contact transitions may instead invalidate smoothness or active-set
stability. These conditions must be checked for the actual time-discrete
system; static buckling does not by itself imply a singular dynamic-step
Jacobian.

\paragraph*{B3, rederived.}
Construct custom operator-level adjoints using the structure of the
differentiated model. In the A1/A1* operation-mirroring case, let
$\bs^{(j)}=U_j(\bs^{(j-1)};\bth)$ for $j=1,\ldots,m$, and define
$U_j^*(\adj{\bs};\bs^{(j-1)})=
 [D_{\bs}U_j(\bs^{(j-1)};\bth)]\transp\adj{\bs}$.
The reverse sweep applies these Jacobian-transpose actions in the order
$j=m,\ldots,1$:
\begin{equation}
\adj{\bs}^{(0)} = [D_{\bs}U_1]\transp\cdots[D_{\bs}U_m]\transp
                  \adj{\bs}^{(m)},
\label{eq:reversesweep}
\end{equation}
with each Jacobian evaluated at its own forward input. Each reverse update
also accumulates $[D_{\bth}U_j]\transp\adj{\bs}^{(j)}$ into the parameter
gradient. For block-local updates this can avoid assembling an additional
global adjoint system. Custom operators implementing a continuous adjoint
form a separate A3 case and need not mirror discrete forward operations.

For the discrete operation-mirroring case, B1 and B3 can differentiate the
same object while differing in how the backward pass is constructed. B1
obtains it from an AD framework or compiler; B3 supplies custom operator
adjoints. The latter can avoid a generic primitive-level tape but may still
store states, replay data, and control-flow decisions. McNamara
\etal~\cite{McNamara2004} provide an example: they derive the adjoints of
fluid-step stages by hand and apply them in reverse order, treating projection
and diffusion as exact linear operators. Under our convention this is
A1*$\times$B3. A compiler-generated version with the same operators and
derivative conventions can target the same derivative; finite-iteration and
finite-tolerance implementations need to be checked separately.
ADD~\cite{Geilinger2020ADD}, by contrast, uses an equation-level adjoint and
is B2.

B3 is broader than the block composition.
Most of the entries Tables~\ref{tab:grid}, \ref{tab:methods} and~\ref{tab:methods2} place
under B3 mirror some other structure: the operation sequence of a fluid step, the
recursion of an articulated-body algorithm, or custom continuous-adjoint
operators for flow maps. Many graphics solvers have no block structure. Newton
with a sparse direct factorization, preconditioned conjugate gradients and
multigrid are global by design, and they remain the norm offline. For these
A1 still exists, as the derivative of the Newton or Krylov iterations that ran,
and B3 still applies, as the adjoint of that iteration. Christianson's reverse accumulation instead iterates the adjoint equation at the converged point and discards the executed iterates, so it sits with A2~\cite{Christianson1994}. A Krylov recurrence or a multigrid cycle is a
composition of matrix-vector products, smoothers and transfers, each with a
transpose. The memory advantage is absent: the inputs of each
iteration's adjoint are global vectors, so there is no block-local object to
shrink the stored state to, and the saving that block mirroring offers is
specific to solvers that decompose, which Section~\ref{sec:grid}
returns to. If an inner linear system is solved accurately and differentiated as an exact
operator, A1* records that idealization. A direct factorization still incurs
floating-point error. With an accurate inner solve, the truncated outer Newton
iteration may dominate the difference between the finite-solver and root
sensitivities, but its magnitude also depends on conditioning, residual
sensitivity, and backward-solve accuracy.

For the block-local operation-mirroring case, a custom adjoint can remove the
need for a backward-only global system, but it does not remove the need for
state. Each local Jacobian must still be evaluated at the input seen by its
forward update, so that input must be retained, recomputed, or recovered. The
reverse sweep for vertex block descent~\cite{Shu2026SolverAdjoint} stores nine
doubles, or 72 bytes, of replay data per active vertex per sweep. The reported
approximately 4 KB per vertex per sweep uses a different accounting: it is an
empirical peak of total allocations, including mesh and other data. The
reported storage still scales linearly with $K|V|$. Removing that $K$ dependence
requires a state-recovery strategy, which a block decomposition does not supply. McNamara
\etal~\cite{McNamara2004} likewise hand-derive their operator adjoints while
storing the forward trajectory.

Custom adjoints are not tied to one referent. In the discrete case, the fluid
adjoints of McNamara \etal~\cite{McNamara2004} and Takahashi
\etal~\cite{Takahashi2021} mirror the stages of a fluid step, while Qiao
\etal~\cite{Qiao2021Articulated} differentiate the executed Gauss--Seidel
iterations of an articulated-body contact solver. A transposed pressure solve
treated as an exact linear operator is recorded by A1*, rather than by an
unqualified claim of exactness for the finite solver; McNamara \etal\ are an
explicit example. Custom continuous adjoints instead belong to A3.

An equation-level adjoint is classified as B2, whether its ingredients are
hand-derived or generated automatically. DiffPD \cite{Du2021DiffPD} and
DiffCloth~\cite{Li2022DiffCloth} are examples: DiffPD reuses the factorized
forward matrix in its local--global adjoint iteration. In DiffCloth, the
contact-free adjoint uses the elastic step Hessian, while frictional contact
adds contact-response derivative terms to the global system.
ADD~\cite{Geilinger2020ADD} solves an adjoint system built from its integrator
residual; Nimble~\cite{Werling2021Nimble} differentiates its contact problem;
and DiffXPBD~\cite{Stuyck2023DiffXPBD} assembles derivative blocks and uses
conjugate gradients, with a positive-definite modification of those blocks.
Solving a system iteratively does not change its equation-level target,
although solver tolerances and matrix modifications affect the returned
approximation.

All of these methods use the solver's structure, and they differ in which
object they differentiate.

The cost of a custom adjoint is implementation and maintenance effort. It
needs derivative actions for the operations it uses, supplied analytically, by
local AD, or by another suitable interface. Data-dependent branches and line
searches require retaining or reproducing the executed decisions and
differentiating the corresponding smooth branch; they do not by themselves
rule out a custom adjoint. A black-box routine without a derivative interface
remains an obstacle. Changing an integrator can require new custom derivative
code, whereas a framework-generated realization can often reuse the
framework's differentiation support.

\subsection{The cells of the grid}
\label{sec:grid}

Table~\ref{tab:grid} crosses the two axes, and the following observations draw on it.

\emph{Which cells are occupied, and why.} Table~\ref{tab:grid} reflects both
the methods surveyed and our operational labeling convention. Equation-level
adjoints are assigned to B2 even when AD constructs their Jacobian products;
automatic implicit differentiation already exists~\cite{Blondel2022}.
Discretized continuous adjoint solves can also occupy A3$\times$B2, as in
Real2Sim~\cite{Hahn2019Real2Sim}. B3 may instead organize a discrete or
continuous adjoint as custom operator-level updates. An empty cell is
therefore not, by itself, evidence of mathematical impossibility or an
unimplemented construction. Methods should be compared by differentiated object, backward
organization, and state requirements. The remainder
of this section focuses on discrete custom adjoints that mirror forward
operations.

\emph{Exactness is assessed relative to a referent.} A1 targets the finite executed
computation, A1* an idealized program with exact linear primitives, A2 a
selected root of the discrete equations, and A3 the continuous dynamics. A
target does not guarantee that an implementation attains it: finite
tolerances, discretization, modified Jacobians, and nonsmooth operations may
introduce additional qualifications. The corresponding sensitivities generally
differ, but can coincide in special cases. When two methods return different
gradients for the same nominal simulator, both the referent and the accuracy
of its implementation should therefore be checked.

\emph{Memory is decided by state recovery.} A1 does not imply a tape; that it usually has meant one is
an artifact of how A1 has been realized. Nor does the realization settle
memory on its own. A discrete operation adjoint
needs the state its forward operation saw; continuous adjoints likewise need
an appropriate forward-state representation. These states can be obtained by
several policies: retain them, recompute them from checkpoints, or recover
them by running the map backward. The state-recovery policy helps determine
the storage asymptotics and is reported separately from both axes. Conflating it with the
realization makes a constant-factor saving read as an asymptotic one.

Discrete custom adjoints can mirror the forward solver's operations without
assembling an additional global system. They may retain the trajectory, as in
the early fluid adjoints, or recover the required states by recomputation or
reversibility. McNamara \etal~\cite{McNamara2004} provide an A1*$\times$B3
example because their linear stages are differentiated as exact operators;
Qiao \etal~\cite{Qiao2021Articulated} reverse executed contact iterations. Two
useful structures are reversible updates and block-local
compositions, but neither replaces the need to specify the state-recovery
policy.

The first structure is reversible reconstruction. If an update has a usable
inverse, its intermediate states can be recovered during the backward pass
instead of retained. Recovering the same linearization states and using the
same local derivative conventions can reproduce the derivative of the
corresponding stored-state computation. Numerical instability, rounding, and
non-injective updates can prevent this in practice. Related constructions have
been developed in machine learning, numerical integration, and molecular
dynamics.

Reversible residual networks recompute activations instead of keeping
them~\cite{Gomez2017RevNet}, which makes activation storage independent of the number of reversible blocks,
while activations at non-reversible layers must still be stored. The symplectic adjoint method~\cite{Matsubara2021Symplectic} takes a related
but distinct route for neural ODEs: it retains time-step states as
checkpoints, discards the per-step computation graph, recomputes the required
stages, and uses a matching discrete adjoint to recover the gradient of the
numerical forward scheme. Its memory reduction comes from checkpointing and
recomputation rather than from requiring the forward Runge--Kutta method to be
reversible.

Stam~\cite{Stam2022Reversible} applies it to the adjoint of a physical simulation, with a bitwise-reversible integrator he traces to Levesque and Verlet and to Rein and Tamayo, one
that retraces its own path bitwise, with the reverse step of the adjoint
method as the stated application. His paper mixes fixed-point with floating-point
arithmetic because a pure float trajectory is not exactly reversible. Greener's reversible molecular simulation~\cite{Greener2025Reversible}
reconstructs molecular-dynamics states by reverse-time simulation and matches
taped reverse-mode AD in the reported gradient checks at small step counts. Its
force-field training experiments, however, use 200-step gradient truncation.
This both limits exploding gradients and reduces reverse computation; for the
water-model setup, losses spaced 2000 steps apart require reversing only one
tenth of the forward steps under that truncation policy. The reported training
cost should therefore not be read as the cost of exact full-trajectory
differentiation.

The reversible methods above are one construction that recurs across fields for the
same reason. It was developed in machine learning, numerical integration
and molecular dynamics rather than in graphics.

The second is \emph{mirroring the solver's block structure}, as Shu \etal~\cite{Shu2026SolverAdjoint} do on Vertex Block
Descent~\cite{Chen2024VBD}; analogous operator-level reverse implementations can be constructed for projective
dynamics~\cite{Bouaziz2014} and XPBD~\cite{Macklin2016XPBD}, provided the derivatives of their local updates are defined. Projective dynamics additionally requires differentiating its global linear solve.

The two structures have different requirements. Reversing an update requires
both an inverse and sufficient numerical stability in the backward direction.
A contraction $s_{t+1}=a s_t$, with $0<a<1$, is invertible, but its inverse
amplifies state errors by $1/a$ per step. Some projections and dissipative
contact maps are non-injective; branching alone does not imply
non-invertibility, and smooth block updates may also be singular. These
properties must be checked for the actual update map. Periodic snapshots can
limit drift during reverse reconstruction, as in
Greener~\cite{Greener2025Reversible}. Block-local differentiation instead
requires derivative actions for the local updates, which can be derived by
hand or generated by AD; data-dependent decisions may need to be retained or
replayed. The largest reported reverse-sweep VBD run also checkpoints across
time steps.

Their memory behavior depends on state recovery. Reversible reconstruction can
keep a small active reverse workspace, but
Greener~\cite{Greener2025Reversible} also retains periodic coordinate/velocity
snapshots and random seeds. With a fixed snapshot interval, total snapshot
storage grows with the trajectory length; the method should not be described
as strictly constant-memory for arbitrary horizons. Block-local adjoints
reduce the data needed per update but do not remove it: the reported
reverse-sweep replay storage remains linear in
$K|V|$~\cite{Shu2026SolverAdjoint}. Both structures can avoid a backward-only
global system in suitable implementations, while their total memory must be
compared under explicitly matched checkpointing, truncation, and replay
policies.

For the stiff, contact-rich solvers emphasized in this report, stable reverse
reconstruction can be difficult, while block-local adjoints provide another
way to reduce storage
constants~\cite{Shu2026SolverAdjoint}, and per-step checkpointing with recomputation yet another~\cite{Qiao2021Articulated}. More generally, A1
and low memory are compatible through recomputation, at potentially
substantial additional computational cost. Block structure and reversibility
can improve this trade-off, but neither is necessary for an
executed-computation derivative to use little memory. The
breadth of efficient implementations across graphics solvers remains an open
question.

A custom adjoint need not mirror a discrete block decomposition. An
A3$\times$B3 method implements a discretized continuous adjoint, potentially
exploiting structures already maintained by its forward solver;
Section~\ref{sec:models-fluids} discusses the flow-map example. Its
consistency and state requirements must be assessed for that continuous
referent rather than inferred from the discrete block-local case.

\begin{table*}[t]
\centering
\caption{The two axes crossed. Rows identify derivative targets; columns are
operational categories for the organization of the backward pass. Exactness
must be assessed against the row's target and the implementation's numerical
qualifications. Construction may be automatic, manual, or hybrid within
these categories. Memory depends on retained states, recomputation,
checkpointing, and solver workspace, not on the column alone. Discrete custom
adjoints can exploit forward operator or block structure without necessarily
eliminating trajectory storage. A3 includes both equation-level solves and
custom continuous-adjoint operators. Empty entries reflect the labeling
convention and the examples selected, not impossibility claims.
Entries are representative, not exhaustive.}

\label{tab:grid}
\small
\begin{tabularx}{\textwidth}{@{}L{3.1cm} >{\hsize=.78\hsize}X >{\hsize=.92\hsize}X >{\hsize=1.3\hsize}X@{}}
\toprule
& \textbf{B1 recorded} & \textbf{B2 resolved} & \textbf{B3 rederived} \\
& generated reverse operations; fully stored unrolling $O(TKn)$
& equation-level sensitivity or adjoint solves; memory depends on target and state recovery
& custom operator-level adjoints; memory depends on state recovery \\
\midrule
\textbf{A1 executed computation} \newline targets the finite program on a differentiable branch; A1* treats internal linear solves as exact
& DiffTaichi~\cite{Hu2020DiffTaichi}, Brax~\cite{Freeman2021Brax},
  JAX~MD~\cite{Schoenholz2020}, SPNets~\cite{Schenck2018SPNets},
  NeuralSim~\cite{Heiden2021TDS}
& no representative listed: the equation-level solve for an implicit step normally targets A2
& \emph{via the operator sequence, trajectory stored or checkpointed:} adjoint fluid
  control~\cite{McNamara2004} (A1*), Takahashi \etal~\cite{Takahashi2021} (A1*),
  Qiao \etal~\cite{Qiao2021Articulated}, ChainQueen~\cite{Hu2019ChainQueen}, symplectic adjoint~\cite{Matsubara2021Symplectic} \newline
  \emph{via reversibility:} reversible nets~\cite{Gomez2017RevNet},
  bitwise-reversible integrator~\cite{Stam2022Reversible} (state recovery),
  reversible MD~\cite{Greener2025Reversible} (untruncated formulation;
  training uses 200-step truncation) \newline
  \emph{via block structure:} reverse-sweep VBD~\cite{Shu2026SolverAdjoint} \\
\addlinespace[3pt]
\textbf{A2 converged equation} \newline targets a regular root; accuracy also depends on forward/adjoint solves and Jacobian modifications
& automatic construction is possible; an equation-level adjoint is labeled B2
& OptNet~\cite{Amos2017}, cvxpylayers~\cite{Agrawal2019},
  DEQ~\cite{Bai2019}, Le~Lidec \etal~\cite{Lidec2024}, DiffPD~\cite{Du2021DiffPD},
  ADD~\cite{Geilinger2020ADD}, Nimble~\cite{Werling2021Nimble},
  DiffXPBD~\cite{Stuyck2023DiffXPBD} (modified Jacobian)
& equation-level adjoints are assigned to B2, including hand-constructed ones \\
\addlinespace[3pt]
\textbf{A3 continuous dynamics} \newline differentiate then discretize
& generated components are possible; classification follows the organizing adjoint computation
& Real2Sim~\cite{Hahn2019Real2Sim}: continuous sensitivity/adjoint equations discretized with BDF2; neural ODEs~\cite{Chen2018NODE}: augmented adjoint ODE
& flow-map adjoints~\cite{Li2025FlowMapAdjoint}; consistency and state requirements depend on the custom continuous-adjoint operators \\
\bottomrule
\end{tabularx}
\end{table*}

\subsection{Using the taxonomy}
\label{sec:tax-compare}

The referent should follow the intended optimization problem and the solver
regime. A2 is attractive when a root of the discrete equations is the intended
target and the forward and adjoint solves approximate its sensitivity
accurately. This requires checking conditioning and derivative accuracy rather
than relying on a tight forward tolerance alone. When a fixed finite iteration
budget is part of the model being optimized, A1 directly targets that
computation. An approximate A2 gradient may still be useful, but its target
and approximation should be stated and evaluated at the production solver
configuration.

The realization should follow the solver's structure, development constraints,
and memory budget. A custom adjoint can exploit regular structure but requires
implementation and maintenance effort. Automatically generated differentiation
is often convenient for rapidly changing solvers, and checkpointing can reduce
its storage requirements. An equation-level sensitivity or adjoint solve is attractive when its
target is appropriate and the required adjoint system can be solved
efficiently, especially when forward factorizations or preconditioners can be
reused. None of these labels alone determines locality or memory.

\paragraph*{Classifying an engine.}
First identify the derivative target: executed iterations, an idealized
program with exact linear operators, a regular root of discrete equations,
continuous dynamics, or a stated surrogate. Then identify the organizing
computation: equation-level sensitivity or adjoint solves (B2), otherwise
framework- or compiler-generated differentiation (B1), or custom operator-level
adjoints (B3). Record automatic, manual, or hybrid construction and the relevant
local or global operators separately. Finally, state what is retained or
recovered: a tape, trajectory states, checkpoints, replay data, or
reverse-reconstructed states. Mixed components and finite-tolerance
qualifications belong in the notes to Tables~\ref{tab:methods}
and~\ref{tab:methods2}. The diagnostic in Section~\ref{sec:eval} measures
the resulting behavior at the production configuration.

We do not offer a ranking. Every published comparison we examined that
appears to establish one sets a particular implementation against another at a
particular problem size, and the axes above suggest such results are evidence
about a cell rather than about a method.

\section{Discontinuity: Contact, Friction, and Collision}
\label{sec:contact}

\begin{figure*}[t]
\centering
\begin{minipage}[t]{0.5891\linewidth}\includegraphics[width=\linewidth]{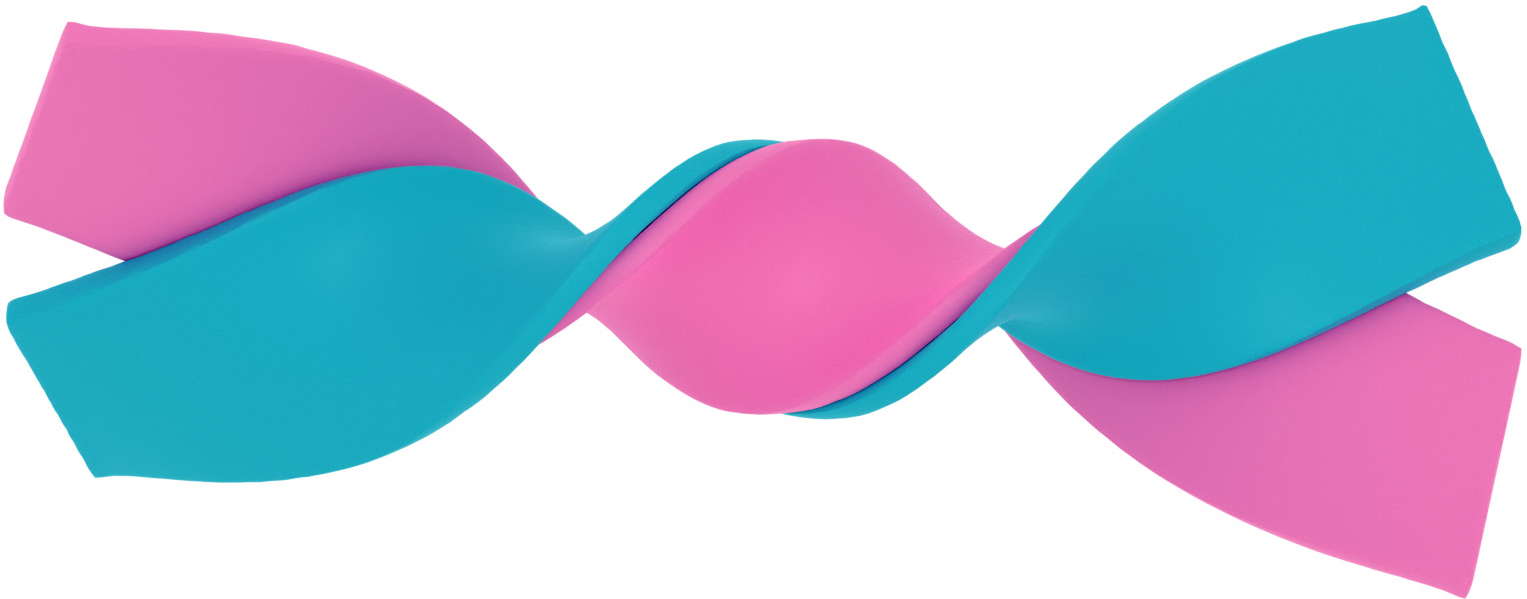}\end{minipage}%
\begin{minipage}[t]{0.4108\linewidth}\includegraphics[width=\linewidth]{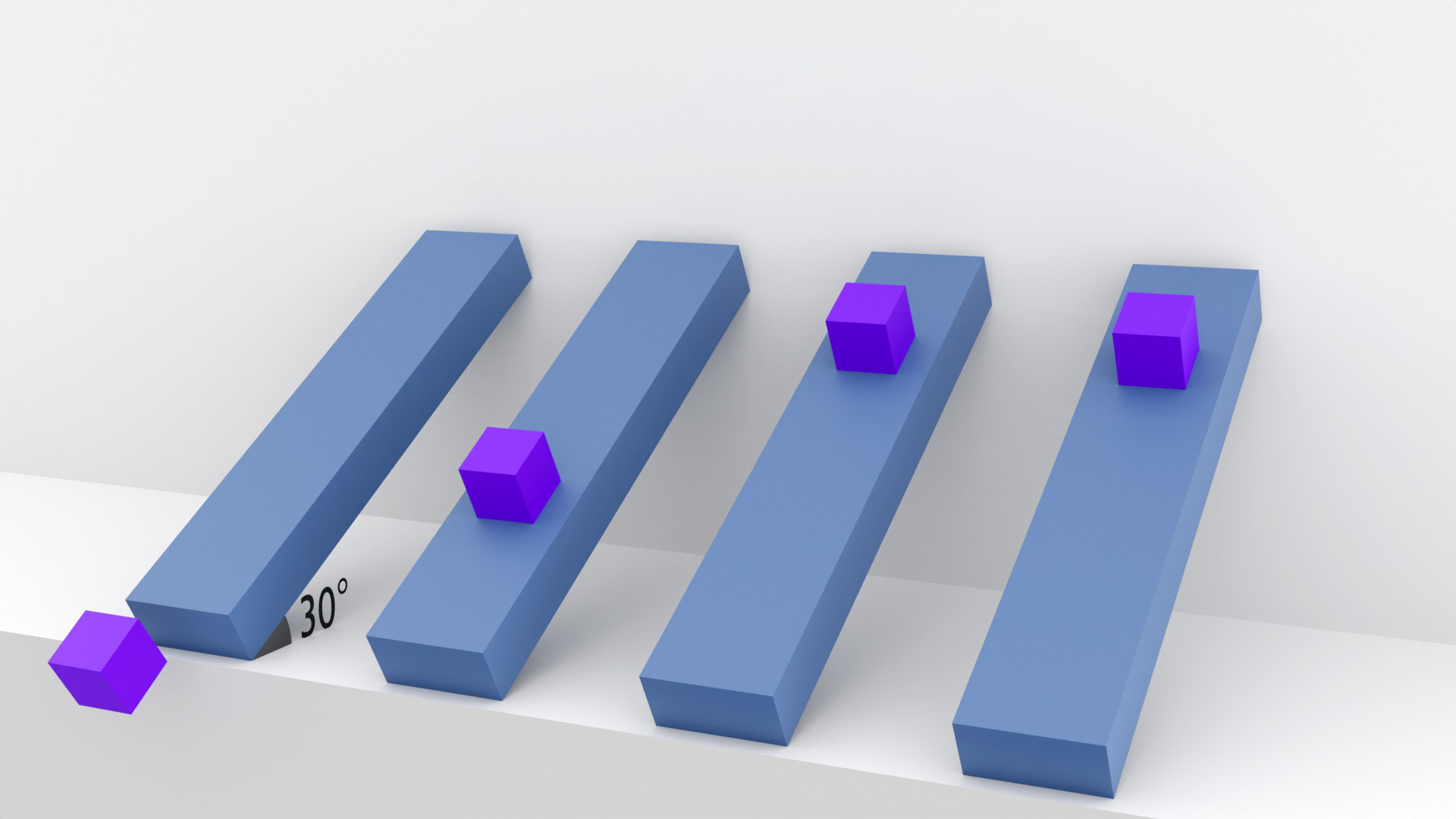}\end{minipage}
\caption{What a contact solver is asked to resolve, and why contact modes
matter. \emph{Left}, two elastic beams twisted around one another, 97
thousand vertices between them, in sustained frictional contact along the whole
interface while they buckle. The active pairs change along the trajectory.
Local branch derivatives can remain valid despite these temporal changes, but
may poorly predict finite parameter updates that change the contact sequence. \emph{Right}, an elastic cube released
on an incline at four friction coefficients. Whether it sticks or
slides is decided by the friction cone. At a regular stick--slip transition,
the branch derivatives can differ (Section~\ref{sec:contact-why}). Images reproduced, cropped but
otherwise unaltered, from \cite{Chen2024VBD} under the authors' CC~BY~4.0
license.}
\label{fig:contactload}
\end{figure*}

The taxonomy of Section~\ref{sec:taxonomy} concerns \emph{how} a derivative is
obtained from a computation, assuming the relevant local derivatives exist.
Contact can violate that assumption. This section develops an additional,
orthogonal axis: what a method does where the simulation map is non-smooth.
An engine makes a choice on each axis, and conflating them is a
recurring source of confusion, because a paper that improves contact gradients
and a paper that improves solver adjoints may both be described as ``better
gradients'' while addressing different deficiencies.
Figure~\ref{fig:contactload} shows what a solver is asked to resolve in
this regime, where individual contact laws and changes in the active set both
matter. The section is restricted to discontinuity arising
from contact, friction and collision in solids. Fluids have discontinuities of
their own, in the merging and splitting of free surfaces and in the discrete
operators that track them; McNamara \etal~\cite{McNamara2004} met one in the
fast-marching redistancing of a level set and derived its adjoint by observing
that the algorithm is locally continuous, and we do not treat that class
further.

\subsection{Sources of non-smoothness in contact}
\label{sec:contact-why}

At the position level, a discrete rigid-contact constraint can be written
using the gap $\phi$ at the constrained configuration and the normal impulse
$\lambda$ over the step:
${0 \le \phi \perp \lambda \ge 0}$, meaning that at each contact either the bodies
are separated with zero normal impulse, or they are touching with a
non-negative normal impulse. Velocity-level time-stepping formulations instead
pair the impulse with a scheme-specific normal-velocity residual, which may
include gap stabilization or restitution; the gap formula is not a universal
time-stepping law.
Coulomb friction couples tangential impulse to normal impulse through a cone,
together with a maximum-dissipation or equivalent stick--slip condition.
The resulting time-stepping schemes, in the tradition of
Stewart and Trinkle~\cite{Stewart1996} and Anitescu and
Potra~\cite{Anitescu1997}, are linear or nonlinear complementarity problems. The
same formulation underpins contact-implicit trajectory
optimization~\cite{Posa2014}, which optimizes through contact by treating the
complementarity conditions as constraints rather than by differentiating a
forward solve, and is a major alternative for contact-rich trajectory
optimization.

This creates three distinct obstacles, and different methods address
different ones.

\begin{enumerate}
\item \textbf{Non-smoothness at fixed topology.} Even with a fixed set of
candidate contacts, the solution has distinct branches (separating, sticking,
sliding). Locally unique, non-degenerate solutions admit branch derivatives,
but derivatives can differ or fail to exist at mode boundaries.
\item \textbf{Discrete events.} Contact modes and event times can change
under parameter perturbations. Differentiating each step at its current active
set can give the correct local derivative of the discrete trajectory when the
branch sequence is stable. Matching continuous-time impact sensitivities also
requires event-time dependence when the discrete contact treatment omits it.
The earliest rigid-body control work
already shows the effect: Popovi\'c \etal~\cite{Popovic2000} plot the
simulation function of a particle bouncing off a polygonal obstacle as
piecewise smooth, discontinuous wherever the impulse switches facet, and
vary the surface normals to narrow the gaps between the smooth pieces of the feasible exit set, and use curvature-adapted polygonal approximations of curved obstacles so that a first-order
prediction still serves. Positions may remain continuous in time while impact
velocities jump; facet or event-sequence changes can also make the
parameter-to-output map discontinuous. Saltation matrices and event-time
sensitivities address isolated transversal events with a locally stable event
sequence; grazing and simultaneous impacts require separate treatment.
\item \textbf{Gradient collapse in the model.} Some physically meaningful
sensitivities are zero or undefined regardless of the numerical treatment. A
ball resting on a table has no gradient of final position with respect to
initial height, once dropped from any height that leads to rest at the same
place. The information is absent from the dynamics themselves rather than lost in
discretization, so this is a modeling limit and no differentiation scheme
recovers it.
\end{enumerate}

Much of the literature's disagreement about ``correct'' contact gradients
comes from prioritizing different obstacles.

\subsection{Complementarity branch derivatives}
\label{sec:contact-complementarity}

The most direct approach differentiates the complementarity formulation on a
stable active set. For a locally unique, non-degenerate solution, the
corresponding active-set equations define a smooth local solution map. De~Avila Belbute-Peres
\etal~\cite{Peres2018} embedded an LCP solver as a differentiable layer,
computing gradients by differentiating the LCP's complementarity conditions at the solution in the manner
of~\cite{Amos2017}. Nimble~\cite{Werling2021Nimble} develops analytic LCP
derivatives that exploit active-set structure, together with optional
\emph{complementarity-aware} heuristic directions. These heuristics depart
from the local derivative in some contact regimes, including cases where the
true local gradient is zero, to provide an optimization signal.

Zhong \etal~\cite{Zhong2022Correct} ask directly whether these gradients are
correct in the sense their users assume. They examine what differentiable
contact simulators return with respect to position, velocity and control, and
find discrepancies that ordinary use does not reveal. Their checks complement
the diagnostics of Section~\ref{sec:quality}.

The forward simulation in this family is a hard-contact simulation with no
artificial compliance, an advantage when the downstream task depends on
accurate contact mechanics. Velocity-level
complementarity does not eliminate interpenetration in practice, since the
constraint is imposed on velocities and drift is corrected by stabilization,
but the penetration that remains is a discretization error rather than a
modeling parameter traded against gradient smoothness.
Under the stated regularity assumptions, the returned object is a branch
Jacobian. At a mode boundary, the method must specify its generalized
derivative or heuristic selection. A valid local derivative need not provide
an effective direction for crossing into another contact mode.

\subsection{Compliant and penalty relaxation}

Another strategy relaxes hard contact. Penalty and spring-damper laws are simple
examples; their differentiability depends on the chosen activation and friction
regularization. The convex contact solver of MuJoCo~\cite{Todorov2012} uses a relaxed optimization model, while the original Brax~\cite{Freeman2021Brax} uses
Baumgarte-stabilized velocity-level collision impulses. These are distinct
contact models, rather than interchangeable spring-damper laws. Le~Lidec
\etal~\cite{Lidec2023contact} compare the models of MuJoCo, Drake, Bullet,
DART, ODE and RaiSim and the qualitative differences they induce. Analytically differentiable formulations such as
ADD~\cite{Geilinger2020ADD} use compliant contact with explicit friction
smoothing to obtain analytical gradients by adjoint sensitivity analysis for multibody systems.

The trade-off between compliance and gradient behavior depends on the system.
Increasing stiffness can worsen conditioning in multiscale problems, with the
effect depending on the time step, masses, constraint geometry, and
preconditioning, so stiffness alone does not determine the condition number.
Constraint formulations with compliance~\cite{Tournier2015,Macklin2016XPBD}
soften that trade for the forward solve by writing the constraint with its
compliance $\alpha$, the inverse of its stiffness, rather than with the
stiffness itself: the multiplier system carries $\alpha/h^2$ on its diagonal,
which tends to zero as the constraint stiffens. At fixed time step and with
bounded constraint and mass operators, this avoids the penalty matrix's growing
entries. The condition number can still be unbounded: with redundant
or degenerate constraints the system approaches singularity as
$\alpha\to0$, although that ill-conditioning comes from the constraint
Jacobian rather than from growing penalty entries. DiffXPBD~\cite{Stuyck2023DiffXPBD} constructs
an adjoint for compliant constraint dynamics, but the works reviewed here do
not establish a systematic comparison of backward conditioning between
compliance-form and penalty-form equations. Paulus \etal~\cite{Paulus2025}
address a different aspect: DiffMJX improves numerical gradient accuracy with
adaptive integration, while contacts from distance (CFD) supplies modified
contact derivatives. Section~\ref{sec:contact-decouple} discusses this
forward/backward decoupling.

\subsection{Barrier and interior-point formulations}
\label{sec:contact-barrier}

A third family approximates contact with barrier potentials on the strictly
feasible, non-intersecting domain. Incremental Potential Contact
(IPC)~\cite{Li2020IPC} combines these potentials with continuous collision
detection (CCD) and a feasibility-preserving line search to maintain
intersection-free and inversion-free configurations. The codimensional
extension~\cite{Li2021CIPC} covers cloth, rods, and points.

For differentiable simulation this is a convenient starting point, with the
following regularity. On the feasible domain, the contact
potential is $C^1$ and piecewise $C^2$, with pieces determined by geometric
feature configurations. Away from feature-transition boundaries, a stationary
point with a nonsingular Hessian admits implicit differentiation. At a smooth
local minimizer the true Hessian is positive semidefinite, and it is positive
definite if the minimizer is non-degenerate; the barrier alone does not ensure
this. Nor does piecewise $C^2$ regularity alone guarantee Newton convergence.
IPC modifies element Hessians to obtain descent directions and uses a
CCD-filtered line search to keep iterates feasible~\cite{Li2020IPC}; local
convergence additionally depends on regularity and non-degeneracy.

Those devices make the forward solve robust. They do not change the derivative
of the minimizer, which is still given by~\eqref{eq:ift} with the unprojected
Hessian, and nothing in the backward pass calls for the projected one. The
adjoint system is a linear solve rather than a descent step, so it needs the
true Hessian to be non-singular rather than positive definite, and a pivoted
symmetric or LU factorization handles an indefinite matrix, at the price of not
reusing the forward solver's Cholesky factors. The adjoint system should therefore be
assembled from the unprojected Hessian. Huang
\etal~\cite{Huang2024DiffIPC} do this: their adjoint system is built from the
stiffness and mass matrices of the discretization; projection is confined to the
Newton update and applied only when the unprojected direction fails to descend,
and the barrier stiffness is held fixed, since adapting it would require
differentiating through the update. Under those choices, the conditions for
exactness are the ones~\eqref{eq:ift} always carries, a converged and
non-degenerate minimizer. A backward pass that instead reuses the projected
matrix the forward solver factorized returns an approximation to that
derivative. With no projection it uses the original derivative system;
otherwise the result is generally perturbed, with error depending on the
projection, the right-hand side, and the conditioning of the original and
modified systems. Reading that
result as the exact derivative of some other object would require naming the
residual whose Jacobian the projected matrix is, and we are not aware of a
formulation that does so. The shortcut returns an approximation to the A2
derivative, and the reporting checklist of
Section~\ref{sec:eval} asks for this to be stated. Barrier stiffness
and small gaps can create widely separated curvature scales and difficult
linear solves. IPC's discussion of unnecessary ill-conditioning concerns the
choice of barrier stiffness, rather than a universal loss of invertibility as a
gap closes. Related barrier-based work on rigid bodies~\cite{Ferguson2021RigidIPC}, on
affine bodies~\cite{Lan2023DiffIPC}, on projective dynamics with the barrier recast as a positional constraint~\cite{Lan2022}, on a geometric contact potential with localized support~\cite{GCP2025},
and on multibody systems in which contact, friction and nonlinear joint constraints are all
potentials of one minimization~\cite{Chen2022UnifiedBarrier}, uses related
contact potentials on the feasible domain, subject to the geometric
regularity qualifications above.

Barrier width and stiffness affect both contact response and its sensitivities.
Adaptive stiffness is common, while some differentiable implementations hold
it fixed. If a parameter-dependent stiffness update is part of the
computation being differentiated, its derivative must be included. Quantitative
errors in contact-time sensitivities cannot be bounded by barrier width alone;
impact transversality and conditioning also matter. Motivated by the locking of IPC's CCD-filtered line search as much as by conditioning costs,
recent work drops the barrier altogether, enforcing non-penetration as a
constraint through an augmented Lagrangian solver fed by continuous collision
detection~\cite{Zheng2026NoBarrier}; that recovers the solver's efficiency but
reintroduces an explicit active set, and with it the question of
Section~\ref{sec:contact-complementarity} for whoever differentiates the
step. Its sensitivities remain branch-wise under a stable active set,
including when the constrained optimality conditions are differentiated
implicitly; mode boundaries require an additional derivative convention or
event treatment.

\subsection{Randomized smoothing and stochastic estimators}

Randomization can smooth an objective, a transition map, or a collision query;
these choices generally define different optimization problems. For an
objective $f$, one may consider $\mathbb{E}_{\epsilon}[f(\bth+\epsilon)]$, which
can be smooth even when $f$ is discontinuous. A score-function or
likelihood-ratio estimator~\cite{Williams1992} differentiates this expectation
without differentiating the simulator. A pathwise estimator instead uses
simulator derivatives and is a distinct construction: it is unbiased only
when differentiation and expectation can be interchanged without missing
boundary contributions. Section~\ref{sec:quality} discusses this
comparison~\cite{Parmas2018PIPPS,Suh2022}.

Le~Lidec \etal~\cite{Lidec2022} apply randomized smoothing to optimal control of
non-smooth dynamical systems, and Montaut
\etal~\cite{Montaut2023} apply it specifically to collision detection, obtaining
useful derivatives of a quantity that is combinatorial in nature. Pang
\etal~\cite{Pang2023} study both randomized and deterministic smoothing of
quasi-dynamic contact models, including a log-barrier formulation, for
contact-rich manipulation planning. Their analysis illustrates how smoothing
can supply informative slopes in otherwise flat regions. Smoothing local
dynamics before rollout is generally different from smoothing the full
rollout objective.

Under suitable integrability conditions and a regular smoothing distribution,
score-function estimators require no simulator derivatives. Sampling introduces
computational cost and variance, whose magnitude depends on noise scale,
dimension, objective, estimator, and baseline. Variance need not increase
monotonically with the smoothing scale.

\subsection{Event-aware and time-of-impact gradients}

A smaller body of literature addresses event-time dependence directly. For an
isolated transversal impact with a locally stable event sequence, trajectory
sensitivity includes a contribution from $\mathrm{d}\tau/\mathrm{d}\bth$ and the
impact update. An event-time model or detector makes this contribution
available; geometric collisions often use CCD, while simpler systems can use
analytic event equations. Grazing and simultaneous impacts require separate
treatment. Zhong \etal~\cite{Zhong2023} differentiate collision timing and
geometry, and Werling \etal~\cite{Werling2021Nimble} include an approximation
to continuous-time elastic-collision sensitivities. Cloth
methods~\cite{Liang2019,Li2022DiffCloth} that differentiate the detected
contact system can still have valid local discrete derivatives; agreement
with continuous-time event sensitivities is a separate question.

The works reviewed here provide fewer event-aware implementations than
fixed-branch or smoothed contact derivatives. This describes the reviewed
coverage only and does not rank the families' maturity.

\subsection{Decoupling the forward and backward contact model}
\label{sec:contact-decouple}

One option cuts across the families above: the contact model used in the
forward pass and the contact model differentiated in the backward pass need not
be the same. Surrogate directions
can provide optimization signals even when they differ from the gradient of
the forward objective; their convergence properties require separate analysis.
For example, CFD modifies contact derivatives while leaving the forward
contact model unchanged~\cite{Paulus2025}. Dojo~\cite{Howell2022Dojo}
provides a related equation-based construction: the forward solve approaches
hard contact, while smooth sensitivities can be evaluated at a cached, more
relaxed central-path state. These differentiate the relaxed equations rather
than the final harder-contact step, so both complementarity parameters should
be reported. The approach has reached hardware. Song
\etal~\cite{Song2024Quadruped} run the forward pass in a non-differentiable
simulator and take the gradient from a simplified single-rigid-body model aligned
to its state, and transfer the resulting quadrupedal policy to a real robot.
Schwarke \etal~\cite{Schwarke2024Locomotion}, who reach hardware with a single
analytically smoothed simulator instead, note that decoupling separates the
choices of forward realism and backward smoothing, but adds the cost of
maintaining and aligning two models.

The idea predates its contact-focused instances by two decades. Wojtan
\etal~\cite{Wojtan2006} introduce an \emph{integrator-swapping} approximation
so that the adjoint of a linearized implicit cloth step becomes
tractable: the backward pass differentiates a different scheme than the forward
pass runs, and is accepted on a consistency argument, the two schemes agreeing to second order with a local $O(\Delta t^3)$ discrepancy the authors quantify, rather than
because it is the adjoint of the executed forward scheme. This earlier instance
indicates that the decoupling is a general structural option, and one to expect
in settings other than contact.

The returned vector need not be the gradient of the forward objective, and it
need not be a conservative field. A scalar surrogate objective should therefore
not be assumed without establishing one. Standard exact-gradient convergence
results do not apply directly: guarantees require additional conditions, such
as sufficient descent alignment or controlled bias. Decoupling can improve
optimization in the regimes discussed in Section~\ref{sec:quality}, while
requiring separate checks of derivative accuracy and optimization behavior.

\begin{table*}[t]
\centering
\caption{Treatments of contact discontinuity. This axis is orthogonal to both axes of
Table~\ref{tab:grid}: an engine can combine a cell there with one or more
treatments here.
``Obstacle addressed'' refers to the enumeration in
Section~\ref{sec:contact-why}.}
\label{tab:contact}
\small
\begin{tabularx}{\textwidth}{@{}L{2.7cm} L{3.5cm} C{1.6cm} L{3.4cm} X@{}}
\toprule
\textbf{Family} & \textbf{Mechanism} & \textbf{Obstacle addressed} & \textbf{Principal cost} & \textbf{Representative work} \\
\midrule
Complementarity branch derivatives & differentiate a regular active-set LCP, NCP or constrained solution map & (1) & mode-boundary selection; flat contact regimes & \cite{Peres2018,Werling2021Nimble,Qiao2020Scalable} \\
Compliant / penalty & relax hard constraints or regularize contact forces & (1) & stiffness--smoothness trade-off; altered physics & \cite{Todorov2012,Geilinger2020ADD,Degrave2019} \\
Barrier / interior point & barrier potential on the feasible domain; piecewise smooth geometry & (1) & stiffness and conditioning; derivatives of any stiffness updates & \cite{Li2020IPC,Li2021CIPC,Huang2024DiffIPC} \\
Randomized smoothing & smooth an objective, transition map, or collision query & (1), (2) & sampling cost; variance; smoothing bias & \cite{Lidec2022,Montaut2023,Pang2023,Suh2022} \\
Event-aware / time-of-impact & differentiate regular event times and impact updates (approximated in Nimble) & (2) & event detection and sensitivity; often CCD for collisions & \cite{Popovic2000,Liang2019,Hu2020DiffTaichi,Zhong2023,Werling2021Nimble} \\
Forward/backward decoupling & backward differentiates a different model than the forward runs & (1), (3, partially) & may not be the forward objective's gradient & \cite{Paulus2025,Song2024Quadruped,Howell2022Dojo} \\
\bottomrule
\end{tabularx}
\end{table*}

\section{Physical Model Families}
\label{sec:models}

\begin{figure*}[t]
\centering
\begin{minipage}[t]{0.5222\linewidth}\includegraphics[width=\linewidth]{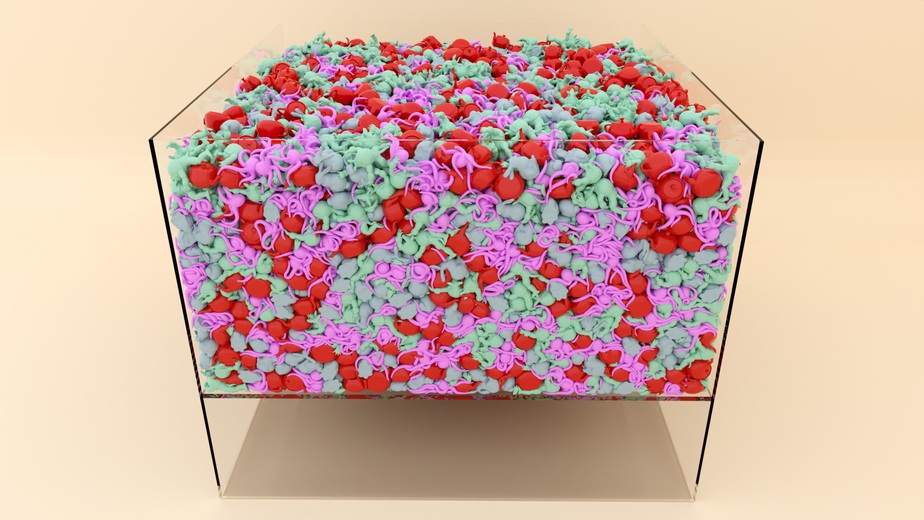}\end{minipage}%
\begin{minipage}[t]{0.4777\linewidth}\includegraphics[width=\linewidth]{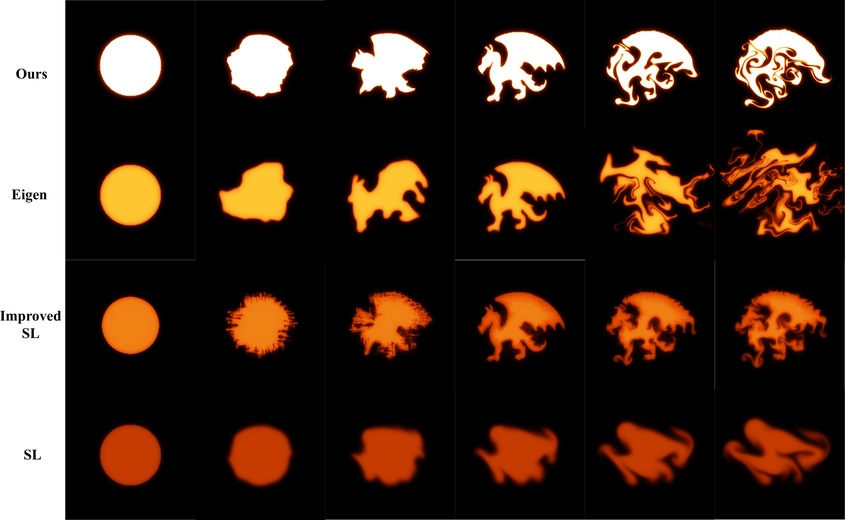}\end{minipage}
\caption{The two model families where the choice of referent bites hardest, and
why. \emph{Left}, deformables at the scale the field now works at: every one of
these bodies is soft, they are in sustained mutual contact, and the solver is
truncated to a fixed iteration budget, which is the regime where the converged equation and the executed computation stop agreeing
(Section~\ref{sec:models-deformable}). \emph{Right}, the same smoke plume under
four schemes. The differences are large and come from the
numerics rather than from the physics being modeled, which is the sense in which
a fluid solver's structure decides what its adjoint can be
(Section~\ref{sec:models-fluids}). Images reproduced, cropped but otherwise
unaltered, from \cite{Chen2024VBD,Li2025FlowMapAdjoint}.%
}
\label{fig:families}
\end{figure*}

With the taxonomy and contact treatments in place, we survey the model classes. Each
subsection asks what the model's structure implies for differentiation. The
physics itself is covered by the standard simulation literature. Figure~\ref{fig:families} shows the two
families where that structure bites hardest.

\begin{table*}[tp]
\centering
\caption{Representative methods grouped by physical model. \textbf{Ref.}:
A1 executed computation, A2 converged equation, A3 continuous dynamics
(Section~\ref{sec:referent}). \textbf{Real.}: B1 framework-generated reverse;
B2 equation-level adjoint, including a discretized continuous adjoint; B3
custom operator-level adjoint composition (Section~\ref{sec:realization}).
For hybrid engines these labels identify the main documented backward
construction; custom solves and recorded remainders are noted and do not imply
exactness of the full engine. A1* treats an internal linear solve as exact;
finite primal and adjoint tolerances still cause error. A2$^{a}$ marks an
additional approximation in the derivative system. \emph{surr.}\ marks a
returned direction that need not differentiate the forward objective; it can
still have a specified surrogate target. \textbf{State} describes recovery:
\emph{tape}, recorded reverse-pass intermediates; \emph{trajectory}, states and
required auxiliary data per step; \emph{checkpoint}, selected states with
recomputation; \emph{recompute}, per-step intermediates regenerated from saved
states; \emph{final state}, converged state of each step plus required auxiliary
data, not just the rollout endpoint; \emph{replay}, sweeps rerun from saved
inputs; \emph{reverse}, state recovered by reversing the map. These choices do
not specify a unique memory cost. \emph{none} and \emph{not stated} indicate no
reverse-state store or an unreported strategy. \textbf{Contact} follows
Table~\ref{tab:contact} where applicable. Alternatives are noted in the final
column. The realization is n/a when there is no backward pass, including
forward sensitivities; shooting alone does not determine the realization.
Methods that only consume a simulator appear in
Sections~\ref{sec:quality} and~\ref{sec:applications}.}
\label{tab:methods}
\footnotesize
\setlength{\tabcolsep}{3.5pt}
\renewcommand{\arraystretch}{0.88}
\begin{tabularx}{\textwidth}{@{}L{3.7cm} C{0.7cm} C{0.7cm} L{2.4cm} L{1.9cm} X@{}}
\toprule
\textbf{Method} & \textbf{Ref.} & \textbf{Real.} & \textbf{State} & \textbf{Contact} & \textbf{Notes and primary use} \\
\midrule
\multicolumn{6}{@{}l}{\textit{Control of simulation before general-purpose AD}}\\[1pt]
Spacetime constraints~\cite{WitkinKass1988} & n/a & n/a & n/a & n/a & dynamics posed as constraints, not as a forward map \\
Interactive spacetime~\cite{Cohen1992}       & n/a & n/a & n/a & n/a & spacetime windows; same formulation \\
Smoke keyframes~\cite{Treuille2003}          & A1* & n/a & none & n/a & forward sensitivities of the discrete operators, treating linear solves as exact \\
Adjoint fluid control~\cite{McNamara2004}    & A1* & B3 & trajectory & n/a & discrete adjoint, chosen over the continuous one; trajectory stored \\
Cloth keyframes~\cite{Wojtan2006}         & surr. & B2 & trajectory & penalty (repulsive springs) & adjoint of the full backward Euler scheme swapped in for the linearized implicit step the forward runs (Section~\ref{sec:contact-decouple}) \\
Rigid-body sketching~\cite{Popovic2003}      & A1 & n/a & not stated & impulse & multiple-shooting parameter estimation; derivatives by automatic differentiation of the simulation function; no backward pass described \\
Neural ODE~\cite{Chen2018NODE}               & A3 & B2 & reverse & n/a & continuous equation-level adjoint integrated backwards; reconstructs states with finite-tolerance error \\
ANODE~\cite{Gholaminejad2019}                & A1 & B1 & checkpoint & n/a & checkpoints and differentiates the discretization to avoid inconsistencies from continuous backsolve \\
Reversible MD~\cite{Greener2025Reversible}   & A1 & B3 & reverse & n/a & reconstructed states; snapshots and seeds stored; training truncates gradients \\
Reversible integrator~\cite{Stam2022Reversible} & A1 & B3 & reverse & n/a & bitwise state recovery by fixed-point updates; the adjoint differentiates the underlying unrounded Verlet map \\
\addlinespace[3pt]
\multicolumn{6}{@{}l}{\textit{Rigid and articulated bodies}}\\[1pt]
Degrave \etal~\cite{Degrave2019}             & A1 & B1 & tape & softened LCP, Gauss--Seidel projection & early end-to-end engine in an AD framework; the projected iterations are differentiated as executed \\
LCP layer~\cite{Peres2018}                   & A2 & B2 & final state & complementarity & 2D engine; LCP embedded as a differentiable layer by the method of OptNet; explicit remainder of the step recorded in PyTorch \\
Qiao \etal~\cite{Qiao2020Scalable}           & A2$^{a}$ & B2 & final state & nonlinear constraints / CCD & impact-zone derivatives locally linearize rigid-body geometry; remaining operations recorded \\
Nimble~\cite{Werling2021Nimble}              & A2 & B2 & final state & complementarity & analytic active-set LCP derivatives; optional heuristic contact directions and approximate elastic-impact sensitivities \\
Qiao \etal~\cite{Qiao2021Articulated}        & A1 & B3 & recompute & complementarity & module-by-module adjoint, including the executed Gauss--Seidel contact iterations; recomputes rather than tapes \\
ADD~\cite{Geilinger2020ADD}                  & A2 & B2 & final state & compliant, smoothed & analytic adjoint of a specific implicit multibody integrator \\
Brax~\cite{Freeman2021Brax}                  & A1 & B1 & tape & impulse (Baumgarte) & massively parallel rollouts on accelerators \\
Dojo~\cite{Howell2022Dojo}                   & surr. & B2 & final state (relaxed, $\kappa>0$) & barrier / IP & cached central-path sensitivity; relaxed A2 target may differ from the harder forward step \\
Zhong \etal~\cite{Zhong2023}                 & A1 & B1 & tape & impulse, TOI & corrects gradients when the contact normal is not fixed, by continuous collision detection; PyTorch and Taichi implementations \\
NeuralSim~\cite{Heiden2021TDS}               & A1 & B1 & tape (CppAD path) & penalty / NCP & unrolled engine with learned augmentations embedded inside it \\
Paulus \etal~\cite{Paulus2025}               & A1 & B1 & checkpoint & penalty / decoupled & DiffMJX: adaptive integration with checkpointed AD; optional CFD supplies surrogate contact derivatives \\
Song \etal~\cite{Song2024Quadruped}          & surr. & B1 & tape & decoupled & non-differentiable forward, single-rigid-body model differentiated in PyTorch for the gradient; reaches hardware \\
Le Lidec \etal~\cite{Lidec2024}              & A2 & B2 & final state & complementarity & highly efficient differentiable robotics simulation at speed \\
\addlinespace[3pt]
\bottomrule
\end{tabularx}
\end{table*}

\begin{table*}[!tp]
\centering
\caption{Table~\ref{tab:methods} continued: deformables, cloth and fluids. Columns are as in Table~\ref{tab:methods}.}
\label{tab:methods2}
\footnotesize
\setlength{\tabcolsep}{3.5pt}
\renewcommand{\arraystretch}{0.88}
\begin{tabularx}{\textwidth}{@{}L{3.7cm} C{0.7cm} C{0.7cm} L{1.9cm} L{2.6cm} X@{}}
\toprule
\textbf{Method} & \textbf{Ref.} & \textbf{Real.} & \textbf{State} & \textbf{Contact} & \textbf{Notes and primary use} \\
\midrule
\multicolumn{6}{@{}l}{\textit{Deformable solids}}\\[1pt]
ChainQueen~\cite{Hu2019ChainQueen}           & A1 & B3 & trajectory & grid-mediated & hand-derived CUDA MPM backward; cached per-step states; TensorFlow reference is an alternative B1 path \\
DiffTaichi~\cite{Hu2020DiffTaichi}           & A1 & B1 & tape & impulse with TOI (rigid, mass-spring); grid boundary (MPM) & source-transformation AD preserving megakernel parallelism; custom primitive derivatives must be declared \\
Real2Sim~\cite{Hahn2019Real2Sim}             & A3 & B2 & trajectory & n/a & continuous adjoint discretized with BDF2 for viscoelastic fitting; direct-sensitivity alternative \\
DiffPD~\cite{Du2021DiffPD}                   & A2 & B2 & final state & penalty / complementarity & reuses the projective-dynamics factorization in the backward \\
Qiao \etal~\cite{Qiao2021Soft}               & A1 & B1 & recompute & dry friction & coupled soft multi-body systems; projective-dynamics step differentiated with CppAD, per-step state stored, intermediates recomputed \\
DiffXPBD~\cite{Stuyck2023DiffXPBD}           & A2$^{a}$ & B2 & trajectory & compliant & implicit-step adjoint with derivative blocks projected to positive definiteness; states and derivative data stored \\
Huang \etal~\cite{Huang2024DiffIPC}          & A2 & B2 & final state & barrier / IP & differentiable time-dependent deformation with contact \\
Musculotendons~\cite{Chen2022}               & not stated & not stated & not stated & n/a & adjoint method over a musculotendon integrator (BDF1 and SDIRK2 available), differentiating joint torques; the adjoint itself is not described \\
Reverse-sweep VBD~\cite{Shu2026SolverAdjoint}& A1 & B3 & replay, $O(K|V|)$ & barrier / penalty & local ${3{\times}3}$ adjoint solves; no tape and no global system \\
\addlinespace[3pt]
\multicolumn{6}{@{}l}{\textit{Cloth and codimensional objects}}\\[1pt]
Liang \etal~\cite{Liang2019}                 & A2 & B2 & final state & constrained QP \newline CCD & QP collision response and linear solve differentiated implicitly, remainder recorded \\
DiffCloth~\cite{Li2022DiffCloth}             & A2 & B2 & final state & dry friction & explicit treatment of the non-smooth friction cone \\
\addlinespace[3pt]
\multicolumn{6}{@{}l}{\textit{Fluids and hybrid methods}}\\[1pt]
SPNets~\cite{Schenck2018SPNets}              & A1 & B1 & tape & SDF projection, one-way & differentiable position-based fluid dynamics \\
DiffFR~\cite{Li2023DiffFR}                   & surr. & n/a & none & penalty & forward-mode sensitivity hand-derived through the DFSPH pressure iterations, kept only for fluid particles adjacent to the rigid body; two-way SPH coupling; rigid-body control \\
$\Phi$Flow~\cite{Holl2020PhiFlow}            & A1* & B1 & tape & n/a & recorded Eulerian operators with an equation-level pressure adjoint; finite solve tolerances cause error \\
Solver-in-the-loop~\cite{Um2020}             & \shortstack{A1/\\A1*} & B1 & tape & n/a & learned corrections through time stepping; the CG-initialization experiment differentiates a finite solver iteration sequence \\
Takahashi \etal~\cite{Takahashi2021}         & A1* & B3 & trajectory & one-way coupling & hand-derived adjoint of each executed step through a variational pressure solve; states stored; the advection adjoint applies the interpolation transpose at the forward back-trace and drops the back-trace's dependence on velocity \\
\addlinespace[3pt]
\bottomrule
\end{tabularx}
\end{table*}

\subsection{Rigid and articulated bodies}

Rigid-body dynamics is widely studied in robotics. Its state is often smaller
than a deformable state, making recorded AD practical for many tasks, although
long or batched rollouts can still be memory-bound. Brax~\cite{Freeman2021Brax}
and Degrave \etal~\cite{Degrave2019} use recorded differentiation. Qiao
\etal~\cite{Qiao2020Scalable} combine recorded operations with equation-based
derivatives of the integration and impact-zone solves; the latter locally
linearize the rigid-body geometry. Contact is a major source of difficulty,
while smooth articulated dynamics also offers substantial opportunities for
structured differentiation.

Three structural points recur in every engine in this family.

The first is the choice of coordinates. \emph{Maximal} coordinates carry a full
pose per body and impose joints as constraints, giving a large but sparse and
uniformly structured system; \emph{reduced} coordinates parameterize the
unconstrained degrees of freedom of a kinematic tree and satisfy its joints by
construction. The mass matrix may be dense if formed explicitly, but recursive
algorithms can avoid assembling or solving a dense system. The choice propagates into the backward pass.
Maximal coordinates make the constraint Jacobian explicit, so the object
equation-level differentiation needs is explicit, which suits
formulations such as Dojo's~\cite{Howell2022Dojo}. The unified Newton barrier method~\cite{Chen2022UnifiedBarrier} goes the other way and makes an articulated step take the form of an IPC step, with nonlinear joint constraints as potentials beside contact and friction in one minimization, so the regularity qualifications of Section~\ref{sec:contact-barrier} also
apply. Reduced coordinates reduce the number of unknowns, while introducing
chain-rule dependencies through the joint parameterization; the resulting
cost depends on how these dependencies are evaluated.

The second is that articulated systems have a recursive structure, the
articulated-body and recursive Newton--Euler algorithms of Featherstone's spatial
vector algebra, whose adjoint is itself recursive. The recursion assumes a
kinematic tree; a loop is closed by an explicit constraint and falls back to
the constrained formulation. Qiao
\etal~\cite{Qiao2021Articulated} exploit this, deriving the adjoint of the
spatial-algebra recursion to obtain gradients in time linear in the number of
links rather than through a recorded tape, and recomputing
intermediates during the backward pass instead of taping them. Their construction is the adjoint of the executed computation taken module by
module, including the Gauss--Seidel iterations of the contact solve as they
ran, with per-step checkpoints recomputed in the backward pass.

The third point concerns rotations, whose local variations lie in a tangent
space. Generic AD differentiates the chosen coordinate program; it does not
automatically supply a manifold-aware gradient. Quaternion normalization,
orthogonality constraints, retractions, or Lie-algebra coordinates must be
represented explicitly and interpreted consistently. Romany{\`a}-Serrasolsas
\etal~\cite{Rotation2025} derive rotation-dynamics derivatives using
Lie-algebra structure, simplifying their implementation and improving
conditioning.

\subsection{Deformable solids}
\label{sec:models-deformable}

Deformables are where the choice of referent matters most, because the state
dimension is large and the solvers are iterative and truncated. The exception
is explicit integration, as in ChainQueen~\cite{Hu2019ChainQueen}
and the MPM examples of DiffTaichi~\cite{Hu2020DiffTaichi}, where there is no implicit time-step root solve. The explicit-step equation and executed program have matching derivatives only when their constitutive and decomposition primitives are differentiated consistently; custom rules, such as those for an iterative SVD, require a separate declaration; the continuous
referent A3 remains distinct, since the discretization error is untouched, and
only the realization and the tape remain to be chosen.

\emph{Finite element} formulations with implicit integration~\cite{Sifakis2012}
give the residual form~\eqref{eq:residual} directly, and the equation-level referent applies cleanly
at a regular converged solution of the Newton solve. \emph{Projective
dynamics}~\cite{Bouaziz2014} replaces the Newton solve with a local--global
alternation whose global step is a prefactorized linear solve;
DiffPD~\cite{Du2021DiffPD} exploits this structure, reusing the
prefactorized operator within the adjoint solve, reducing the cost of backward
linear algebra without implying equal forward and backward runtimes. \emph{Position-based dynamics}~\cite{Mueller2007PBD} and their
compliant extension XPBD~\cite{Macklin2016XPBD} are sequences of constraint projections that avoid assembling and solving a global system, and
DiffXPBD~\cite{Stuyck2023DiffXPBD} differentiates that formulation through a
trajectory adjoint targeting the implicit step equation rather than the finite
projection sequence. Its backward pass projects derivative blocks to positive definiteness, so it can approximate that target derivative even when
the forward solve has converged. Qiao \etal~\cite{Qiao2021Soft} extend differentiable
simulation to coupled soft multi-body systems. \emph{Vertex Block
Descent}~\cite{Chen2024VBD} performs coupled per-vertex updates in
Gauss--Seidel order, with independent parallel updates within suitable color
classes. This local update structure supports the reverse-sweep adjoint of
Shu \etal~\cite{Shu2026SolverAdjoint}.

Local update structure creates opportunities for efficient operator-level
reverse passes. Equation-level adjoints expose the coupling of the step
equations, but can also be solved matrix-free or with block preconditioners.
The comparison therefore depends on the solver and state-recovery strategy,
rather than on whether a global matrix is explicitly assembled.

Local projection and stencil updates appear in contact and hyperelasticity
solvers~\cite{Lan2022,Lan2023}, and in unified particle constraint
solvers~\cite{Macklin2014}. Their coupling differs: penetration-free projective
dynamics retains a global step solved by aggregated Jacobi iterations,
stencil descent combines local results through a sweep, and Newton or
multigrid methods~\cite{Wang2020,Larionov2021} require treatment of global or
interlevel dependencies. A differentiable musculoskeletal simulation with inertial musculotendons~\cite{Chen2022} sits on
the same axes, as an adjoint method over integrators that include BDF1 and SDIRK2, with joint torques rather than muscle excitations as the differentiated parameters.

Two recent graphics results extend the scope of differentiable deformable
solvers. Daviet \etal~\cite{NIFE2025} learn quadrature rules for implicit grid
cells and combine them with mixed finite elements, enabling differentiation as
the represented domain evolves. This extends beyond fixed mesh topology;
earlier mesh-based differentiable FEM already supports geometry and shape
parameters~\cite{Huang2024DiffIPC}. Shen \etal~\cite{ElasticLoco2025} carry
mixed second-order information through the simulation for locomotion. The
report's axes concern first derivatives, and the same questions recur, largely
unexamined, for second derivatives.

Material-model identification is a related thread: Hahn
\etal\ \cite{Hahn2019Real2Sim} estimate viscoelastic parameters from captured
motion, and more recent work learns constitutive laws
directly~\cite{Ma2023NCLaw}, using the simulator's differentiability to supervise
a neural material model from observed dynamics.

\subsection{Cloth and codimensional objects}

Cloth combines large state, stiff implicit integration~\cite{Baraff1998}, and
frequent collision-topology changes in contact-rich tasks, so it can stress
both taxonomic axes at once.

Wojtan \etal~\cite{Wojtan2006} gave an early differentiable cloth simulator, with an approximate adjoint obtained by swapping the integrator, later followed by Liang \etal~\cite{Liang2019}, differentiating through a QP-based collision response.
DiffCloth~\cite{Li2022DiffCloth} descends from a different pair, the projective
dynamics with dry frictional contact of Ly \etal~\cite{Ly2020PDContact} and the
DiffPD backward~\cite{Du2021DiffPD}, and adds an explicit treatment of the
non-smooth friction cone, which matters because
sticking and sliding are branches of Section~\ref{sec:contact-why}'s first
obstacle and transitions between them can affect cloth optimization. DiffAvatar~\cite{li2024diffavatar} demonstrates differentiability with respect to the cloth pattern space, and PhysAvatar~\cite{Zheng2024PhysAvatar} recovers garment material from video through a C-IPC solver with finite-difference gradients.

The persistent difficulty is self-collision. A garment's active contacts depend
on its configuration and the parameters being optimized, and their number can
vary substantially across steps. The candidate count is bounded by the finite
mesh but can grow quadratically with its complexity. This variability
complicates sparse assembly and state storage. Branch derivatives remain
locally valid under regularity and stable contact sequences, although they may
poorly predict finite mode changes or omit continuous-time event sensitivities.
Whether adjoint solves dominate, and whether fixed-capacity allocation is
practical, depends on the implementation and workload.

Codimensional formulations~\cite{Li2021CIPC} address the forward problem for
cloth, rods, and points within one intersection-free framework, and inherit the
feasible-domain regularity that makes the barrier family attractive to
differentiate. Changing contacts does not by itself destroy this regularity;
conditioning, geometric degeneracies, and implementation costs still require
assessment at the resolutions of interest.

\subsection{Fluids}
\label{sec:models-fluids}

Fluids split along two axes that the practice of the field tends to merge:
the view taken of the continuum, Eulerian at fixed locations or Lagrangian on
moving material, and the spatial discretization, grids, meshes, particles or
particle-grid hybrids. The pairings that dominate graphics are Eulerian solvers
on grids and Lagrangian solvers on particles, but neither is forced; Eulerian
solvers have been built on unstructured Voronoi cells with Delaunay
tetrahedral duals~\cite{Brochu2010}. \emph{Eulerian}
solvers often compose advection, projection, and diffusion operators, providing
natural units for derivative implementations. Projection and diffusion may be
linear for fixed coefficients and boundaries, while advection, limiters, and
free-surface handling can introduce nonlinear or nonsmooth dependencies.
Derivative-based fluid control arrived early in graphics, by propagating derivatives of the discrete fluid operators forward, treating linear solves as exact~\cite{Treuille2003}
and, a year later, by the adjoint~\cite{McNamara2004}, though the topology changes of thin features~\cite{Wojtan2007} introduce discrete decisions that require separate treatment when differentiating a fluid solver. $\Phi$Flow~\cite{Holl2020PhiFlow}
provides a modern differentiable Eulerian solver, and the solver-in-the-loop
line of work~\cite{Um2020} uses differentiability to train corrections that
interact with the solver over multiple steps rather than in isolation. Beyond
supplying gradients, the differentiable solver there trains a learned component
through the time-stepping loop.

One recent fluid result is a realization of the kind Section~\ref{sec:realization}
describes under B3. Li
\etal~\cite{Li2025FlowMapAdjoint} observe that a bidirectional flow map solver
and its adjoint share the same map: the forward pass transports impulse
variables along it, and the backward pass transports adjoint variables back
along the same map. Solving the adjoint equations directly on that map avoids
both differentiating the intermediate numerical steps and storing them, which is
a rederived realization reached through the structure of the flow map rather
than through a block decomposition.

\emph{Particle} methods may use explicit updates or iterative constraint and
pressure solves, as in SPNets~\cite{Schenck2018SPNets}. Neighbor lists are
discrete data structures, but list changes need not make the physical map
nonsmooth when contributions vanish smoothly at the support boundary.
Cutoff rules, truncation, and neighbor-capacity limits determine the relevant
derivative behavior. DiffFR~\cite{Li2023DiffFR} differentiates
an SPH solver with two-way fluid-rigid coupling in forward mode, by hand
through the DFSPH pressure iterations, and reports that the full derivative is
unstable after contact; it keeps only the fluid particles adjacent to the rigid
body in the gradient, a localization its authors note departs from the
physical ground truth, so Table~\ref{tab:methods2} lists the
returned vector as a surrogate direction. \emph{Hybrid} methods, the material point method (MPM) in
particular~\cite{Jiang2016MPM,Stomakhin2013}, combine particle-grid transfers
with constitutive updates and are amenable to differentiation on regular
branches. Transfer smoothness depends on the interpolation kernels and
support treatment. Yield transitions, contact, and spectral degeneracies may
require custom or regularized derivatives, as illustrated by the custom SVD
gradients in DiffTaichi~\cite{Hu2020DiffTaichi}.
ChainQueen~\cite{Hu2019ChainQueen} demonstrates differentiable MPM for
elastic solids, DiffTaichi~\cite{Hu2020DiffTaichi} ships MPM examples for solids and a liquid, and PlasticineLab's
benchmark~\cite{Li2021PlasticineLab} is likewise soft bodies. The particle-grid
machinery carries over to fluids, and
FluidLab~\cite{Xian2023FluidLab} takes it to liquids. Recorded MPM
differentiation can store per-particle and per-grid data at each
step, making long rollouts memory-intensive unless checkpointing or other
state-recovery strategies are used.

\subsection{Coupling and multiphysics}
\label{sec:coupling}

Solid--fluid coupling adds interface constraints whose differentiation is not
simply the composition of the two media's derivatives: even one-way coupling
with sub-grid boundary conditions needs its own adjoint
terms~\cite{Takahashi2021}, and the reason generalizes to every coupled system.
DiffFR~\cite{Li2023DiffFR} meets the same difficulty from the two-way side,
with SPH boundary particles carrying the coupling forces, and answers it by
localizing the derivative to the fluid around the body (Section~\ref{sec:models-fluids}).

A \emph{monolithic} scheme treats both media as one coupled system. Its
derivative is an instance of Section~\ref{sec:taxonomy} applied to that system,
with interface terms included in the coupled Jacobian. \emph{Partitioned}
schemes retain separate solvers for the media. In a sequential Gauss--Seidel
scheme, for example, one solves the fluid with the solid held fixed and then
the solid with the fluid held fixed, possibly repeating the alternation;
other partitioned schemes use different update schedules.

For an iterated partitioned scheme, the question of Section~\ref{sec:taxonomy}
recurs at the coupling level. Differentiating the converged coupled equations
requires the coupled state to satisfy those equations, whereas differentiating
a finite executed sequence targets the composition of its block updates.
Those block derivatives may be recorded, generated, or derived; an exact
derivative of the composition does not require two runtime tapes.
Partitioned multiphysics adjoints already have an established literature,
including discrete adjoints for block-coupled
solvers~\cite{Burghardt2022MultiphysicsAdjoint}. A more specific direction
suggested by the taxonomy is to study exact derivatives of coupling
iterations truncated by design while preserving the structure of each
constituent solver.

\subsection{Model class and cell}
\label{sec:models-synthesis}

Read by group rather than by row, Tables~\ref{tab:methods} and~\ref{tab:methods2}
show a pattern across model classes.

Rigid and articulated systems often have fewer state variables per instance
than deformable or fluid models, but long horizons and large batches can still
create substantial memory pressure. Contact is a major source of difficulty
in many rigid-body applications. Deformable models often combine large states
with truncated iterative solves, making both storage and the distinction
between executed and equation-level derivatives important. Cloth adds changing
contact and friction states, which complicate both global adjoint solves and
recorded computations. Eulerian fluid solvers often compose explicit operators
with a pressure solve, whose derivative may be obtained implicitly or by
differentiating its iterations. Their memory cost still grows with grid
resolution and rollout length. Hybrid particle--grid methods require substantial
per-step information, while transfers, constitutive models, plasticity, and
contact can introduce additional differentiability issues.

The choice of referent and realization should therefore follow the solver's
convergence, memory, and accuracy requirements rather than the physics label
alone. Opportunities for local reverse updates depend on the actual update
dependencies, including any global solves, rather than on whether a model
unifies several material types. A tightly converged deformable solve can
support an equation-level derivative; for a rigid solver truncated for real
time, an equation-level adjoint evaluated at the last iterate generally need
not equal the derivative of the executed computation.

\section{Gradient Quality}
\label{sec:quality}

\begin{figure}[!tb]
\centering
\includegraphics[width=\columnwidth]{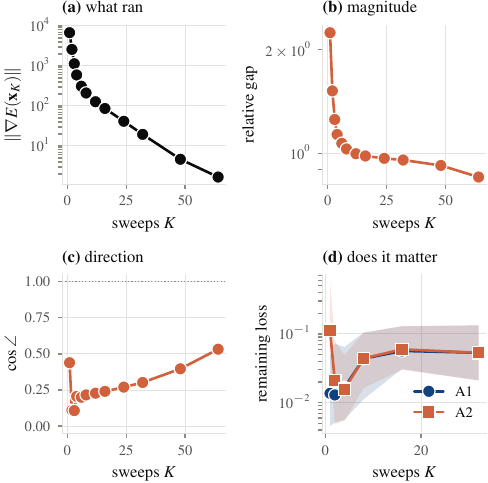}
\caption{The report's running example, one implicit step followed from what is
computed to what it costs. Two mass-spring chains stepped
with implicit Euler and solved by $K$ Gauss--Seidel sweeps; panels (a) to (c) take one step at ${k=2{\times}10^{5}}$, panel (d) fits random stiffness fields near ${6{\times}10^{3}}$ on a shorter, pinned chain over twelve steps. \emph{(a)} what ran:
the residual falls with the budget but does not vanish, so the iterate the
sweeps reached (A1) and the root they approach (A2) are two different objects to
differentiate. \emph{(b)} the relative norm of the vector difference between the two
gradients is of order one across the plotted range. \emph{(c)} they disagree in direction too, the
cosine sitting between $0.1$ and $0.55$, so the two are nearly orthogonal.
\emph{(d)} and, for a target the same solver generated, it makes almost no difference to the optimization: from two sweeps onward no stable advantage is separable, median and interquartile range over forty seeds. Panels (b) and (c) are large at the stiff setting; (d) shows no separation between A1 and A2 for its fitting problem from two sweeps on. For this reason the report separates \emph{which derivative} a method returns from \emph{whether it
helps}.}
\label{fig:running}
\end{figure}

The sections above concern obtaining a derivative, and this section asks whether the derivative helps. There is now
substantial evidence that gradient-based optimization through a differentiable
simulator can be \emph{outperformed} by zeroth-order methods that never touch the
simulator's derivatives~\cite{Metz2021,Suh2022}, on problems where the gradients
are computed correctly by any of the definitions in
Section~\ref{sec:taxonomy}. A survey that cataloged differentiation techniques
without confronting this would misrepresent the state of the art.

We find it useful to separate four mechanisms. They have different causes,
diagnostics, and remedies, although the literature frequently groups them
under \emph{bad gradients}. Figures~\ref{fig:bias}, \ref{fig:horizon},
and~\ref{fig:contactmech} illustrate the associated diagnostics on a small
testbed, with the literature providing broader evidence. In particular, the
contact experiment demonstrates sensitivity amplification associated with
repeated wall encounters; it does not establish chaos or an asymptotic law for
gradient variance. The testbed illustrates these effects rather than
determining their prevalence or magnitude at graphics scale.

\subsection{Mechanism 1: non-convergence bias}

If the forward solver is truncated and the backward pass evaluates an
equation-level adjoint (referent A2) at the final iterate, the returned vector
generally differs from the derivative of the executed computation. For a fixed
problem this mismatch is deterministic: repeated evaluation of the same solve
does not remove it, and optimizing the outer objective does not ensure that
the inner solve converges. The mismatch can decrease as the forward solve
converges, but a small forward residual alone does not certify gradient
accuracy. Bounds also depend on the conditioning of the residual Jacobian,
the parameter sensitivity of the residual, and the regularity of the solver's
convergence. Tight-tolerance validation can therefore miss discrepancies that
appear at production iteration budgets~\cite{Shu2026SolverAdjoint}.

\begin{figure}[!tb]
\centering
\includegraphics[width=\columnwidth]{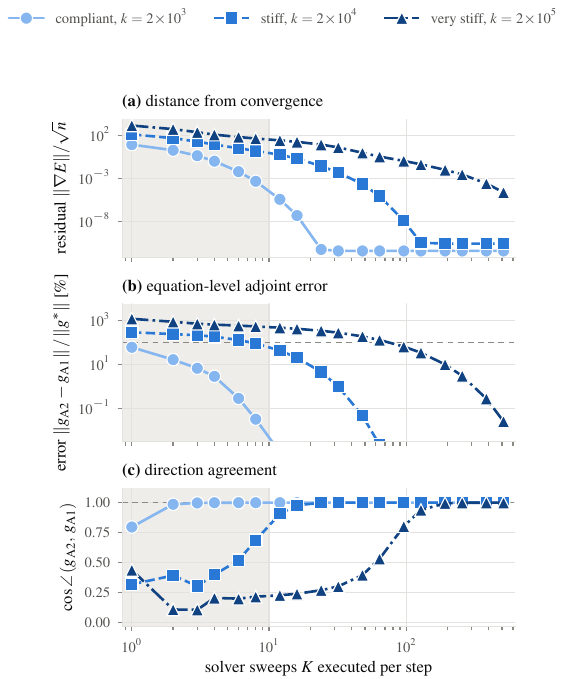}
\caption{Mechanism~1 measured. A mass-spring chain is stepped with implicit Euler
and solved by $K$ Gauss--Seidel sweeps of per-vertex Newton updates, a
one-dimensional instance of vertex block descent~\cite{Chen2024VBD}; the
parameter is the vector of per-spring stiffnesses. We compare the exact
derivative of the $K$ sweeps that actually executed, referent A1, against the equation-level adjoint, referent A2, evaluated as an
implementation evaluates it: at the iterate the $K$ sweeps reached, as if the
residual there had vanished. \textbf{(a)} how far the truncated solve is from
convergence. \textbf{(b)} the norm of the difference between the two gradients,
relative to the gradient of the same solve run to convergence; above the dashed line the discrepancy is
larger than the quantity being estimated. \textbf{(c)} the cosine of the angle
between them, which indicates whether the negative returned vector is a
local descent direction for the executed objective. Actual progress also
depends on step size, curvature, and the optimizer. The
shaded band marks the one to ten sweeps a real-time deformable solver actually
runs. Within that band the two derivatives agree well for compliant material from two sweeps on,
but for the stiffest material (${k=2{\times}10^{5}}$) the equation-level adjoint is nearly orthogonal to the
true derivative of the executed computation, and agreement is recovered only
after hundreds of sweeps; at ${k=2{\times}10^{4}}$ a few dozen suffice. The forward sensitivity was differentiated by hand
through the executed arithmetic and checked against central differences of the
same pipeline to a relative error below $4\times10^{-6}$.}
\label{fig:bias}
\end{figure}

We measured this mechanism directly, and Figure~\ref{fig:bias} reports the result. The experiment is small so that the exact derivative of the
executed computation can be derived by hand rather than taken on trust from a
differentiation framework. A chain of ${n=24}$ point masses (${m=1}$, rest length
$1$) sits under gravity ${-9.8}$ with no vertex pinned and is stepped with
implicit Euler at ${h=1/60}$; the parameter is the vector of 23 per-spring
stiffnesses, the loss is squared distance to a target configuration, and the
solver is a sequential Gauss--Seidel sweep of per-vertex Newton updates.
Stiffness sensitivity in this unpinned test depends on the initial deformation
and relative velocities. An unstressed chain with a common initial velocity
under uniform gravity undergoes stiffness-independent free fall. Relative
gradient diagnostics therefore require a nonzero reference sensitivity;
normalization by the converged gradient would be undefined in that degenerate
case.
The optimization of Table~\ref{tab:optcost} uses ${n=16}$ with the first vertex pinned, a horizon of 12 steps, 120
steps of normalized gradient descent with a step of 3\% of $\|\bth\|$, and 40
random instances. Code is provided as supplementary material. Within the
iteration budget a real-time solver actually runs, the discrepancy exceeds the magnitude of
the gradient itself once the material is stiff. The disagreement is also directional, which matters more
for optimization. At one to ten sweeps the
equation-level adjoint of the stiff chain is nearly orthogonal to the true
derivative of the executed solve, with $\cos\angle$ falling to between $0.1$
and $0.3$. A positive cosine against the true derivative makes the \emph{negative} of the
returned vector a descent direction, which guarantees only that a sufficiently
small step along it decreases the objective, and says nothing about the step
sizes an optimizer actually takes or about convergence. At these cosines it is a very
inefficient direction. Repeatedly averaging the same deterministic biased
vector does not remove the discrepancy. Across the stiffnesses we swept the effect is governed by stiffness, so
it is easy to miss on the soft test cases that gradient checks typically use; we
did not vary the problem size and so claim nothing about how it scales with
that.

A near-orthogonal gradient may still be usable, so we also measured
what the bias costs an optimizer. Fitting random stiffness fields on a shorter chain by gradient
descent through the same solver over twelve steps, changing only which gradient the optimizer is
given, produces Table~\ref{tab:optcost} over 40 random problem instances.

\begin{table*}[t]
\centering
\caption{What the equation-level referent costs an optimizer, and how much that
depends on where the target came from. Loss remaining after 120
gradient-descent steps, as a fraction of the initial loss, median over 40
random instances of the fitting problem (a 16-vertex chain over twelve steps), with the
fraction of instances on which A2 finished lower; lower is better. Three
constructions of the target, everything else identical: the endpoint the same
solver at the same $K$ produces from the true stiffness field, so that the fit
can reach it exactly; the endpoint a solve run to convergence (2000 sweeps)
produces, fitted with $K$ sweeps; and the rest configuration, which no
stiffness field reaches under gravity. Within-row comparisons isolate the
choice of gradient for each target construction. Across-budget comparisons for
same-$K$ targets also change the target-generating solver; comparisons for
converged or external targets isolate the budget only if instances and target
data are held fixed.}
\label{tab:optcost}
\small
\begin{tabular}{@{}c rrc rrc rrc@{}}
\toprule
& \multicolumn{3}{c}{target generated at the same $K$} & \multicolumn{3}{c}{target generated converged} & \multicolumn{3}{c}{external target} \\
\cmidrule(lr){2-4}\cmidrule(lr){5-7}\cmidrule(lr){8-10}
$K$ & \textbf{A1} & \textbf{A2} & A2 wins & \textbf{A1} & \textbf{A2} & A2 wins & \textbf{A1} & \textbf{A2} & A2 wins \\
\midrule
1 & $1.4\times10^{-2}$ & $1.1\times10^{-1}$ & 10\,\% & $1.5\times10^{-1}$ & $9.7\times10^{-1}$ & 0\,\% & $8.2\times10^{-4}$ & $4.7\times10^{-1}$ & 0\,\% \\
2 & $1.3\times10^{-2}$ & $2.1\times10^{-2}$ & 38\,\% & $3.7\times10^{-3}$ & $8.9\times10^{-1}$ & 0\,\% & $5.9\times10^{-4}$ & $1.2\times10^{-3}$ & 0\,\% \\
4 & $1.7\times10^{-2}$ & $1.6\times10^{-2}$ & 48\,\% & $2.3\times10^{-3}$ & $4.6\times10^{-1}$ & 0\,\% & $3.9\times10^{-4}$ & $7.4\times10^{-4}$ & 0\,\% \\
8 & $4.4\times10^{-2}$ & $4.4\times10^{-2}$ & 45\,\% & $2.6\times10^{-2}$ & $4.3\times10^{-2}$ & 28\,\% & $4.0\times10^{-4}$ & $6.1\times10^{-4}$ & 0\,\% \\
16 & $5.7\times10^{-2}$ & $6.0\times10^{-2}$ & 40\,\% & $5.2\times10^{-2}$ & $5.3\times10^{-2}$ & 35\,\% & $6.7\times10^{-4}$ & $7.0\times10^{-4}$ & 12\,\% \\
32 & $5.3\times10^{-2}$ & $5.3\times10^{-2}$ & 38\,\% & $5.3\times10^{-2}$ & $5.3\times10^{-2}$ & 38\,\% & $6.9\times10^{-4}$ & $6.7\times10^{-4}$ & 100\,\% \\
\bottomrule
\end{tabular}
\end{table*}

The result is narrower than Figure~\ref{fig:bias} suggests, and we report the gap between the two. At a single sweep the
exact gradient is decisively better: the equation-level adjoint leaves roughly
eight times more loss and wins on only four of forty instances. From two sweeps onward, with the target generated at the same $K$, we could not separate them, win rates hovering around a coin flip, which
with a median and a win rate over forty seeds is consistent with a modest
advantage either way rather than evidence of equivalence. Figure~\ref{fig:running}
puts that beside the angle: the gradients disagree in direction at budgets where the outcomes already coincide, with a median cosine over the forty instances of $0.51$ at two sweeps (interquartile range $0.38$ to $0.66$) and $0.61$ at four ($0.55$ to $0.78$); from eight sweeps on, where the median is $0.95$, both directions and outcomes agree. A gradient can be substantially wrong in
direction and still drive the optimization about as well.

Two limits on that reading are built into the design. Normalized steps remove
the magnitude of the gradient from the update, so the comparison measures the
usefulness of two directions under one update rule, and an optimizer that uses
the magnitude, a line search or a quasi-Newton method, could weigh them
differently. For the same-$K$ target columns of Table~\ref{tab:optcost},
changing $K$ also changes the target-generating solver. Comparisons across those
columns therefore do not isolate the effect of a larger budget on fixed target
data. Within this target construction, the two referents produced similar
reported outcomes from two sweeps upward, but not at one sweep.

The third limit is the target itself, and the two further groups of columns in
Table~\ref{tab:optcost} measure it. When the target is the endpoint of a solve
run to convergence and the fit runs $K$ sweeps, so that the model that
generated the target and the model being fitted differ at the true parameters
by the forward truncation error, the exact derivative wins on every
instance up to four sweeps, by two orders of magnitude at two and four, and the
two referents become indistinguishable only from about sixteen sweeps on.
Against an external target that no finite stiffness field reaches under
gravity, A1 wins on every
instance up to eight sweeps, by a smaller margin from two sweeps on, while at thirty-two sweeps A2 finishes lower on every instance, by about two percent. On
this testbed, these results suggest that the equation-level
direction is adequate when the fitted model can match the target at the
parameters it is heading for, and misleading when matching the target means
compensating the truncation bias; the single-step
instance of Appendix~\ref{app:gradcheck}, where A2 loses on every seed at
four sweeps, is the same effect. Which case an application is in depends on
where its target comes from, which is a reason to measure item (iii) of the checklist rather than assume it.

\emph{Diagnostic}: compare against
finite differences of the executed simulator, not of an idealized one, at the production iteration count. \emph{Remedy}: converge the forward solve, or adopt
the executed-computation referent.

\subsection{Mechanism 2: horizon-induced ill-conditioning}

\begin{figure}[!tb]
\centering
\includegraphics[width=\columnwidth]{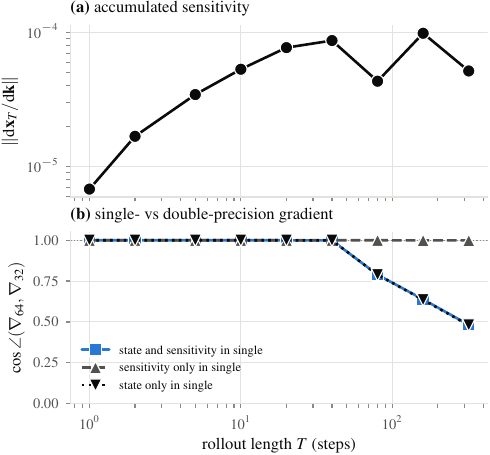}
\caption{Mechanism~2 measured, on the mass-spring chain of
Figure~\ref{fig:bias} stepped with implicit Euler and eight Gauss--Seidel
sweeps. Per-spring stiffnesses are the parameters and the sensitivity is
propagated forward through the executed arithmetic, so the experiment tests
sensitivity to the forward states and arithmetic precision without introducing
a reverse-mode implementation. \emph{(a)}, the accumulated
sensitivity $\|\mathrm{d}\x_T/\mathrm{d}\mathbf{k}\|$ varies by a factor of fifteen over three hundred steps: it does not explode, and the loss-gradient
norm, not shown, rises by less than three orders of magnitude while the loss
rises by eight and a half, so neither would alarm a monitor. \emph{(b)}, the cosine between derivatives obtained from the same sensitivity recurrence
in double and in single precision, with two controls that
separate the two places rounding can enter. With state and sensitivity both in
single, agreement holds to four figures out to twenty steps and three at forty, and then falls:
$0.79$ at eighty, $0.64$ at a hundred and sixty, $0.48$ at three hundred and
twenty (the last two vary with hardware; see text). With the trajectory in double and only the sensitivity recurrence in
single, the cosine stays above $0.999$ throughout. With the trajectory in
single and the sensitivity in double, the fall is reproduced to within $0.003$ in cosine. The
disagreement therefore comes from the single-precision trajectory, which
differs from the double one by a relative $10^{-4}$ at eighty steps while the
loss differs by about two parts in ten thousand (at most $2.2\times10^{-4}$): the gradient is far more sensitive
to the trajectory than the objective is. The final contraction of the
sensitivity with the loss gradient is done in double in every variant, so
that it does not enter the comparison.}
\label{fig:horizon}
\end{figure}

Equation~\eqref{eq:chain} contains products of $T-t-1$ state-transition
Jacobians. Uniform contraction can make these products decay, while sustained
expansion along relevant directions can make them grow. The resulting
vanishing or growing sensitivities are properties of the differentiated
dynamics and cannot be removed merely by changing the differentiation
implementation. Stiffness introduces disparate time scales, but its effect on
discrete sensitivities depends on the integrator, time step, scaling, and
spectrum. It does not by itself imply growth of the condition number of $A_t$
or of the gradient; on this report's converged implicit-Euler chain the condition number of $A_t$ does grow roughly linearly with $k$ ($5.5\times10^{3}$, $7.6\times10^{4}$ and $7.9\times10^{5}$ at $k = 2\times10^{3}$, $2\times10^{4}$ and $2\times10^{5}$). Figure~\ref{fig:horizon} examines horizon-dependent
sensitivity and precision disagreement on the chain of Figure~\ref{fig:bias};
it does not measure every Jacobian product or establish a condition-number
bound. The accumulated sensitivity does \emph{not} explode: over three hundred
steps it varies by a factor of fifteen, so ``exploding gradients'' is the
wrong diagnosis here. Instead, the agreement between the double- and single-precision derivatives degrades as the horizon grows. When the same differentiated real-arithmetic recurrence is evaluated in double
and in single precision, with the implementation differing only in arithmetic
precision, their cosine is one to four figures out to twenty steps and to three at forty (the vectors differ by one and five percent in norm), and then they separate, the cosine between them falling
to $0.79$ at eighty steps, $0.64$ at a hundred and sixty and $0.48$ at three
hundred and twenty on the run reported; a rerun on other hardware gives $0.65$ and $0.25$ at the last two, so the collapse is reproducible and its exact value is not. Two controls locate the source of the disagreement. Running the trajectory in
double and only the sensitivity recurrence in single leaves the cosine above
$0.999$ at every horizon: sensitivity arithmetic alone is not the dominant
source of the directional disagreement in this experiment.
Running the trajectory in single and the sensitivity in double reproduces the
collapse to within $0.003$ in cosine. The single-precision forward pass follows a trajectory that
differs from the double-precision one by a relative $10^{-7}$ at the first
step and $10^{-4}$ by eighty, a difference the loss barely registers, at most two
parts in ten thousand, while the derivative evaluated along its rounded states points in a different
direction. The horizon enters twice: the dynamics amplify a
rounding-level perturbation of the trajectory by three orders of magnitude,
and the gradient map amplifies the result to an order-one change of direction
while the objective moves by $10^{-4}$. For this parameterization, the derivative is more sensitive to the
rounded forward states than the objective value is. The mixed-precision
controls associate the observed disagreement primarily with the forward
trajectory; the size of the total parameter sensitivity alone does not bound
every state-transition Jacobian product. The gradient norms here are small
in the units of the problem, between $5\times10^{-8}$ and $3\times10^{-5}$.
We use cosine to isolate directional disagreement. Relative norm errors are
also scale-invariant, and both directional and relative diagnostics require
care when a gradient is close to the numerical noise floor.

In single precision, the sensitivity recurrence approximates the derivative
of the real-arithmetic operations evaluated along rounded forward states; it
is not the exact derivative of the floating-point rounding map. The horizon
at which small perturbations of those states substantially change the
derivative's direction marks a precision limitation for this implementation.
Whether optimization still benefits past
that point is a separate question, which a change of direction does not settle
on its own, as Table~\ref{tab:optcost} shows for Mechanism~1. On this testbed
the point is near eighty steps; a production simulator
in single precision, the common case on the GPU, has its own trajectory
sensitivity and will reach it earlier or later, and the diagnostic carries over where the number does
not. A gradient-norm monitor shows nothing unusual in this experiment.
\emph{Diagnostic}: compare the gradient against the same gradient computed in
higher precision, and treat the horizon at which they part as the point beyond
which the gradient is precision-limited and should be tested before it is relied on.
\emph{Remedy}: truncate the horizon, which trades bias for conditioning; this is
the core of short-horizon actor--critic
methods~\cite{Xu2022SHAC}, which learn a value function to
terminate the backpropagation window, and of adaptive-horizon
variants~\cite{Georgiev2024AHAC} that choose the truncation point from the
observed dynamics.

\subsection{Mechanism 3: chaos and variance amplification}

Some contact-rich and turbulent systems exhibit chaotic sensitivity. In such
systems, a derivative of an individual rollout can be large and mathematically
correct while being difficult to use for optimizing an expectation over initial
conditions or noise. Under suitable assumptions, pathwise gradient variance
can grow rapidly, including exponentially with horizon, even when the gradient
of the expectation remains modest. Metz \etal~\cite{Metz2021} document this across
several domains and give it its clearest statement; Parmas
\etal~\cite{Parmas2018PIPPS} identified the same effect in model-based policy
search earlier and proposed combining pathwise and likelihood-ratio estimators to
control it. Suh \etal~\cite{Suh2022} sharpen the analysis for
simulators specifically, showing that the first-order estimator's advantage over a
zeroth-order one depends on the smoothness of the underlying landscape and can be
reversed in the discontinuous or stiff contact examples they study.

Figure~\ref{fig:contactmech}(b) measures gradient-spread amplification on the
chain, and the conditions under which it appears are part of the result. Launched at a single wall, the chain bounces once and departs; an ensemble of
launches perturbed by $10^{-6}$ returns gradients whose spread stays at its
forty-step level for as long as we ran it. With a second wall behind it, so that
it bounces repeatedly, the spread climbs roughly seventy-fold between forty
and thirteen hundred steps, doubling every two hundred. In this testbed, the two-wall trajectories show increasing gradient spread
over the tested horizons, whereas the single-wall trajectories show no
comparable growth. This comparison associates the observed amplification with
repeated wall encounters under these settings; it does not establish that
recurrence is necessary for amplification, or that a single contact cannot
amplify sensitivities. It also does not establish chaos: we have not measured
a Lyapunov exponent. Nor does it establish the exponential variance growth
of Metz \etal's analysis, which requires an appropriate statistical and
horizon-dependent assessment.
\emph{Diagnostic}: compare the variance of the pathwise estimator against a
zeroth-order baseline at the horizon in use. If the former is larger, treat
it as a warning that is not by itself conclusive, since bias, the quantity each estimator
targets and the cost of each evaluation enter the comparison too. \emph{Remedy}: interpolate between first- and zeroth-order
estimators~\cite{Suh2022,Parmas2018PIPPS}, or smooth the objective
(Section~\ref{sec:contact}).

\subsection{Mechanism 4: contact-induced gradient collapse}

\begin{figure}[!tb]
\centering
\includegraphics[width=\columnwidth]{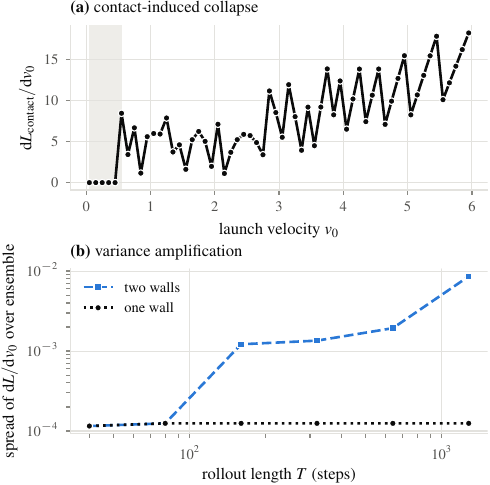}
\caption{Contact-gradient support and sensitivity amplification in the
mass-spring contact testbed ($k=2{\times}10^{4}$, eight sweeps, no gravity,
wall stiffness $2{\times}10^{5}$), with launch velocity as the single parameter
and accumulated contact energy as the objective. Panel (a) uses a 120-step
rollout; panel (b) varies the horizon. \emph{(a)} the derivative against launch
velocity: below the velocity at which a vertex reaches the wall within the
horizon (shaded), no contact occurs and the derivative is zero; gradients can
be nonzero in the active-contact region. This is a derivative-support
diagnostic, not evidence that minimizing contact energy fails in the
zero-contact region. \emph{(b)} spread (standard deviation) of the same
derivative over an ensemble of launches perturbed by $10^{-6}$. Between two
walls, repeated bounces are associated with roughly seventy-fold growth in
spread over the tested horizons. With one wall, the chain bounces once and
the spread remains approximately flat. These observations do not establish
chaos or show that repeated contact is necessary for sensitivity amplification.}
\label{fig:contactmech}
\end{figure}

Distinct from variance, a contact-only objective can have a region in which
its local derivative is zero because no contact occurs within the horizon.
This does not imply that every task loss is flat before contact, or that
contact information is confined to a measure-zero set. In the penalty model
of Figure~\ref{fig:contactmech}(a), with accumulated contact energy as the
objective and launch velocity as the parameter, the derivative is identically
zero for launches that do not reach the wall: five of the sixty sampled
velocities. Gradients can be nonzero throughout the active-contact region,
rather than only at its boundary.

This plot illustrates the local support of the contact-energy derivative.
If the nonnegative contact energy is minimized, its zero-contact plateau is
already a global minimum; the plot alone therefore does not demonstrate
optimization failure. For a task whose successful solution requires
establishing contact, a locally constant contact-dependent term can
nevertheless leave an optimizer without a useful local signal. Changing the
derivative referent or its implementation cannot change the exact derivative
of a locally constant function.
\emph{Diagnostic}: examine near-zero gradients in relation to the task
objective, including whether successful solutions require leaving such
regions. \emph{Remedy}: possible responses include
smoothing~\cite{Pang2023,Lidec2022}, global or population-based
search~\cite{Antonova2022}, and forward/backward decoupling
(Section~\ref{sec:contact-decouple}). Non-local repulsive terms, such as those
in Repulsive Shells~\cite{Repulsive2024}, can extend geometric sensitivity
beyond a contact neighborhood, but do not guarantee informative gradients for
every task or parameterization.

\subsection{Summary of the evidence}
\label{sec:quality-reading}
\label{sec:quality-read}

For contact-rich, long-horizon tasks, improving derivative exactness alone
may not improve optimization when estimator variance or a lack of useful local
information is the limiting factor. In smooth, short-horizon inverse problems
with deliberately truncated forward solves, non-convergence mismatch can
instead be important. These task labels do not determine the dominant
mechanism: identifiability, objective conditioning, model discrepancy, and
noise can also limit optimization. Our
own measurement bounds how much follows from that: exactness was decisive at the tightest iteration budget and, for a target the same truncated solver generated, not above it, despite a large angle between the gradients; for a target from a converged solve or from outside the simulator it stayed decisive up to four and eight sweeps. A substantial derivative mismatch is therefore not sufficient for exactness
to decide the outcome.

The comparison in Suh \etal~\cite{Suh2022} is further qualified by
Onoda \etal~\cite{Onoda2026Revisit}, who emphasize per-step inverse-variance
weighting on standard robotics control benchmarks. In their experiments,
controlling variance was more consequential than explicit empirical-bias
detection. This qualifies a narrative centered on discontinuity-driven
estimator switching, but variance reduction alone does not identify chaos as
the cause or establish a conclusion about deployed robotics in general.
The result suggests diagnosing the limiting source of gradient error or
variability on the task, rather than inferring it from the presence of contact.

The consequence for how results are reported is that a
differentiable simulator paper demonstrating improved gradient accuracy and one
demonstrating improved task performance make different claims, and the second does not follow from the first. We take
this up in Section~\ref{sec:eval}.

\section{Systems and Implementation}
\label{sec:systems}

Turning a differentiation strategy into a usable simulator is largely a
systems problem, and a substantial fraction of the engineering effort in
this field has gone into it.

\subsection{Memory}

As discussed in Section~\ref{sec:fwdrev}, reverse mode can evaluate scalar-loss
gradients efficiently but may require substantial storage or recomputation.
Memory is often a binding constraint for large graphics simulations. Three
responses recur.

\emph{Checkpointing}~\cite{Griewank2000} stores state at selected steps and
recomputes intervening segments during the backward sweep, and is common wherever memory binds. As Section~\ref{sec:realization} notes, the achievable
point on the memory--recomputation curve follows from the schedule rather than
from a single law: storing $O(\sqrt{T})$ states costs a constant number of extra
forward sweeps, whereas $O(\log T)$ states costs a recomputation factor that
grows logarithmically as well, under the usual uniform-step-cost model.
ACA~\cite{Zhuang2020} supports adaptive ODE stepping by storing accepted states
and step sizes and recomputing local differentiation graphs. This is distinct
from choosing an optimal sparse checkpoint schedule under a fixed memory
budget.

\emph{Recomputation of derived quantities} can reduce storage for quantities
reconstructible from the state, at the cost of additional work. The tradeoff
depends on reconstruction cost, memory traffic, and the hardware. Hand-derived
adjoints can expose these opportunities, but AD compilers, custom derivative
rules, and rematerialization policies can exploit them as well, since a
recorded computation can still be recomputed.

\emph{Avoiding a general tape} does not imply avoiding all trajectory storage.
Equation-level differentiation can avoid retaining inner solver iterates,
although a time-dependent adjoint still requires access to forward states and
other information needed to evaluate its local derivatives. A structural
adjoint of the executed updates instead retains or reconstructs the inputs
required by those updates. Reverse-sweep VBD~\cite{Shu2026SolverAdjoint}
reports a $71\times$ memory reduction relative to its unrolled-AD baseline on
the stated benchmark. Its replay record contains nine doubles, or 72 bytes,
per vertex per sweep; the approximately four-kilobyte normalized peak
allocation reported in the same study includes other allocations and is a
different memory measure. The reported allocation scales linearly with
$K|V|$, and the largest rollout is checkpointed every five steps. The backward
computation, the data retained within a step, and the checkpoint schedule
across time should therefore be reported separately. The referent and
realization constrain these choices, but do not uniquely determine the
complete memory budget.

\subsection{Parallelism and locality}

A mathematically correct backward pass may have less favorable memory
locality or parallelism than its forward pass and therefore achieve lower
throughput. Preserving useful structure can help, but performance depends on
the implementation and must be measured. DiffTaichi's central technical
contribution~\cite{Hu2020DiffTaichi} is a source-transformation reverse mode
that keeps the megakernel structure intact, and the reverse-sweep
formulation~\cite{Shu2026SolverAdjoint} applies the same principle to a
hand-derived backward, making it a mirror of the forward sweep, so a colored
Gauss--Seidel forward pass yields a colored Gauss--Seidel backward pass with
identical parallelism.

Nondeterministic floating-point reductions can perturb both the forward
trajectory and the recomputation of checkpointed segments. If replay differs
from the original execution, the resulting derivative may change, especially
in sensitive or contact-rich dynamics. The size of that change should be
measured. Depending on the required replay fidelity, deterministic reductions
or saved branch and state information may be needed.

\subsection{Precision}

Derivative computations can be more sensitive to finite precision than
objective values. For compatible nonsingular matrices, the condition number
of a product is bounded by the product of the individual condition numbers,
but need not equal it; sensitivity also depends on the directions excited and
on the forward states at which derivatives are evaluated. Single precision
may therefore be adequate for a forward observable while giving a
substantially different gradient. Mixed precision should target the dominant
error source. In Figure~\ref{fig:horizon}, double-precision sensitivity
accumulation along a single-precision trajectory does not remove the observed
disagreement, whereas a double-precision trajectory with single-precision
sensitivity propagation retains close agreement. These controls motivate
reporting forward and derivative precision separately, together with
task-specific error measurements.

\subsection{Engines}

Tables~\ref{tab:methods} and~\ref{tab:methods2} place the methods on the two
taxonomic axes. Table~\ref{tab:engines} instead compares particular software
implementations: how the backward pass is constructed, what information is
retained or recomputed, and where the reported implementation runs. These
categories should not be read as a necessary equivalence between a
hand-derived adjoint and low memory use. Generated and traced derivatives can
also use custom rules, code generation, and checkpointing, while hand-derived
trajectory adjoints still require forward information.

\begin{table*}[!t]
\centering
\caption{Differentiable simulation systems viewed as software, complementing
Tables~\ref{tab:methods} and~\ref{tab:methods2}. \textbf{Backward construction}
describes how derivative computations are implemented. \textbf{Backward memory}
lists principal retained information, not a complete allocation bound: forward
states, contact data, and linear-solver data may also be required or recomputed.
The absence of an inner-iteration tape does not imply constant rollout memory.
Backend and AD coverage depend on the solver and version. Entries summarize
the cited descriptions and supported derivative paths.}
\label{tab:engines}
\small
\begin{tabularx}{\textwidth}{@{}L{3.5cm} L{4.0cm} L{3.6cm} X@{}}
\toprule
\textbf{System} & \textbf{Backward construction} & \textbf{Backward memory} & \textbf{Execution backend} \\
\midrule
DiffTaichi~\cite{Hu2020DiffTaichi} & compiler source transformation & tape, with checkpointing & Taichi; CPU and CUDA \\
Warp~\cite{Macklin2022Warp} & compiler source transformation & kernel-launch tape and required data & CPU and CUDA \\
Brax~\cite{Freeman2021Brax} & framework tracing (JAX/XLA) & saved residuals; optional rematerialization & CPU, GPU and TPU; parallel envs \\
JAX MD~\cite{Schoenholz2020} & framework tracing (JAX/XLA) & saved residuals; optional rematerialization & CPU, GPU and TPU \\
$\Phi$Flow~\cite{Holl2020PhiFlow} & backend AD; custom implicit solve derivatives & backend- and solver-dependent data & pluggable array backend \\
Genesis~\cite{Genesis2024} & kernel source transformation (Taichi lineage) & saved state and derivative data & GPU and CPU backends; differentiable MPM path \\
Newton~\cite{Newton2025} & Warp-generated derivatives for supported solvers & solver-dependent saved data & CUDA (CPU on macOS); MuJoCo Warp is not differentiable \\
NeuralSim~\cite{Heiden2021TDS} & templated AD; taped reverse mode or CppADCodeGen & backend-dependent tape or generated-code state & CPU \\
Nimble~\cite{Werling2021Nimble} & hand-derived analytic LCP gradients & per-step Jacobians from the step's inputs and LCP partition; no inner-iteration tape & CPU \\
Dojo~\cite{Howell2022Dojo} & implicit differentiation at a relaxed central-path point & per-step states and implicit-derivative data; no inner-iteration tape & CPU \\
ADD~\cite{Geilinger2020ADD}        & hand-derived integrator adjoint & per-step state, no tape & CPU \\
DiffPD~\cite{Du2021DiffPD}         & hand-derived, applied through the reused forward factorization & per-step state, no tape & CPU \\
DiffCloth~\cite{Li2022DiffCloth}   & adjoint vector from a global solve, with per-step contact handling & per-step state, no tape & CPU \\
DiffXPBD~\cite{Stuyck2023DiffXPBD} & hand-derived trajectory adjoint & forward states and derivative quantities; sparse adjoint solve & CPU, multi-core \\
DiffIPC~\cite{Huang2024DiffIPC} & hand-derived integrator adjoint & per-step state, no tape & CPU \\
Reverse-sweep VBD~\cite{Shu2026SolverAdjoint} & hand-derived reverse sweep of block updates & nine-double replay record per vertex per sweep; total allocation scales as $K|V|$; the $10^6$-body rollout is checkpointed every 5 steps & CUDA \\
\bottomrule
\end{tabularx}
\end{table*}

\begin{figure*}[!tb]
\centering
\begin{minipage}[t]{0.2601\linewidth}\includegraphics[width=\linewidth]{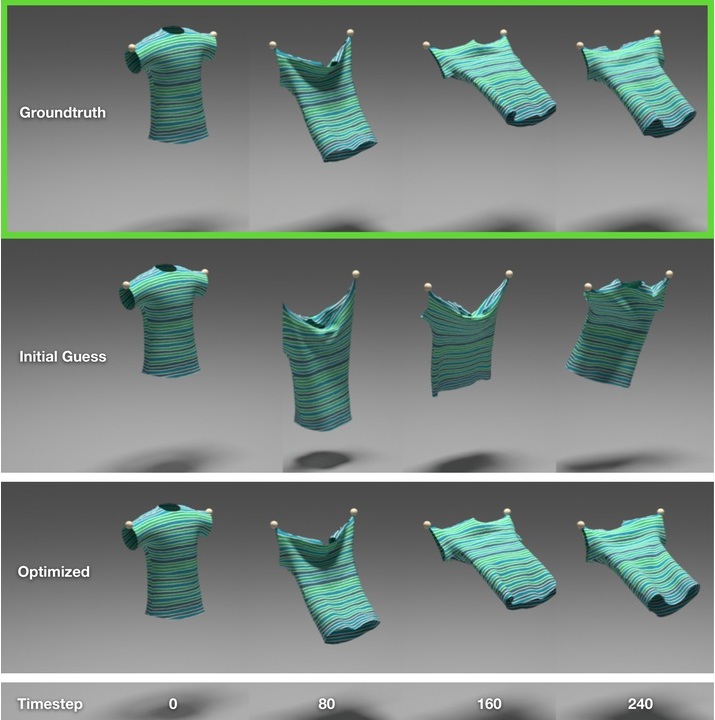}\end{minipage}%
\begin{minipage}[t]{0.3518\linewidth}\includegraphics[width=\linewidth]{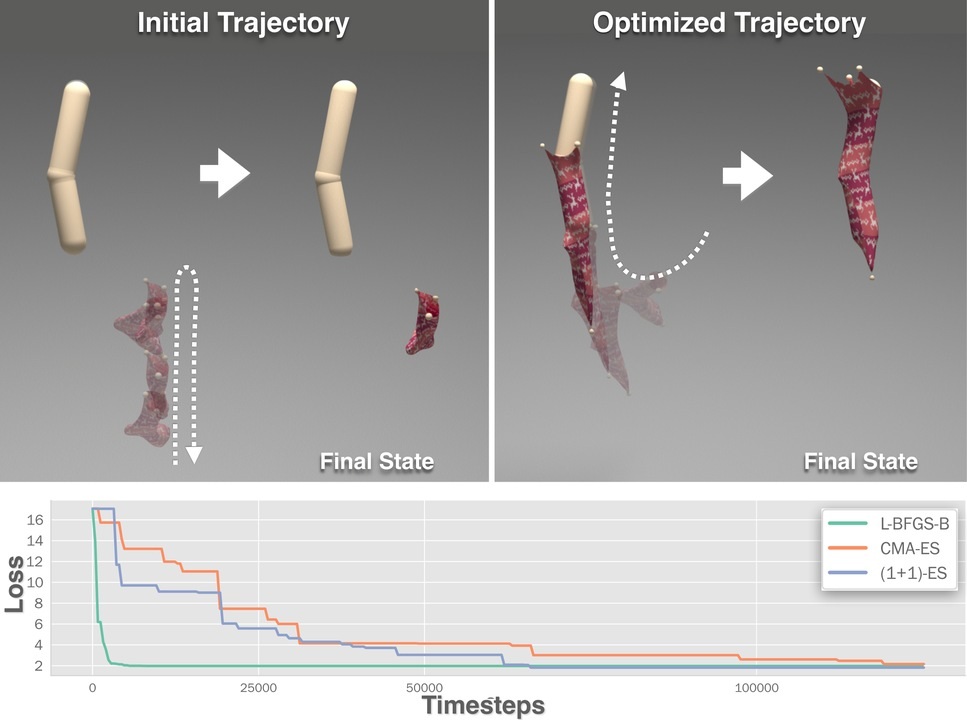}\end{minipage}%
\begin{minipage}[t]{0.3881\linewidth}\includegraphics[width=\linewidth]{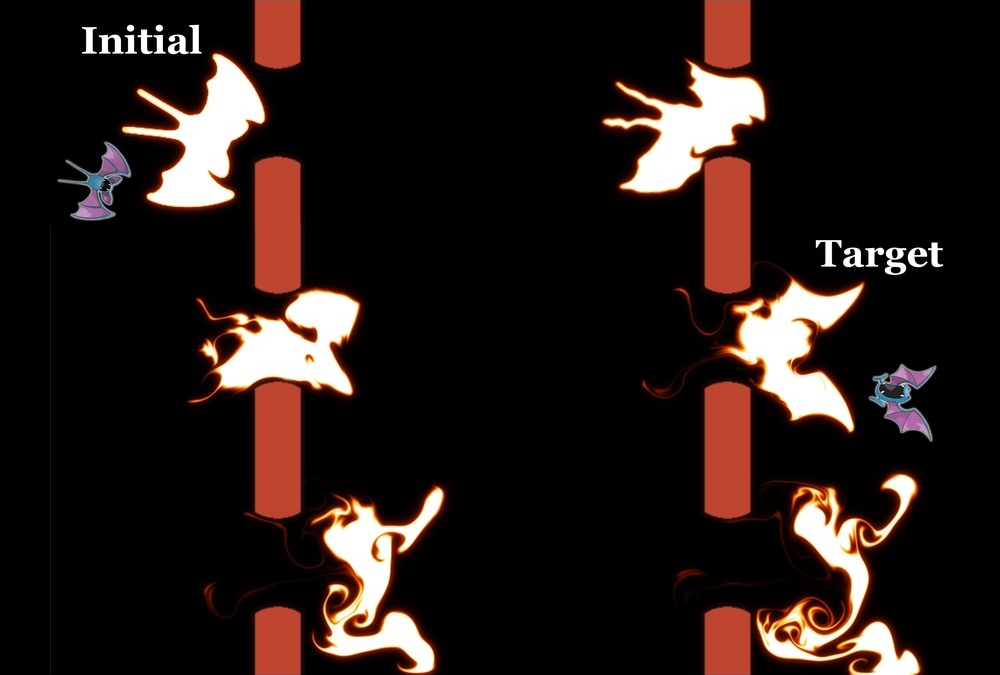}\end{minipage}
\caption{Three uses of a simulator gradient, as their authors report them.
\emph{Left}, material parameters recovered from observed cloth motion by
optimizing through the simulator, the inverse problem of
Section~\ref{sec:app-sysid}. \emph{Middle}, a trajectory optimized through
frictional contact; the loss curve beneath compares the analytic gradient
against two zeroth-order optimizers on the same task, which is the comparison
item~7 of our checklist asks every paper to report.
\emph{Right}, smoke steered from one silhouette through fixed obstacles into
another, by optimizing the forces with an adjoint taken on the flow
map. Keyframe control of fluids is an early application of differentiable
simulation. The illustrated result uses the recent continuous-adjoint flow-map
method~\cite{Li2025FlowMapAdjoint}, corresponding to referent A3 in
Section~\ref{sec:referent}. None of the three results is ours. Images reproduced, cropped or letterboxed but otherwise unaltered,
from \cite{Li2022DiffCloth,Li2025FlowMapAdjoint}.%
}
\label{fig:applications}
\end{figure*}

\section{Applications}
\label{sec:applications}

This section reads the literature through the two axes rather than by task.
Figure~\ref{fig:applications} shows three uses of simulator gradients
corresponding to the application classes below.

\subsection{System identification and real-to-sim}
\label{sec:app-sysid}

System identification inverts the simulator against observation. Given a
recording of a real object deforming or colliding, differentiable simulation
can be used to fit model parameters to the observations. Hahn
\etal~\cite{Hahn2019Real2Sim} estimate viscoelastic parameters from captured
deformation; Huang
\etal~\cite{Huang2024DiffIPC} estimate hyperelastic parameters and friction coefficient from captured animation sequence; gradSim~\cite{Jatavallabhula2021gradsim} composes a differentiable
simulator with a differentiable renderer to identify physical parameters from
video alone, without intermediate state supervision; PhysAvatar~\cite{Zheng2024PhysAvatar} fits the density, stretching and bending stiffness of a garment to multi-view video of a dressed person by running a C-IPC cloth simulator inside the loop, with the three-parameter gradient taken by finite differences because no analytic gradient of that solver was available.

Many system-identification tasks admit short-horizon objectives and smooth
parameterizations, but contact, long observation windows, and poorly
identifiable parameters can also make these problems difficult. For
forward solves truncated by design, the derivative mismatch of
Section~\ref{sec:quality} is one relevant source of error.
Table~\ref{tab:optcost} fits a material field and separates the two referents
up to four or eight sweeps when the target comes from a converged solve or
from outside the simulator, respectively, and only at a single sweep when the
same truncated solver generated the target. The usefulness of derivative exactness therefore has to be measured at the
solver budget and target construction used by the application.

The first of two recurring issues is \emph{identifiability}: an observed trajectory
can constrain some parameter combinations tightly while leaving others weakly
determined. A point estimate alone does not express that uncertainty.
Transient observations can help distinguish effects such as stiffness and
damping, but the result depends on the excitation and observation model.
Derivative access supports local Gauss--Newton or Fisher-information
approximations to parameter uncertainty. Computing and interpreting these
approximations still requires a noise model, sufficient identifiability or
explicit regularization, and additional Jacobian products and linear solves;
a scalar-loss adjoint does not provide a covariance matrix almost for free.
The second is that the model class bounds the
achievable fit: when a real material is not in the family the simulator can
express, optimization can at best fit the observations within that model
class; convergence to a global best fit is not guaranteed. Learning the
constitutive law itself~\cite{Ma2023NCLaw} attacks this
directly, using differentiability to supervise a neural material model from
observed motion rather than only fitting coefficients of a prescribed
constitutive family.

\subsection{Control and policy learning}

Control is a prominent application of simulator gradients, which supply a
pathwise policy gradient through the dynamics in place of a score-function
estimator. Results in practice are mixed, for the reasons of
Section~\ref{sec:quality}. Successful methods address different limitations:
PODS~\cite{Mora2021PODS} uses a Gauss--Newton policy update assembled from
first derivatives of the simulator and reward curvature, rather than second
derivatives of the dynamics; SHAC~\cite{Xu2022SHAC} truncates backpropagation
with a learned value function; AHAC~\cite{Georgiev2024AHAC} adapts its horizon using a
dynamics-Jacobian threshold intended to avoid stiff contact gradients; and GIPPO combines an analytic-gradient-informed target-policy update with
PPO updates~\cite{Son2024}. Wiedemann \etal~\cite{Wiedemann2023}
train trajectory-tracking controllers for aerial vehicles on analytic policy
gradients, and Li \etal~\cite{Li2023DiffFR} control rigid bodies through
two-way fluid coupling, optimizing release states and training neural
controllers on the localized forward-mode gradient of
Section~\ref{sec:models-fluids}. Antonova \etal~\cite{Antonova2022} use local
gradients within a broader global-search strategy. Evidence of practical
transfer includes Schwarke \etal~\cite{Schwarke2024Locomotion}, who train a
quadrupedal locomotion policy with short-horizon analytic gradients and a
learned critic and transfer it to hardware zero-shot. This demonstrates
transfer for their task and training setup; it does not by itself establish
that the simulator gradients are unbiased or that model exploitation is
absent.

These methods use simulator gradients in different ways: curvature
approximations improve the update, truncation and learned values control the
backpropagation horizon, combined analytic and PPO updates manage stochastic gradient
quality, and global search reduces reliance on a single local basin. They
therefore cannot all be reduced to horizon control or to Mechanisms~2 and~3.
A practitioner has to determine which limitation dominates the
task and which part of the optimization procedure addresses it. Improving
derivative accuracy can help, but does not replace these choices.

\subsection{Inverse design and fabrication}

Differentiable simulation makes design an optimization over shape, material, and the actions that use them.
Applications include co-design of robot morphology or material and
control~\cite{Spielberg2019,Xu2021RobotDesign}, cavity design of pneumatic actuators~\cite{Gjoka2024Pneumatic}, shape interpolation for underwater
swimmers~\cite{Ma2021DiffAqua}, garment pattern design and recovery~\cite{wang2018rule, li2024diffavatar, chen2025dress, Dress2025}, knife-trajectory optimization for cutting~\cite{Heiden2021DiSECt}, grasp synthesis~\cite{Turpin2022GraspD},
deployable structures~\cite{Panetta2019, Ren2022Umbrella}, shock protecting structures~\cite{Huang2024Shock}, trajectory optimization for
cable-driven and underwater soft robots~\cite{Bern2019,Du2021Underwater}, and
locomotion benchmarks~\cite{Wang2023SoftZoo}. Two further instances optimize
through an MPM equilibrium rather than a dynamic rollout: a topology
optimization method that computes adjoint sensitivities through the
force-equilibrium solve~\cite{Li2021LETO}, and a folding-wing aerial vehicle
whose bistable mechanism is designed by optimizing through a quasi-static MPM
equilibrium and validated in dynamics by forward simulation
only~\cite{Li2021BistableHAV}. Differentiating with respect to
shape~\cite{Tozoni2021,Huang2024DiffIPC} rather than material is a distinct problem, since the parameter changes the
domain and therefore the discretization; Ye \etal~\cite{SDRS2026} and PhysiOpt~\cite{PhysiOpt2025} sidestep it by keeping the discretization fixed and moving vertices or densities within it. The equilibrium-design examples are quasi-static, whereas swimming and
locomotion involve dynamic trajectories. Shock-protecting microstructures
are optimized through nonlinear equilibrium homogenization and evaluated
with separate transient drop simulations and physical tests. Short horizons
and smooth equilibria can simplify sensitivity analysis, but gradient reliability
still depends on conditioning, contact transitions, and solver convergence.

Co-design evaluates morphology and control against a shared objective.
Evaluating each body only with a controller tuned for another body can confound
their contributions. Joint updates, alternating optimization, and nested
controller optimization are all ways to account for this dependence.
Differentiable simulators provide sensitivities for both sets of
variables~\cite{Spielberg2019,Xu2021RobotDesign}; the gains reported by
Xu \etal~\cite{Xu2021RobotDesign} reflect their parameterization and
optimization method as well as the co-design formulation, and cannot be
attributed to joint optimization alone.

Fabrication imposes requirements beyond simulated performance, including
available materials, minimum feature sizes, assembly, and tolerances. Some
admit continuous differentiable constraints; others involve discrete choices
or uncertain manufacturing processes. Physical validation extends beyond
deployable structures and microstructures~\cite{Panetta2019,Chen2021Bistable,Ren2022Umbrella,Huang2024Shock}, not all of it gradient-based:
examples include the bistable aerial vehicle~\cite{Li2021BistableHAV},
pneumatic actuators~\cite{Gjoka2024Pneumatic}, and garment
design~\cite{wang2018rule}. These examples show that fabrication can be
included in differentiable design, while robustness to tolerances and the
combination of discrete and continuous manufacturing constraints remain
important practical challenges.

\subsection{Animation authoring}

An early application closely tied to the graphics community is making
simulation controllable. Spacetime
constraints~\cite{WitkinKass1988,Cohen1992} posed animation as a constrained
optimization over the whole trajectory; differentiating the solver itself made
keyframe control of fluids and particle systems tractable, first with derivatives propagated
forward~\cite{Treuille2003} and then with the
adjoint~\cite{McNamara2004,Wojtan2006}. A closely related use is reconstruction rather than
authoring: enforcing physical consistency on motion estimated from video, either
through an explicit contact-and-dynamics optimization~\cite{Rempe2020} or by
driving a simulated character to track the estimate~\cite{Yuan2021SimPoE}. Modern
differentiable simulators are a direct continuation of the spacetime line with
vastly greater capacity.

Little recent work is aimed at authoring. Most recent work addresses control
and design, where the objective is a scalar that an algorithm can minimize
unattended. An animator's
objective is instead partly unstated, revised on seeing the result, and expressed as
constraints at particular moments rather than as a quantity to minimize over a
trajectory. This motivates three authoring requirements that are not consistently
evaluated in recent differentiable-engine papers.

The first is latency. Spacetime constraints were interactive in
ambition~\cite{Cohen1992}; a differentiable solve taking minutes is a batch tool,
whereas an authoring loop needs a useful partial answer in the time it takes to
drag a keyframe. The second is locality of edit. An animator who fixes frame~40
may expect earlier frames to survive. This locality can be enforced by
freezing a trajectory prefix, restricting optimization variables, or adding
constraints; it does not follow from the gradient alone. The third is that the objective is revised in response to what the artist
sees, so the optimization is repeated against a changing objective. Motion sketching~\cite{Popovic2003} and interactive rigid-body
manipulation~\cite{Popovic2000} already addressed these concerns, the first locality of edit and the second latency. Connecting
such authoring interfaces to modern differentiable engines, while measuring
latency and edit locality, remains a useful research direction. These
requirements complement the gradient-quality questions of
Section~\ref{sec:quality}: a well-conditioned gradient alone does not specify
an interactive \mbox{workflow}.

\subsection{Coupling with rendering and neural scene representations}
\label{sec:app-video}

The most recent thread composes differentiable simulation with differentiable
rendering, so that physical parameters can be optimized against raw video. Gradient-based optimization requires usable derivatives or declared
surrogate rules through the chosen simulation--rendering interface. The
conditioning of their composition affects numerical stability and optimization
performance, rather than being a prerequisite for the chain rule itself. $\nabla\text{Sim}$~\cite{Jatavallabhula2021gradsim} demonstrated 
the pipeline on synthetic data. Early works laid the foundation for recovering 
physical properties from visual data by introducing correspondence-free material 
reconstruction using sparse surface constraints~\cite{weiss2020correspondence} 
and optimizing meshless elastic objects against multi-view 
captures~\cite{chen2022virtual}. PAC-NeRF~\cite{Li2023PACNeRF} recovers geometry and then identifies material parameters from multi-view video by coupling a continuum simulator to a neural radiance
field; PhysGaussian~\cite{Xie2024PhysGaussian} attaches continuum mechanics to
Gaussian splats~\cite{Kerbl2023} for physically grounded dynamics of captured
scenes. PhysDreamer~\cite{Zhang2024PhysDreamer} differentiates through MPM
and Gaussian rendering to fit material properties to video generated from a
learned dynamics prior. Related work simulates neural object representations
directly~\cite{LeCleach2023}.

The composition itself has received little study. Rendering
discontinuities (visibility) and simulation discontinuities (contact) are
different in character, and the strategies for handling discontinuities developed for
one~\cite{Li2018EdgeSampling,Loubet2019} do not obviously transfer to the other.
A gradient that must traverse both is subject to both sets of pathologies, and we
are not aware of a systematic study of the composition.

Composition also creates implementation constraints. When the simulator and
renderer use different frameworks, state transfer and backward interfaces must
be implemented explicitly, and memory and runtime depend on both components
and their coupling. The intermediate mesh, particles, or Gaussian
cloud~\cite{Kerbl2023} needs a defined derivative path from physical state to
rendered observations, with the corresponding vector--Jacobian products in
the backward pass; the representation need not be invertible or differentiable
as an inverse map. Representation choice therefore affects both physical
fidelity and the cost of the composed optimization.

The newest work in this direction inverts the usual order, using a generative
prior to propose a plausible object and a differentiable simulator to make it
physically consistent. Zhan \etal~\cite{PhysiOpt2025} optimize the shape directly in the latent space of the 3D generator to improve physical compliance. Li \etal~\cite{Dress2025} reconstruct a
simulation-ready garment from a single image this way, combining a diffusion
prior with differentiable cloth. PAT3D~\cite{Lin2026PAT3D} applies simulation-based refinement to complete
scene layouts, pairing generative models that propose the objects and a vision-language model that infers their support relations with a differentiable rigid-body simulator that drives the scene
to an intersection-free static equilibrium under gravity. We expect this pairing to become common, because
the two parts are complementary: the generative model supplies a plausible
starting point in a space too large to search, and the simulator supplies the
constraint that the result has to be physical. It also raises the requirements on
gradient quality, since the simulator is now being asked to correct a proposal it
did not produce.

\section{Evaluation, Benchmarks, and a Reporting Checklist}
\label{sec:eval}

\subsection{Existing benchmarks}

Benchmarks have appeared for specific domains, including PlasticineLab for soft-body manipulation~\cite{Li2021PlasticineLab}, FluidLab for fluids~\cite{Xian2023FluidLab} and SoftZoo for locomotion~\cite{Wang2023SoftZoo}, and Newbury
\etal~\cite{Newbury2024Review} tabulate existing engines' design choices and features side by side. Table~\ref{tab:benchmarks} summarizes them. They are valuable, but almost all
measure \emph{task} performance, which confounds the quality of the gradients
with the capability of the optimizer, the expressiveness of the policy class, and
the fidelity of the physics.

\begin{table*}[!t]
\centering
\caption{Task benchmarks and a cross-engine survey for differentiable simulation.
The task evaluations combine gradient behavior with the optimizer, policy
class, and physical model; the Newbury \etal\ row summarizes engine features.
Together they provide useful comparisons, but not a shared cross-engine
protocol for reference-Jacobian accuracy at matched solver settings.}
\label{tab:benchmarks}
\small
\begin{tabularx}{\textwidth}{@{}L{3.0cm} L{3.0cm} L{3.0cm} X@{}}
\toprule
\textbf{Suite} & \textbf{Domain} & \textbf{Underlying simulator} & \textbf{What it reports, and what it does not} \\
\midrule
PlasticineLab~\cite{Li2021PlasticineLab} & soft-body manipulation & MPM elastoplastic, on DiffTaichi & task reward across manipulation goals; no reference derivatives \\
FluidLab~\cite{Xian2023FluidLab} & complex fluid manipulation & differentiable fluid engine & task reward on multi-material fluid tasks; non-smooth landscapes discussed, gradients not measured against a reference \\
SoftZoo~\cite{Wang2023SoftZoo} & soft robot co-design & MPM, multiple environments & co-design performance across morphologies and terrains, with gradient-landscape examples; no shared reference-Jacobian protocol \\
Brax~\cite{Freeman2021Brax} & rigid-body control & its own engine & throughput and policy return at scale; simple physics chosen for speed \\
DiSECt~\cite{Heiden2021DiSECt} & robotic cutting & dedicated cutting simulator & parameter fits against real cutting data; a system-identification benchmark in effect \\
Newbury \etal~\cite{Newbury2024Review} & cross-engine survey & none (engines tabulated) & survey of engines' design choices and features; not a task benchmark or a reference-Jacobian protocol \\
\bottomrule
\end{tabularx}
\end{table*}

\subsection{Gaps}

We see three specific gaps. First, there is no standard \emph{gradient accuracy}
benchmark: a set of problems with reference derivatives against which an engine's
Jacobians can be checked. Second, comparisons rarely control for the
referent, so an equation-level engine with a tightly converged solver and an
executed-computation engine with a loose one are compared as if the difference
were physical.
Third, the memory/accuracy trade-off is almost never reported as a curve, though
it is the quantity a practitioner most needs.

A reference suite can begin with small, controlled problems. Central
differences of the executed simulator provide one reference on a smooth branch.
For a well-scaled smooth scalar loss, the idealized error model
$O(h^2)+O(\epsilon/h)$ gives an optimal-step error of order
$\epsilon^{2/3}$, where $\epsilon$ is the evaluation precision. This is not a
universal relative-error guarantee: derivative scales, cancellation, solver
noise, and event changes matter, and a step-size sweep is needed to assess a
particular check. The chain diagnostic reports agreement of order $10^{-7}$.
Closed-form spring, impact, or elastic-bar examples give another reference,
provided the analytic solution matches the declared continuous or discrete
model and its event convention. Finally, equation-level derivatives can serve
as limiting references for iterative programs when both the primal iterates
and their sensitivities converge under suitable regularity conditions.
Driving the primal residual to machine tolerance alone does not guarantee
that the A1 and A2 derivatives coincide. Sweeping tolerance or iteration count
can reveal their gap, alongside the residual and conditioning.

The supplementary material contains the seed of such a suite:
\texttt{gradcheck}, a small NumPy package that takes any engine exposing a
loss, its returned gradient and a declared referent, and reports the numbers of
the checklist below on the reference problems of Section~\ref{sec:quality};
Appendix~\ref{app:gradcheck} shows its output for the chain under both
referents. A suite along those lines would fix a small number of configurations per family,
publish the reference derivatives, and ask an engine for three numbers at its own
production settings: the relative error of the returned gradient, the cosine of
its angle to the reference, and the memory and wall-clock cost of obtaining it.
The angle matters separately from the magnitude, as
Section~\ref{sec:quality} shows, and the cost matters because tighter primal and adjoint solves can reduce
some convergence errors. Roundoff, finite-difference error, discretization
error, and deliberate smoothing or surrogate bias require separate controls. Implementation checks can use a reference matched to each declared model.
Cross-engine comparisons additionally require compatible physical problems,
discretizations, and settings; declaring the referent in
Table~\ref{tab:grid} is necessary but not sufficient.

\subsection{A reporting checklist}

We propose that papers presenting differentiable simulators report the following.

\begin{enumerate}
\item \textbf{Referent and realization.} Which object is the returned derivative
  exact with respect to, and how is the backward pass obtained
  (Table~\ref{tab:grid})? These are separate answers and both are needed.
\item \textbf{Forward solver tolerance and iteration count} used in every reported experiment. Report the tolerance that ran, not the one the method supports.
\item \textbf{Implementation, referent gap, and usefulness.} A single finite-difference check
  assumes the executed computation is the intended referent and so scores A2,
  A3 and decoupled backward passes against a target they did not choose. At the production configuration and iteration budget, report relative
  errors for (i) and (ii), and task performance for (iii): \emph{(i)} agreement with the method's
  \emph{own} declared referent, in two parts that must not be merged: the
  returned vector against an independent evaluation of the same object at the
  same state, which checks implementation accuracy subject to conditioning and precision, and the
  returned vector against the declared target. For A2 this target is the
  converged-root sensitivity, so evaluation at a truncated iterate introduces
  a convergence gap; for A1 the target remains the finite executed program.
  \emph{(ii)} Agreement with finite differences of the executed simulator
  includes numerical checking error and, for methods that are not A1, a
  possible referent gap. In particular,
  for an A1* method it measures the gap between the idealized program with exact
  linear solves and the finite-tolerance program that ran, subject to the
  numerical accuracy of the check; and \emph{(iii)}
  whether it helps, against a baseline at equal budget, stating the budget in
  both queries and wall-clock, since a backward pass is rarely the one forward
  pass that query counting assumes. A deliberate surrogate should still be checked against an independent
  implementation of its declared backward rule. An exact-referent comparison
  is marked not applicable only where that rule has no such referent; its
  discrepancy from the executed simulator remains useful to report.
\item \textbf{Contact treatment} on the axis of Table~\ref{tab:contact},
  including any smoothing parameter and whether forward and backward use the
  same model. This includes solid--solid contact and fluid--solid or multiphysics
  interfaces where relevant. Fluid and coupled problems also have other
  sources of gradient instability (Section~\ref{sec:models-fluids}).
  Objective slices are complementary diagnostics, rather than substitutes
  for derivative checks or a description of interface treatment.
\item \textbf{Gradient-evaluation memory and time relative to forward}, as measured,
  stating the AD mode and whether the timing includes the primal evaluation,
  together with the state-recovery strategy by name (stored, checkpointed and
  recomputed, or recovered by reversing the map) and how memory scales with
  the solver depth, since that is what the strategy decides and the realization
  label does not.
\item \textbf{Behavior under horizon length.} Estimator variance against $T$
  where the task is stochastic. Gradient norm alone is not enough: in
  Figure~\ref{fig:horizon} the state sensitivity varies by a factor of fifteen and the loss-gradient norm rises by less than three orders while
  the loss rises by eight and a half. These magnitudes alone do not reveal the
  gradient's directional disagreement with a higher-precision evaluation; report
  that agreement too and give the horizon at which it breaks.
\item \textbf{A zeroth-order baseline} on the same task and budget. Given the
  evidence in Section~\ref{sec:quality}, a differentiable simulator paper that
  does not compare against not using the gradients has not established that the
  gradients help.
\end{enumerate}

\section{Open Problems and Research Agenda}
\label{sec:open}

\paragraph*{Convergence-aware adjoints.}
Perturbation bounds for implicit derivatives at approximate solutions
already exist under regularity assumptions, including the analysis of
Blondel \etal~\cite{Blondel2022}. Practical estimates must account for
conditioning, the primal residual, and the adjoint-solve residual. A remaining
challenge is to turn such bounds into inexpensive diagnostics for stiff,
contact-rich simulators and to distinguish error in an approximate A2
derivative from the separate A1--A2 referent gap of a finite solver program.

\paragraph*{Structure-preserving adjoints.}
A solver's structure, block or otherwise, determines what its adjoint can look like; the
reverse sweep on vertex block descent~\cite{Shu2026SolverAdjoint} is one
instance, and DiffTaichi's source transformation~\cite{Hu2020DiffTaichi},
which keeps the megakernel structure through the backward, is an older one
from the other end of the realization axis. A compiler-level challenge is to
extend automatic derivation to broader classes of solver-specific adjoints,
including their state-recovery policies, while retaining efficient operator
structure. This could reduce per-solver implementation effort and combine
automatic construction with explicitly preserved operator locality.
\paragraph*{Combining reversibility and block structure.}
Section~\ref{sec:grid} distinguishes reversible state recovery from block
mirroring. Hybrid recovery and stored snapshots already provide ways to
combine reconstruction with saved states. The boundary does not coincide with the split
between conservative and dissipative physics: a dissipative step can be
invertible but numerically unstable to reverse, while projection can discard
information, and active-set branching need not by itself destroy invertibility.
An open systems question is how to choose reconstruction, recomputation, and
storage automatically for individual components of stiff contact solvers,
with explicit bounds on numerical error and measured memory savings.

\paragraph*{Gradients through changing contact topology.}
Event-aware differentiation must handle changes in event ordering,
simultaneous contacts, and grazing configurations. Making these treatments
robust and efficient at the scale of cloth self-collision remains a challenge.

\paragraph*{Solver design for differentiation.}
Backward-pass requirements already influence simulator design, for example
in DiffTaichi~\cite{Hu2020DiffTaichi} and Dojo~\cite{Howell2022Dojo}. A further
goal is to design solvers with quantitative guarantees on adjoint conditioning
and tunable forward/backward accuracy trade-offs, while retaining useful
physical models and throughput.

\paragraph*{A gradient-accuracy suite.}
Section~\ref{sec:eval} outlines a common reference-gradient protocol.
Its value depends on multiple engines reporting comparable errors, costs, and
referents. Constructing reliable references near events and under poor
conditioning is a technical task as well as a coordination problem.

\paragraph*{Standard gradient diagnostics.}
The four mechanisms of Section~\ref{sec:quality} have distinct diagnostics, which are not consistently reported together. A library that reports which mechanism is
limiting a given optimization would reduce misattributed effort.

\paragraph*{Robotics and embodied AI.}
Robotics motivates both high-throughput simulation and accurate contact
sensitivities. Engines and frameworks such as Brax in its current release~\cite{Freeman2021Brax},
Warp, Genesis, and Newton~\cite{Macklin2022Warp,Genesis2024,Newton2025}
contain different backends with different differentiation and contact
capabilities; they cannot be classified collectively as a tape over one
contact model. In particular, a framework's support for differentiation does
not imply that every backend supports it. Barrier and complementarity
formulations~\cite{Howell2022Dojo,Huang2024DiffIPC} can require coupled solves
within each environment, but this does not preclude parallelism across
independent environments. Assessing their throughput and physical accuracy
requires matched tasks, tolerances, hardware, and backend versions, rather
than a universal ranking by contact family.

Three questions remain open. First, parameters identified through a truncated
solver can compensate for its numerical error, so their transfer to another
solver or to hardware requires validation. The hardware demonstrations of
Schwarke \etal~\cite{Schwarke2024Locomotion} and
Song \etal~\cite{Song2024Quadruped} show successful policy transfer with
smoothing or decoupled forward/backward models. They do not establish that
referent-induced parameter-identification bias has been removed. Second, the
contact-rich examples in Section~\ref{sec:quality} illustrate cases where
local gradients offer limited benefit. A systematic comparison at matched
budgets is needed to identify which tasks benefit from them. Third, deformable
and multi-material
manipulation~\cite{Li2021PlasticineLab,Xian2023FluidLab,Heiden2021DiSECt}
offers a potential advantage for reverse-mode sensitivities: a backward pass
returns the sensitivity of a scalar loss to a large
number of state and material parameters at once, where coordinate finite differences require a number of evaluations
proportional to the parameter dimension. Random-direction estimators can use
a dimension-independent number of queries per sample, but their variance and
sample requirements can grow with dimension. Shape, as distinct from material or control, has been treated
as a differentiable quantity in its own right~\cite{Xu2021RobotDesign,SDRS2026} with open-loop or low-dimensional controllers, while joint material-and-policy optimization already has
a precedent~\cite{Spielberg2019}. Scaling it to robust physical deployment
across diverse whole-body tasks remains a useful target.

\paragraph*{Composition with generative and neural representations.}
Coupling to radiance fields and Gaussian splats is advancing quickly.
PAC-NeRF~\cite{Li2023PACNeRF} differentiates through simulation, whereas
PhysGaussian~\cite{Xie2024PhysGaussian} uses physics to animate captured
Gaussian scenes; PhysDreamer~\cite{Zhang2024PhysDreamer} also differentiates
through simulation and Gaussian rendering to fit dynamics to generated video.
These pipelines motivate joint diagnostics of simulation and rendering
sensitivities, including the effects of visibility, contact, and material
conditioning. Generative proposals also make it important to distinguish
physical consistency from agreement with a learned visual prior.

\section{Conclusion}
\label{sec:conclusion}

Differentiable simulation has moved in a few years from a specialized technique
for controlling simulations to a broad infrastructure layer under inverse problems in
graphics, vision, and robotics. Its literature is correspondingly heterogeneous,
and this report has argued that much of the heterogeneity is organized by a
question that is rarely asked explicitly, namely what is being differentiated.
Answering it separates the derivative's referent from its realization. The executed
computation, a discrete equation, or a continuous model determines the target
of differentiation and the meaning of exactness. Recording operations,
solving adjoint equations, or constructing custom operators determines how
that derivative is evaluated; memory and runtime additionally depend on state
recovery, solver budgets, and hardware. These distinctions complement the
choice of physical model: two engines can compute the same derivative
differently, or report gradients with different mathematical referents.

Exact differentiation of an executed
program does not require storing every intermediate value. Checkpointing and
recomputation provide general memory--time trade-offs; reversible state
recovery and solver-specific block adjoints exploit additional structure.
Reversibility can eliminate some stored states when reconstruction is stable,
although practical snapshots or irreversible operations retain storage costs.
Block mirroring can reduce storage per update without eliminating dependence
on solver depth. Claims of memory efficiency should therefore specify the
state-recovery strategy, scaling variable, and numerical reconstruction error,
rather than infer memory use from the realization label alone.

The field's implicit assumption that more exact gradients
produce better optimization survives only in a narrower regime than it is
usually granted. It can fail when other mechanisms limit progress in
contact-rich, long-horizon tasks. In our smooth, short-horizon chain experiment, designed to isolate
Mechanism~1, exactness was decisive at the tightest solver budgets, and how far up the budget its advantage persisted depended on where the target came from: one sweep for a target the same truncated solver generated, four to eight for a converged or external one. Which regime a given problem is in is currently left to
practitioners' intuition. Making that distinction measurable, through the diagnostics and
reporting standards we have proposed, would be a useful
contribution.

\appendix
\section{The diagnostic on the chain}
\label{app:gradcheck}

\begin{table}[t]
\centering
\footnotesize
\caption{The reporting checklist computed by the released diagnostic for the
chain under both referents. Item (i) is split into an implementation part and
a convergence part. The A1 zero in (i-b) denotes its finite-program target,
not equality with the converged-equation derivative. Item (iii) uses two query
budgets with wall-clock beside them. (v) uses this four-sweep configuration,
distinct from Figure~\ref{fig:horizon}, and is reported for A1 only because the A2 engine provides no
single-precision path; (vi) is the diagnostic's generic flat fraction, which is zero on this chain; the chain has no contact, so this says nothing about contact-induced flatness. Absence of contact does not rule out other
flat directions.}
\label{tab:gradcheck}
\begin{tabularx}{\columnwidth}{@{}Xcc@{}}
\toprule
Checklist item & \textbf{A1} & \textbf{A2} \\
\midrule
(i-a) implementation: relative error to an independent evaluation of the returned vector at the same state & as (ii) & $5\times10^{-16}$ \\
(i-b) convergence gap: relative error to the declared referent's target & 0 & 0.47 \\
(ii) relative error to executed finite differences & $2\times10^{-7}$ & 1.0 \\
(ii) cosine to executed finite differences & 1.000 & 0.184 \\
(iii) median remaining loss after 120 gradient steps & 0.27 & 0.76 \\
(iii) zeroth-order at equal queries, two forwards per gradient step: remaining loss / gradient wins & 0.45 / 100\% & 0.40 / 0\% \\
(iii) zeroth-order at equal queries, measured cost per gradient step (5.8 and 2.2 forwards): remaining loss / gradient wins & 0.36 / 95\% & 0.41 / 0\% \\
(iii) wall-clock, gradient run / zeroth-order at two forwards / at measured cost (s) & 11 / 3 / 10 & 2 / 2 / 3 \\
(iv) gradient-evaluation time / forward & 4.8 & 1.2 \\
(v) cosine float64/float32 at ${T=1}$, 20, 320 & 1.000, 1.000, 0.9998 & n/a \\
(vi) flat fraction of parameter space & 0\% & 0\% \\
\bottomrule
\end{tabularx}
\end{table}

Table~\ref{tab:gradcheck} is the output of the released diagnostic for the
chain of Figure~\ref{fig:bias} under both referents, on the same forward pass:
one implicit step at the moderately stiff setting (${k \approx 2\times10^{4}}$), four
Gauss--Seidel sweeps, with the squared distance to the rest configuration
as the loss. The
A1 engine returns the exact derivative of the four sweeps that ran; the A2
engine evaluates the equation-level adjoint at the iterate they reached. Item
(i) is reported in two parts that a single number would confuse. The
implementation part compares the returned vector with an independent
evaluation of the same object at the same state: for the A2 engine, the same
equation-level linearization at the truncated iterate, evaluated by a dense
forward-sensitivity solve. The reported agreement is of order $10^{-15}$. Attainable agreement
depends on conditioning, arithmetic, and the independent implementation. The convergence part compares the
returned vector with the same adjoint on a solve run to convergence, and the
$0.47$ it reports is the gap between the iterate the engine differentiated at
and the root its referent assumes, not an error in the engine. For an A1
engine the implementation part coincides with item (ii). Item (iii) uses the
normalized descent of Table~\ref{tab:optcost} against a $(1{+}1)$ evolution
strategy over twenty random stiffness fields, at two query budgets and with
wall-clock reported beside them: under the report's convention each gradient
step is charged two forward passes; under the measured budget it is charged
one forward plus the gradient-evaluation-to-forward time ratio of item (iv), a total of
$5.8$ forwards for the A1 engine here and $2.2$ for A2, since forward-mode
sensitivity through four sweeps costs several forwards while the adjoint
costs about one. This comparison uses different AD modes and is not an
intrinsic cost ratio between A1 and A2.

Item (ii) separates numerical agreement for A1 from the A1--A2 mismatch at
this truncated iterate: the A1 result agrees with finite differences to
order $10^{-7}$, whereas the A2 result has a cosine of $0.18$ to that
reference. Item (iii) is a harder instance than
Table~\ref{tab:optcost}, a single stiff step toward a target the simulator did
not generate; there the equation-level direction loses to the zeroth-order
baseline on every seed at either budget, while the executed derivative beats
it on every seed at the convention budget and on nineteen of twenty when the
baseline is given the evaluations the measured cost of a gradient step buys.
The wall-clock row qualifies the query budgets: the A1 gradient run
took eleven seconds against three for the baseline it beat at the convention
budget and ten for the baseline at the measured budget, so the convention
margin is partly bought with time, and the measured budget is, for this
engine, close to a comparison at matched running time as well.
That is the external-target case of Table~\ref{tab:optcost} at a single step. The same-$K$ columns of that table are the milder setting, and the diagnostic reports which of the two regimes a problem is in. Adding an
engine is one adapter exposing a loss, its returned gradient and a declared
referent; whatever an engine cannot provide is reported as not applicable
rather than scored.

{\small
\newcommand{\etalchar}[1]{$^{#1}$}

}

\end{document}